# CLOUD COMPUTING

## ARCHITECTURE AND APPLICATIONS

Edited by **Jaydip Sen**

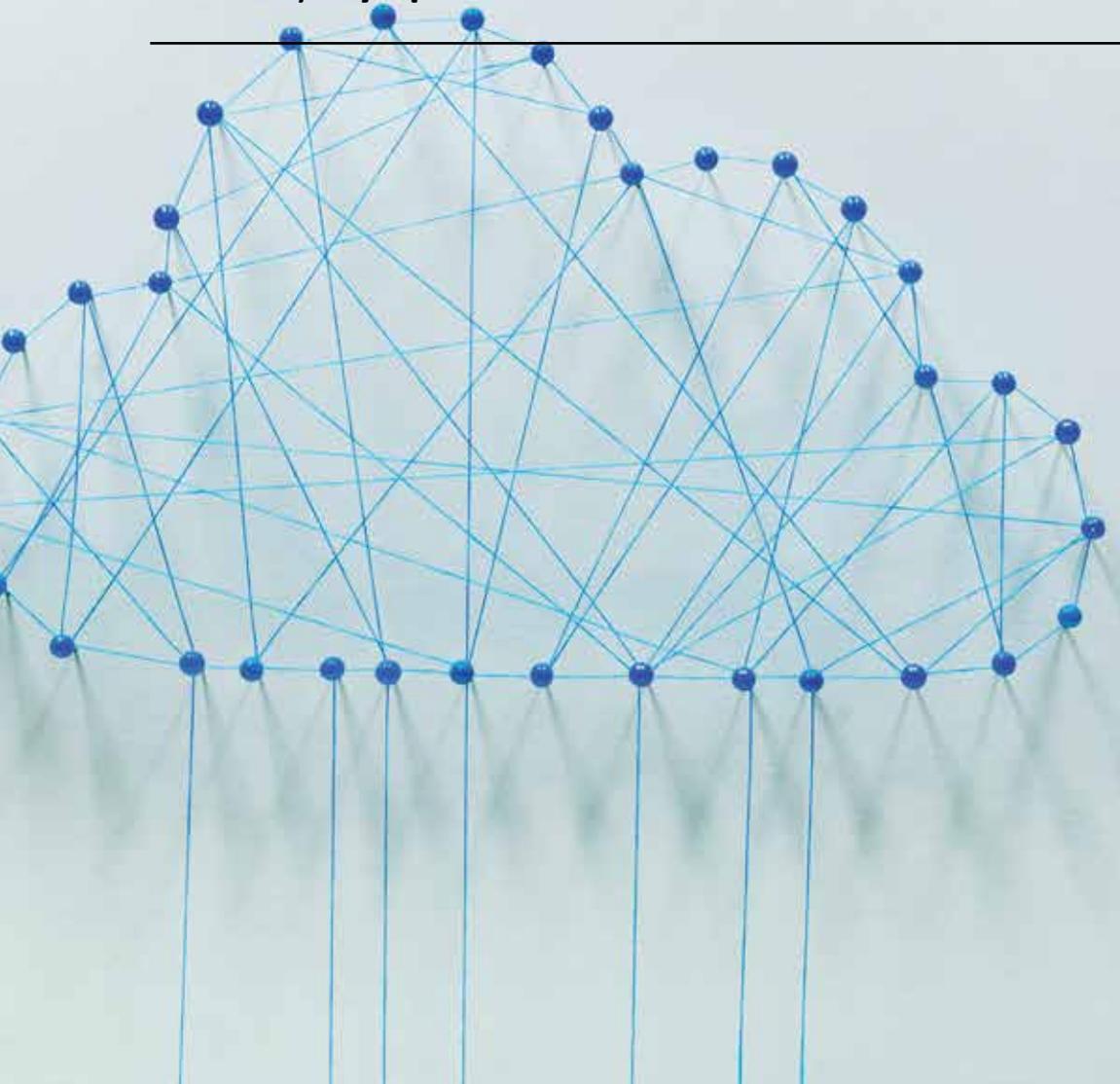

# CLOUD COMPUTING - ARCHITECTURE AND APPLICATIONS

Edited by **Jaydip Sen**

**Notice**

Statements and opinions expressed in the chapters are these of the individual contributors and not necessarily those of the editors or publisher. No responsibility is accepted for the accuracy of information contained in the published chapters. The publisher assumes no responsibility for any damage or injury to persons or property arising out of the use of any materials, instructions, methods or ideas contained in the book.

# Contents

# Preface

Cloud computing transforms the way information technology (IT) is consumed and managed, promising improved cost-efficiencies, accelerated innovation, faster time to market, and the enhanced ability to scale applications on demand. According to Gartner, while the hype grew exponentially during 2008 and continued since then, it is clear that there is a major shift toward the cloud computing model and the benefits may be substantial. With the advent of Internet of Things and big data, humongous amount of data is available today. This extraordinarily large volume of data needs to be analyzed for achieving valuable insights of business. Cloud computing is best suited for handling such volume of data, which is diverse in nature and needs real-time processing. In addition, the cloud services can be hired based on computing requirement without incurring any large fixed overhead cost of IT infrastructure. However, as the shape of the cloud computing is emerging and developing rapidly both conceptually and in real-world applications, the legal/contractual, economic, service quality, interoperability, security, and privacy issues still pose significant challenges. In other words, in spite of the several advantages that paradigm of cloud computing has brought along with it, there are several concerns and issues that need to be addressed before further ubiquitous adoption of it happens. First, in cloud computing, the user may not have the kind of control over his/her data or performance of his/her applications that he/she may need or the ability to audit or change the processes and policies under which he/she must work. Different parts of an application might be in different places in the cloud that can have an adverse impact on the performance of the application. Complying with regulations may be difficult especially when talking about cross-border issues—it should also be noted that regulations still need to be developed to take all aspects of cloud computing into account. It is quite natural that monitoring and maintenance is not as simple a task as compared to what it is for PCs sitting in the intranet. Second, the cloud customers may risk losing data by having them locked into proprietary formats and may lose control over their data since the tools for monitoring who is using them or who can view them are not always provided to the customers. Data loss is, therefore, a potentially real risk in some specific cloud deployments. Third, it may not be easy to tailor service-level agreements (SLAs) to the specific needs of a business. Compensation for downtime may be inadequate, and SLAs are unlikely to cover the concomitant damages. It is sensible to balance the cost of guaranteeing internal uptime against the advantages of opting for the cloud. Fourth, leveraging cost advantages may not always be possible. From the perspectives of the organizations, having little or no capital investment may actually have tax disadvantages. Finally, the standards are immature and insufficient for handling the rapidly changing and evolving technologies of cloud computing. Therefore, one cannot just move applications to the cloud and expect them to run efficiently. Finally, there are latency and performance issues since the Internet connections and the network links may add to latency or may put constraint on the available bandwidth. These challenges among many others provide opportunities for researchers and engineers to further extend the state of the art by developing algorithms and designing architectures that provide higher scalability, improved robustness, extended security and privacy, and enriched user experience while making more innovative applications available to the users.

The purpose of the book is to present some of the critical and innovative applications of cloud computing that are very relevant in today's world of computing where the ability to handle large volume of data in real time is a requirement while guaranteeing security, robustness, and efficiency in computation. With this goal, the book presents a collection of research work of some of the experts in the broad field of cloud computing who have expertise in specific domains such as designing new algorithms for achieving higher computing efficiency, developing innovative applications, implementing more efficient architecture, or testing performance of applications.

In Chapter 1 entitled "State-of-the-Art Antenna Technology for Cloud Radio Access Networks," Sethi et al. have proposed state-of-the-art antenna elements for implementation in cloud radio access network (RAN) radio frequency (RF) front end. The proposed antenna elements are lightweight and low cost, and they are easy to integrate with other microwave and millimeter wave circuits. The authors have also presented detailed design details of the antenna elements.

In Chapter 2 "Cloud Computing for Next-Generation Sequencing Data Analysis," Zhao et al. have discussed various issues in next-generation sequencing data analysis while emphasizing the need of cloud computing infrastructure for data management and analysis for such requirements.

In Chapter 3 "Green-Aware Virtual Machine Strategy in Sustainable Cloud Computing Environments," Wang et al. have presented an energy-aware virtual machine migration strategy for datacenters that are powered by sustainable energy sources. The authors have optimized the overall energy consumption in a datacenter by following an approach of statistical searching. Experimental evaluations have been made in a real-world test bed, which has demonstrated that green energy utilization, if it is done in an optimized manner, can substantially increase the overall revenue of a datacenter by substantially decreasing its operating expenditure.

In Chapter 4 "M-ary Optical Computing," Wang et al. have proposed a scheme of M-ary optical arithmetic operations for high-base numbers. By exploiting degenerate and non-degenerate four-wave mixing, optical computing operations have been demonstrated. The authors have claimed that M-ary optical computing using high-base numbers will facilitate advanced data management and superior network performance in next-generation communication and computing systems.

In Chapter 5 "Networking Solutions for Integrated Heterogeneous Wireless Ecosystem," Florea et al. discuss applications of cloud computing in a heterogeneous wireless network infrastructure. The authors have particularly described a centralized radio resource management framework in their test bed implementation and have also described how device-to-device communications can be achieved in the heterogeneous wireless network.

I am confident that the book will be very useful for researchers, engineers, graduate and doctoral students, and also practitioners in the field of cloud computing. It will also be a very interesting and exciting reading for faculty members of graduate schools and universities. However, since it is not a basic tutorial on cloud computing, it does not contain any chapter dealing with any detailed introductory information on any fundamental concept of cloud computing. It is expected that the readers have at least some basic knowledge on cloud computing architecture and issues related to its deployment. Some of the chapters in the book present in-depth cloud computing architecture-related theories and emerging trends in applications of cloud computing that might be useful to advanced readers and researchers in identifying their research directions and formulating problems to solve.

I express my sincere thanks to the authors of different chapters of the book, without whose invaluable contributions this project could not have been successfully completed. All the authors have been extremely cooperative in all phases of the project—submission of chapters, review, and the editing process. I would like to express my special thanks to Ms. Romina Skomersic of InTech Publishers for her support, encouragement, patience, and cooperation during the entire period of publication of the book. Ms. Romina Skomersic needs to be appreciated for her wonderful gesture and patience that she showed in spite of the delay that has affected the publication schedule of the book. I will be failing in my duty if I do not acknowledge the encouragement, motivation, and assistance that I received from my faculty colleagues in Calcutta Business School and Praxis Business School for this book project. Last but not least, I would like to thank my mother Krishna Sen, my wife Nalanda Sen, and my daughter Ritabrata Sen for being the major sources of my motivation and inspiration during the entire period of publication of this volume.

**Professor Jaydip Sen**
Department of Information Technology and Analytics
Praxis Business School,
Kolkata, India

# State-of-the-Art Antenna Technology for Cloud Radio Access Networks (C-RANs)

Waleed Tariq Sethi, Abdullah Alfakhri,

Muhammad Ahmad Ashraf, Amr G. Alasaad and

Saleh Alshebeili

Additional information is available at the end of the chapter

**Abstract**

The cloud radio access network (C-RAN) is one of the most efficient, low-cost, and energy-efficient radio access techniques proposed as a potential candidate for the implementation of next-generation (NGN) mobile base stations (BSs). A high-performance C-RAN requires an exceptional broadband radio frequency (RF) front end that cannot be guaranteed without remarkable antenna elements. In response, we present state-of-the-art antenna elements that are potential candidates for the implementation of the C-RAN's RF front end. We present an overview of C-RAN technology and different types of planar antennas operating at the future proposed fifth-generation (5G) bands that may include the following: (i) ultra-wide band (UWB) (3–12 GHz), (ii) 28/38 GHz, and (iii) 60-GHz radio. Further, we propose different planar antennas suitable for the implementation of C-RAN systems. We design, simulate, and optimize the proposed antennas according to the desired specifications covering the required frequency bands. The key design parameters are calculated, analyzed, and discussed. In our research work, the proposed antennas are lightweight, low-cost, and easy to integrate with other microwave and millimeter-wave (MMW) circuits. We also consider different implementation strategies that can be helpful in the execution of large-scale multiple-input multiple-output (MIMO) networks.

**Keywords:** 5G antennas, 28/38 GHz antennas, 60 GHz radio, cloud computing, green RAN

## 1. Introduction

Mobile data traffic has grown 4000-fold over the past 10 years, and it is projected to grow by more than 500 times over the next few years [1]. To cope with this large demand for mobile services, the mobile communication industry is currently developing fifth-generation (5G) mobile communication systems with the objective of providing pervasive, ubiquitous, always-connected broadband data communication. Many issues must be addressed to ensure 5G networks' superior performance, such as higher energy efficiency, higher system spectral efficiency, broadened network coverage, user coverage in hot spot and crowded areas, low latency, and better quality of service (QoS). Many key enabling technologies have been suggested for 5G, including millimetric wave transmission, massive multiple-input multiple-output (MIMO) networks, small cellular cells, heterogeneous network architectures, cloud radio access networks (C-RANs), and cognitive radio [2].

Cell densification (i.e., adding more cellular cells to the network) is proposed to increase the capacity, coverage area, and spectral efficiency of 5G networks [3]. However, a major drawback of cell densification is the signal interference between adjacent base stations (BSs), which may diminish the capacity gain. Considering the issues and challenges related to the cell densification in next-generation (NGN) mobile networks, mobile operators have proposed a cost-effective and energy-efficient solution that can provide optimized performance suitable for gigabits per second (Gbps) networks: the C-RAN [2].

The architecture for a general C-RAN system is shown in **Figure 1**. In a C-RAN, the baseband units (BBUs), which consume high power, are separated from the radio access units (also called remote radio heads (RRHs)). The idea in C-RANs is to move the BBUs to a central location (data center) and connect it to the radio access units via optical fibers [4]. At a remote site, the radio access unit (RRH) consisting of the antennas and radio frequency (RF) front end performs digital processing, digital-to-analog conversion, analog-to-digital conversion, power amplification, and signal filtering [2]. Moving the BBUs to a central location improves energy efficiency, since all the baseband processing are done at the central location, called the cloud. Furthermore, the C-RAN network architecture enables inter-BS operations. Coordinated multipoint processing (CoMP) techniques can mitigate the interference between BSs and provide better management and coordination. In addition, CoMP minimizes energy consumption in MIMO systems by enabling coordinated multipoint concepts.

The performance of a 5G RAN strongly relies on an efficient RF front-end transceiver section. In addition to the amplifiers' nonlinearity, in-phase and quadrature-phase imbalance, imperfect timing causing synchronization problems, and channel interference issues, the efficiency of the RF front end is strongly affected by the antenna design, RF impairments, antennas' special dispersion causing signal distortion, mutual coupling, and broadband antennas' nonlinear characteristics. Since wireless transmission involves antennas at both user terminals and the BSs, considerable attention is required in the designing and characterization of the antennas to achieve 5G networks' objectives.

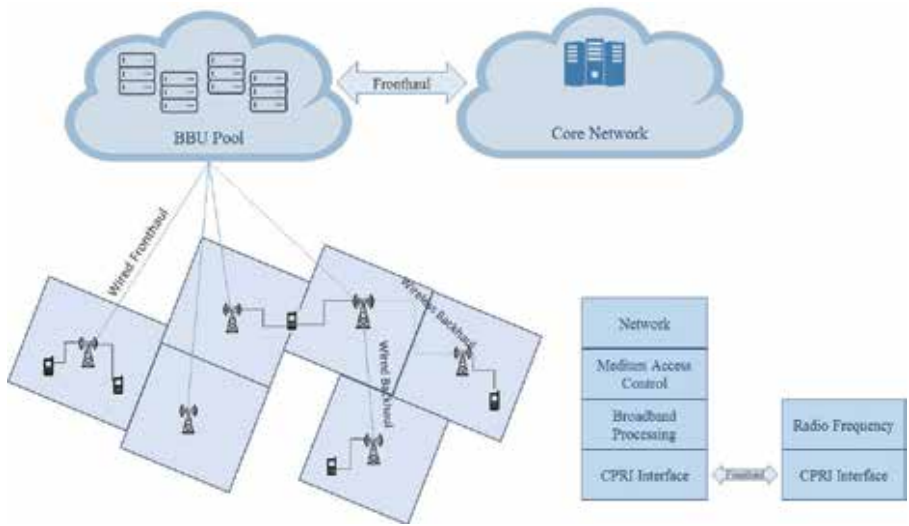

**Figure 1.** Architecture for general C-RAN system.

## 2. State-of-the-art antennas

In addition to the massive bandwidth of the antenna elements required for the implementation of NGN communication networks, several other key parameters such as gain, polarization purity, radiation efficiency, radiation patterns stability over the wide bandwidth, and minimum dispersion to the input signal are required to satisfy the systems' specifications. Antennas are classified into various types based on their key parameters/characteristics and targeted applications. In order to fulfill the ever-growing demand for wireless high-data-rate applications, ultra-wideband (UWB) technology has been considered a comprehensive solution for RF front-end design to enhance channel capacity [5]. UWB technology has drawn considerable attention, especially since the US Federal communication commission (FCC) authorized the use of the 3.1–10.6-GHz frequency band for commercial communication applications in 2002 [6]. Therefore, due to its huge bandwidth and unique feature of spectrum sharing, UWB can be considered one of the leading technologies for the implementation of NGN radio access networks, including the C-RAN.

Recent UWB antenna designs have focused on low cost, small size, and low-profile planar technology because of their ease of fabrication and their ability to be integrated with other components. The planar circuit development technique has brought monopole antennas with different shapes (polygonal, rectangular, triangular, square, trapezoidal, pentagonal, and hexagonal), circular, elliptical, etc.), which have been proposed as suitable candidates for UWB antenna systems [7, 8]. Mainly, the printed antennas consist of the planar radiator and ground plane etched oppositely onto the dielectric substrate of the printed circuit boards (PCBs). In some configurations, the ground plane may be coplanar with the radiator. The radiators can also be fed by a microstrip line or coaxial cable [9].

Numerous microstrip UWB antenna designs have been proposed [10–15]. For instance, a patch antenna has been designed as a rectangular radiator with two steps, a single slot on the patch, and a partial ground plane etched on the opposite side of the dielectric substrate. It provides a bandwidth of 3.2–12 GHz and a quasi-omni-directional radiation pattern [10]. Moreover, a clover-shaped microstrip patch antenna has been designed with a partial ground plane and a coaxial probe feed. The measured bandwidth of the antenna is 8.25 GHz with a gain of 3.20–4.00 dBi. In addition, it provides a stable radiation pattern over the entire operational bandwidth [11]. Another design is a printed circular disc monopole antenna fed by a microstrip line. The matching impedance bandwidth is from 2.78 to 9.78 GHz with an omni-directional radiation pattern, and it is suitable for integration with PCBs [12]. In addition, several elliptical shaped-based antennas have been designed. For example, three printed antennas have been designed starting from the elliptical shape, namely the elliptical patch antenna, its crescent-shaped variant, and the semielliptical patch [13].

Another type of printed antenna is the UWB-printed antenna fed by a coplanar waveguide (CPW). For example, one trapezoidal design and its modified form cover the entire UWB band (3.1–10.6 GHz) and have a notch for the IEEE 802.11a frequency band (5.15–5.825 GHz). The frequency notch function is obtained by inserting different slot shapes into the antenna. The notch frequency can be adjusted by varying the slot's length. The antennas show good radiation patterns as well as good gain flatness except in the IEEE 802.11a frequency band [14]. Another kind of radiating element considered suitable for phased arrays is the class of Vivaldi antennas, also known as quasi-end-fire nonresonant radiator or tapered slot antennas (TSAs) [15]. However, the element is normally fabricated by cutting a notch in a metal plate and backed by a quarter-wave cavity behind the feed point to improve its forward gain. A few examples of designed and fabricated UWB monopole and directional antennas are shown in **Figure 2**.

The next generation of ongoing wireless revolution with the growing demand of wireless facilities in mobiles, the millimeter-wave (MMW) frequency band appears to be a strong candidate for future radio access technologies. In addition to UWB, MMW technology (30–300 GHz) allows the developing of miniaturized and compact antenna sensors to be used in the RF front end, thus reducing the overall size of the system [16–19]. Compared to lower frequency signals, MMW signals can propagate over shorter distances due to their larger attenuation. Therefore, the development of MMW antennas with high gain performance for wireless access networks has attracted the interest of many researchers. In order to improve the spectral efficiency and exploit the benefits of spatial multiplexing, MMW antennas are expected to be used for large-scale MIMO (i.e., massive MIMO) systems. Therefore, it is important that improving a single parameter of an individual antenna will significantly improve the overall performance of a MIMO system, since each branch of the MIMO system will find at least one of them. The following are some well-known architectures and packaging techniques for improving the performance of MMW radios in terms of bandwidth, gain, and directivity.

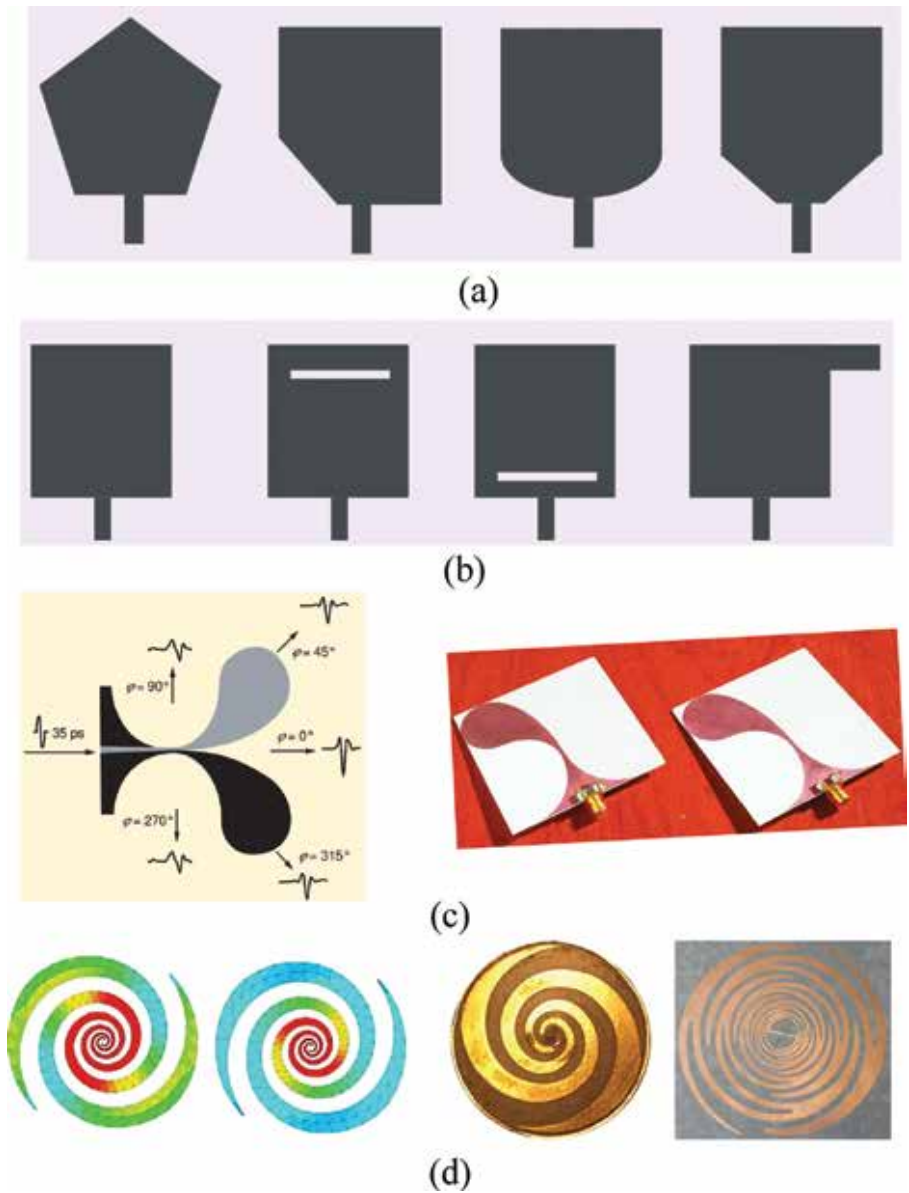

**Figure 2.** Different types of UWB antennas (a) conical antennas, (b) planar monopoles, (c) planar monopoles with band stop filters, (d) Vivaldi antennas and (e) spiral antennas [9].

Vettikalladi et al. [20] explained the significance of the addition of a superstrate on an aperture-coupled antenna at MMW frequencies. It can be seen that with the addition of a superstrate, the bandwidth is noted to be BW = 58.7–62.7 GHz (i.e., 6.7%) with a maximum gain of 14.9 dBi. A new dual-polarized horn antenna fed by a microstrip patch operating in the Ku band was proposed in Ref. [21]. The patch and horn were designed separately and then assembled together. The horn antenna had a reflection coefficient of less than −10 dB and a port isolation greater than 30 dB over 14.6–15.2 GHz and a gain of 12.34 dBi and 10-dB beamwidths of 87° and 88° at 14.9 GHz. The final structure had a gain of 12.34 dBi. The authors in Ref. [22] presented a wideband transition from CPW to horn antenna (CPWHA) based on the slot-coupled stacked-patch antenna technique, while those in Ref. [23] presented a wideband high-efficiency 60-GHz aperture superstrate antenna. It is found by measurement that by using a superstrate above the aperture antenna, we can improve the gain up to 13.1 dBi with a wide bandwidth of 15% and an estimated efficiency of 79%. This good result is higher than that of a classical 2 × 2 array, on an RT Duroid substrate, with a gain of 12 dBi and an efficiency of 60%. In Ref. [24], a new concept of a directive planar waveguide (WG) antenna array for the next generation of point-to-point E-band communication was presented. The proposed antenna consisted of two major parts: first, the array of Gaussian horn radiating elements, and second, the mixed feeding rectangular WG network. A high-gain slot-coupled circular patch antenna with a surface-mounted conical horn for MMW applications at 31 GHz was proposed in Ref. [25]. The design adopted microstrip/conical horn hybrid technology for a 6-dB enhancement over the conventional circular patch antenna. A novel micromachining approach for realizing 60-GHz foam aperture-coupled antennas was presented in Ref. [26]. The foam is indeed an ideal antenna substrate, as its electrical properties are close to those of the air. High-gain compact stacked multilayered Yagi designs were proposed and demonstrated in the V-band in Ref. [27]. This novel design showed for the first time an antenna array of Yagi elements in an MMW-stacked structure. The measured Yagi antenna attained an 11-dBi gain over a 4.2% bandwidth with a size of 6.5 × 6.5 × 3.4 mm². Efficient and high-gain aperture-coupled patch antenna arrays with superstrates at 60 GHz were studied and presented in Ref. [28]. The maximum measured gain of a 2 × 2 superstrate antenna array was 16 dBi with an efficiency of 63%, 4 dB higher than that of a classical 2 × 2 array at 60 GHz.

In order to meet recent requirements of designing large-scale MIMO wireless communication systems, conformal antenna technology enables the development of compact antenna arrays [29, 30]. Moreover, to create high-capacity MMW-MIMO systems, conformal antenna structures can be integrated with modern beam-switching technology, resulting in a data rate of several gigabytes. In cases in which the line-of-sight link is blocked, beam-switching technology allows the dynamic control of the antenna's main beam in order to find the received signal with the highest power. Several antenna arrays with beam-steering and beam-switching capabilities have been developed in Refs. [16, 31, 32]. Recently, a beam-switching conformal antenna array system operating at the 60-GHz mm-wave frequency band offering 1.5-GHz bandwidth was reported in Ref. [33]. However, the size of the developed switched beam array system was 31 × 46.4 mm² rounded around a cylinder with a radius of 25 mm. Second, the simulations resulted in a gain value of 16.6 dBi.

## 3. Design of planar antennas for C-RANs

Among various devices, C-RANs have a good number of highly efficient antennas integrated with their RF front ends. In order to make these antennas more adaptable and fulfill the telecom vendors' requirements, they are expected to operate in one of the future proposed 5G bands: (i) UWB (3–12 GHz), (ii) 28/38 GHz, or (iii) 57–64 GHz suggested for system design and implementation. In this work, we will design, model, and optimize state-of-the-art antenna elements operating over the proposed frequency bands that can be considered suitable candidates for the implementation of a C-RAN's RF front end. The proposed antennas are designed to be efficient, moderate in size, low-profile (i.e., can be implemented using conventional fabrication processes), and cost-effective. In addition, the designed antennas' key parameters such as reflection coefficient, gain, radiation pattern, dispersion effect, radiation efficiency, and pattern stability are calculated and optimized to achieve the C-RAN's high data rate requirements. The following are the design details of our proposed antenna elements suggested for the implementation of the C-RAN's front end.

## 4. UWB antenna element

In this section, we present antipodal tapered slot antennas (ATSAs) with elliptical strips termination modified with elliptical-shaped edge corrugations. The proposed corrugated antenna uses elliptical slots loading to improve the gain by up to 1.9 dB over an operational bandwidth of 0.8–12 GHz. It also improves the front-to-back lobe ratio. The designed ATSA exhibits minimum distortion to ultra-short pulses of 50 ps covering the 3–12-GHz frequency band.

### 4.1. Antenna design

The antenna element shown in **Figure 3(a)** is a traveling wave ATSA developed on Rogers 5880 substrate having dielectric constant $\varepsilon_r$ = 2.2 and thickness $h$ = 1.574 mm. The size of each antenna is 160 × 120 mm$^2$.

The ATSA-EC contains strip conductors on both sides of the substrate. In order to have impedance matching over a bandwidth of more than 10:1, the tapered slot is designed by following the guidelines in Ref. [35]. The exponential taper $C_g$ is used for the ground in order to achieve the broadband microstrip to parallel plate transition. The tapered curve $C_g$ is defined as

$$C_g = W_y - 1 + 0.1\,W_y\,e^{\alpha W_x} \tag{1}$$

where $\alpha$ is the rate of transition for the exponential curve defined as follows:

$$\alpha = \frac{1}{1.92\,W_x}\,ln\left(\frac{W_y + 0.1\,W_t}{0.1\,W_t}\right) \tag{2}$$

where $w_x$ is the $x$-directed length of the curve with $w_y$ and $w_t$ being the $y$-directed initial and final points, respectively. The variation of impedance bandwidth and radiation characteristics against different geometrical parameters of proposed ATSAs are analyzed by full-wave simulation software CST Microwave Studio [36]. **Table 1** presents the geometry of the ATSA, which results in 182% impedance bandwidth with the required radiation performance.

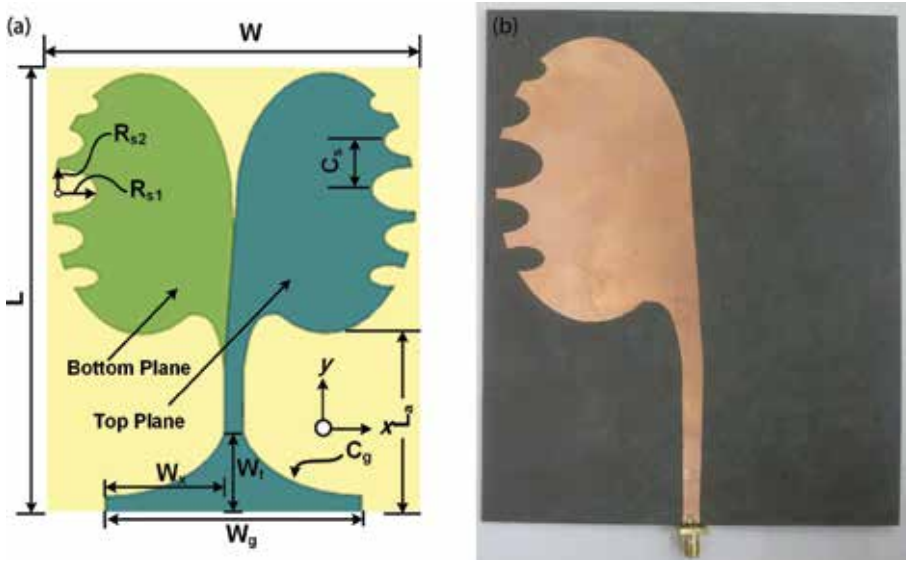

**Figure 3.** Layout diagram of (a) antipodal tapered slot antenna with elliptical-shaped edge corrugation (ATSA-EC) and (b) photograph of fabricated antenna [34].

| $R_1$ | $R_2$ | $D$ | $w_y$ | $w_x$ | $w_t$ | $w_o$ | $w_1$ |
|------|-------|-----|-------|-------|-------|-------|-------|
| **32.5** | 42.25 | 65 | 8 | 43.85 | 35 | 5.95 | 6.12 |

**Table 1.** Optimized geometrical dimensions (mm) of ATSA.

In order to improve the radiation characteristics, elliptical edge corrugations are applied to the ATSA, as shown in **Figure 3(a)**. At each edge of the antenna, unequal half-elliptical slots (UHESs) are loaded with the period $C_s$ = 17 mm. The largest UHES having minor axis and major axis radii $R_{s1}$ = 15 mm and $R_{s2}$ = 8 mm, respectively, is placed at the center of the elliptical fin. Conversely, the major axis radii of the other UHESs are decreased linearly by the factor $C_r$ = 0.7 having the constant ellipticity ratio $e_r$ = 0.533 = $R_{s2}/R_{s1}$.

### 4.2. Results and discussion

The photograph of the fabricated ATSAs is shown in **Figure 3(b)**. The measured return loss of the fabricated ATSA-EC is compared with the simulation results, as shown in **Figure 4**.

The simulation results are in good agreement with the measured performance. Generally, the radiation of an ATSA is a function of length, aperture width, and substrate thickness. The added inductance due to edge corrugation increases the electrical length of the antennas. The loading of the ATSA with UHES can suppress the surface current at both back edges, resulting in improved gain performance compared to un-slotted antenna gain. Similarly, the UHESs increase the effective length of the antenna, resulting in more directive beams in both the E- and H-planes. **Figure 5** presents the simulation results of the ATSAs' gain performance against various corrugation depths compared with un-corrugated ATSAs. The realized gain of the ATSA is found between 3 and 8.5 dBi over the 0.8–6 GHz frequency band. The edge corrugation arranges the current path to be parallel with the desired radiating current and opposite to the undesired surface current. The former enhances the gain, whereas the latter decreases the backward radiation. Therefore, the realized gain of the ATSA-EC is improved over the 0.8–6 GHz band by varying elliptical slots radii $R_{s1}$ and $R_{s2}$. Comparatively, better gain improvement is found for the ellipticity ratio $e_r = R_{s2}/R_{s1}$ less than 0.35, as depicted in **Figure 5**.

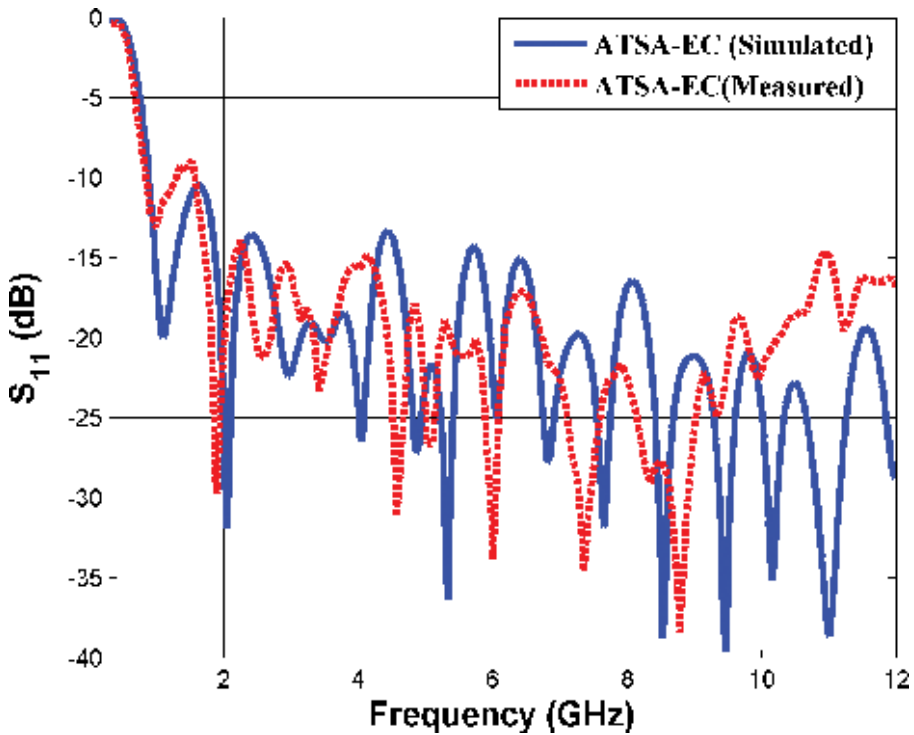

**Figure 4.** Measured return loss characteristics of fabricated ATSA-EC [34].

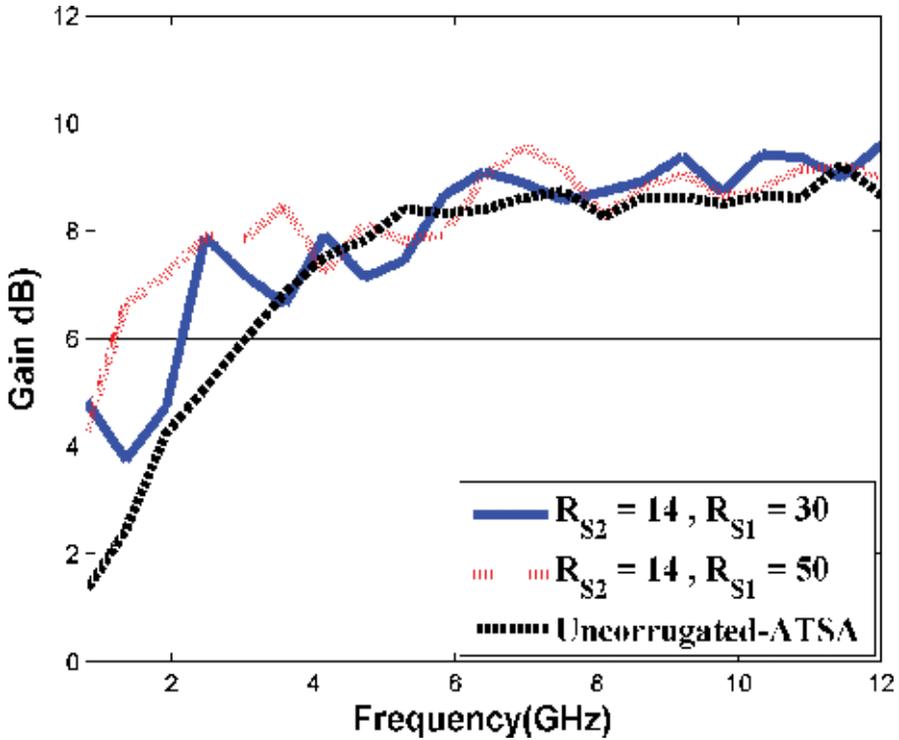

**Figure 5.** Simulated gain characteristics of ATSAs at different edge corrugation values [34].

The simulated time domain response of ATSAs when excited with pass-band Gaussian pulses covering the complete spectrum of operating frequency is shown in **Figure 6**. The received pulses are obtained by placing an *x*-oriented *E*-field probe 10 m along the broadside direction of the antenna. The FWHM of the transmitted pulse is 50 ps, while the received pulses preserve the Gaussian shape having a maximum FWHM of 56 ps related to the ATSA-EC with $R_{S1}$ = 30 mm and $R_{S2}$ = 18 mm. The FWHM of the ATSA without corrugation and ATSA-EC with $R_{S1}$ = 50 and $R_{S2}$ = 14 are found to be 58 and 59 ps, respectively. The fidelity factor is calculated according to the following relation [37].

$$Fidelity = \max_{\tau} \frac{\int_{-\infty}^{\infty} S_t(t) \, S_r(t-\tau) dt}{\sqrt{\int_{-\infty}^{\infty} \left| S_t(t) \right|^2 dt \left| S_r(t-\tau) \right|^2 dt}}, \qquad (3)$$

where $S_t(t)$ and $S_r(t)$ represent the transmitted and received time domain pulses, respectively. The fidelity factors for ATSA-EC at different edge corrugations are presented in **Table 2**.

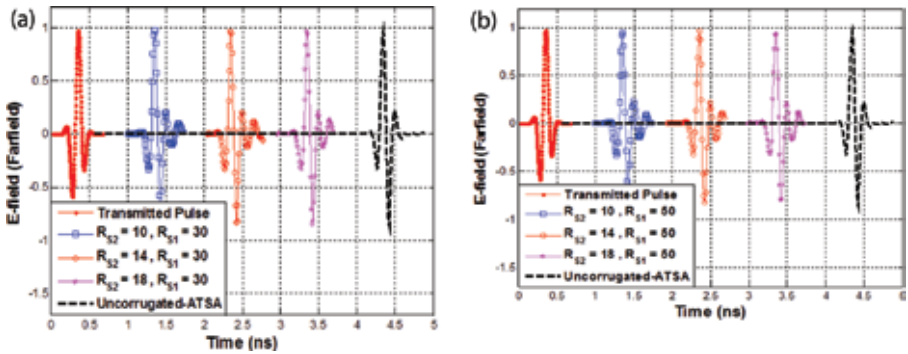

**Figure 6.** Simulated transmitted and received pulses on E-field probe for different edge corrugation parameters [34].

| $R_{s2}$ | 10 | 14 | 18 | 10 | 14 | 18 |
|---|---|---|---|---|---|---|
| $R_{s1}$ | 30 | 30 | 30 | 50 | 50 | 50 |
| **Fidelity** | 0.89 | 0.88 | 0.87 | 0.91 | 0.92 | 0.9 |

**Table 2.** Calculated fidelity factor for ATSA-EC at different edge corrugation.

## 5. Compact switched-beam MMW conformal antenna array system

This section presents a conformal ATSA system designed for future 5G wireless communications. A compact ($25 \times 30$-mm$^2$) ATSA element is designed presenting the reflection coefficient value less than −10 dB over a wide spectrum covering the 14.8–40-GHz frequency band. The MIMO antenna system is comprised of four ATSAs. Antenna elements are placed 90° apart from each other over a small cylinder having a 12-mm radius. The conformal ATSAs are loaded with a dielectric lens for gain enhancement. The optimized dimensions of the dielectric lens are obtained by several full-wave simulations resulting in a gain value of more than 20 dBi from 24 to 40 GHz. The proposed system presents four orthogonal independent beams switched at the angle of ±14° along the coordinate axis.

### 5.1. Broadband MMW ATSA design

The geometry of the ATSA antenna is shown in **Figure 7(a)**. The antenna was designed on RT/duroid® 5880 laminate having a dielectric constant of $\varepsilon_r = 2.2$ and a thickness of 0.254 mm. The top and the bottom plane conductors form an antipodal feed arrangement, enabling the ATSA antenna to exhibit excellent broadband characteristics. The tapered ground plane is obtained by cutting a half-ellipse with the radii of the major axis and minor axis, r1 and r2, respectively.

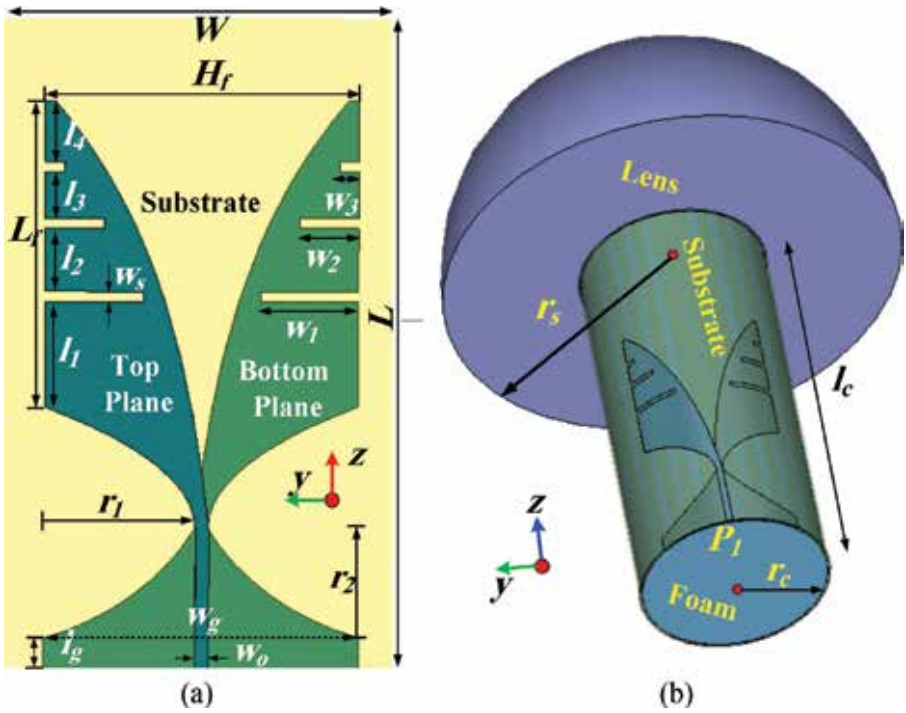

**Figure 7.** Geometry of ATSA, (a) rectangular configuration, (b) lens-loaded conformal configuration [38].

The top plane and the bottom plane conductors are tapered according to the guideline given in Ref. [17]. In order to reduce backward radiations, the linear corrugation is designed and optimized by executing several full-wave simulations using the CST Microwave Studio computer program. The optimized parameters of the ATSA are listed in **Table 3**.

| Parameter | $W$ | $L$ | $H_f$ | $L_f$ | $l_g$ | $W_0$ | $W_g$ | $r_1$ |
|---|---|---|---|---|---|---|---|---|
| Value (mm) | 24 | 32 | 16 | 15 | 1.63 | 0.67 | 16 | 7.6 |
| Parameter | $l_1$ | $l_2$ | $l_3$ | $l_4$ | $w_1$ | $w_2$ | $w_s$ | $r_2$ |
| Value (mm) | 5.15 | 3.07 | 2.27 | 2.99 | 5 | 3 | 0.5 | 5.36 |

**Table 3.** Optimized dimensional parameters of the proposed antenna.

The geometry of the proposed conformal ATSA is shown in **Figure 7(b)**. The ATSA element is designed over a low-thickness flexible substrate, which allows us to round the antenna element over a cylindrical surface. We selected a cylinder of foam material to preserve the electrical characteristics of the designed antenna. The radius of the foam cylinder is 12 mm. In order to enhance the gain, the ATSA is loaded with a half-spherical dielectric lens with a relative permittivity of $\varepsilon_r$ = 2.2 and optimized radius (rs) of 32 mm. The length (lc) of the conformal structure is the same as that of the nonconformal antenna, which is equal to 32 mm. The calculated performance parameters of the proposed antenna structures are discussed in the following subsections.

## 5.2. Results and discussion

The ATSA of the conventional rectangular shape was first optimized to exhibit a −10 dB bandwidth over a wider frequency spectrum. In the conformal antenna design, the radius of the foam cylinder is optimized for a minimum realizable value without compromising the electrical performance of the original ATSA. Finally, a dielectric lens is introduced toward the end-fire direction of the conformal ATSA, and S-parameter values are calculated. The S-parameter curves of the ATSAs having rectangular, conformal, and lens-loaded conformal configurations are presented in **Figure 8**, showing S11 values less than −10 dB from the 14.8 to 40-GHz frequency band. All three curves presenting the reflection coefficient performance of different ATSA configurations are found to be in close agreement with each other. The rectangular-shaped ATSA radiates toward the end-fire direction with E- and H-plane pattern symmetry. **Figure 9(a)** and **(b)** compare the normalized radiation patterns of ATSAs' rectangular, conformal, and lens-loaded configurations at 28 and 38 GHz, respectively. The conformal ATSA without a lens exhibits almost the same radiation pattern as the rectangular ATSA at both frequencies, except there is a shift of 4° in the H-plane pattern. The ATSA configurations without a lens present an average 3-dB beam-width of more than 40° in both planes. On the other hand, the lens-loaded ATSA finds a 14° shift in the H-plane beam with a 3-dB angular width of 12°. The introduction of the dielectric lens toward the end-fire direction enhances the gain of a conformal ATSA due to the focusing of the radiated field in space. The diameter of the dielectric lens is optimized by several full-wave simulations. The results presenting a parametric study of gain versus frequency against different diameters of dielectric lens are shown in **Figure 10**. A significant improvement (i.e., more than 10 dB) in gain parameters is observed by increasing the radius (rs) of the dielectric lens up to 32 mm. **Table 4** presents the comparison of the radiation characteristics among the three configurations of the ATSA .

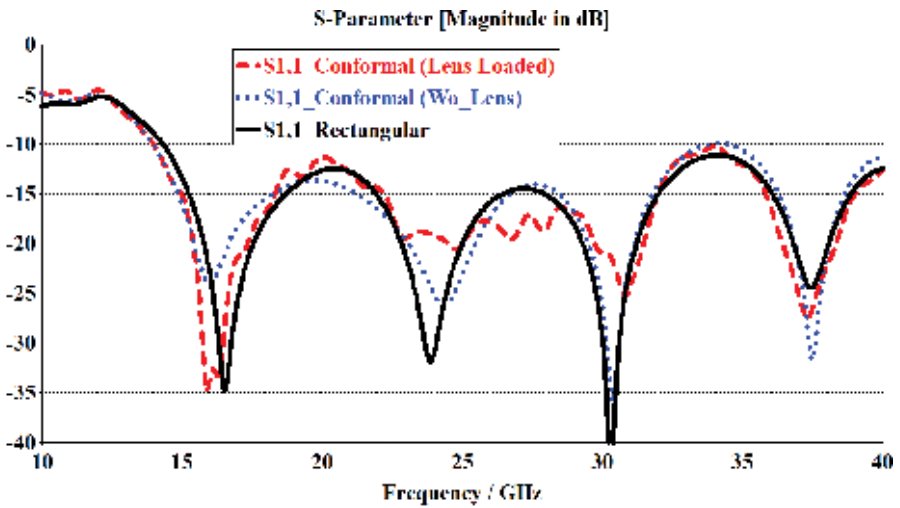

**Figure 8.** Reflection coefficient versus frequency of three different configurations of proposed ATSA [38].

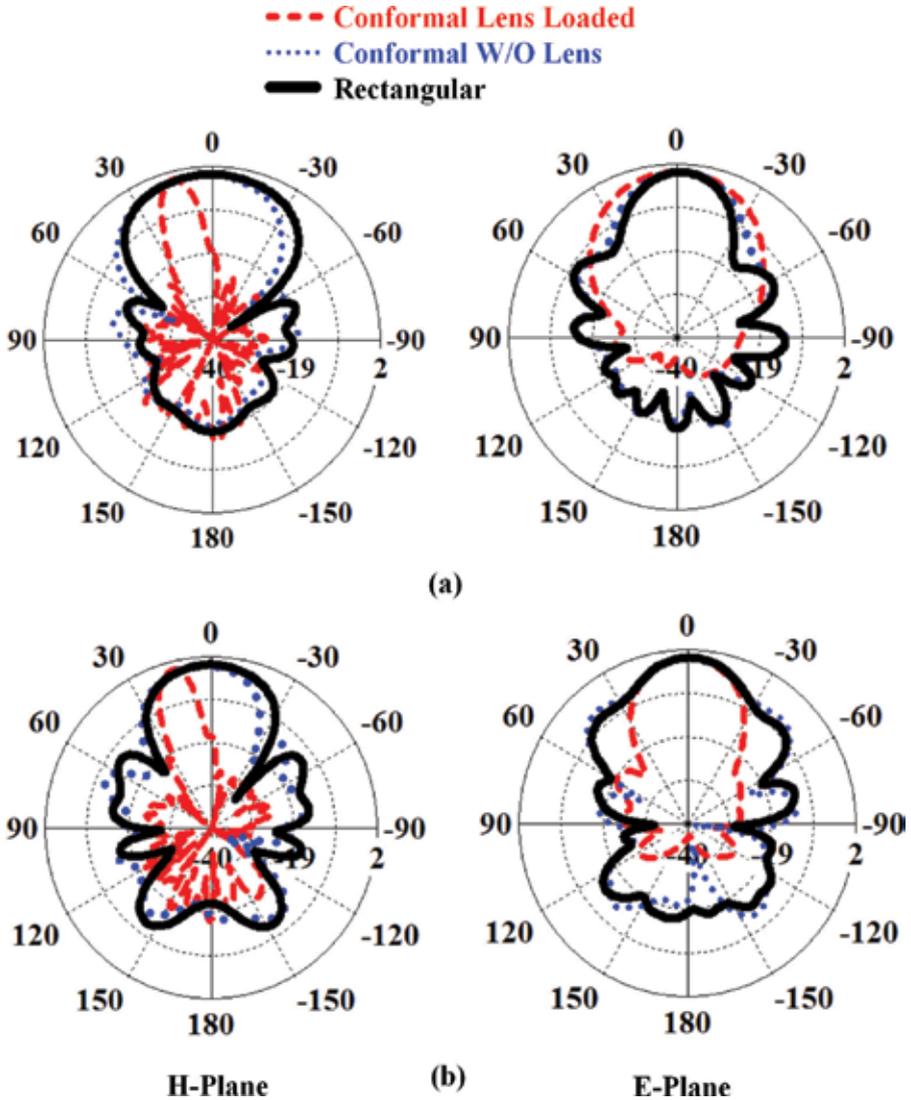

**Figure 9.** Copolarization radiation patterns in $\phi$= 0° (H-plane) and $\phi$ = 90° (E-plane) cuts at (a) 28 GHz and (b) 38 GHz [38].

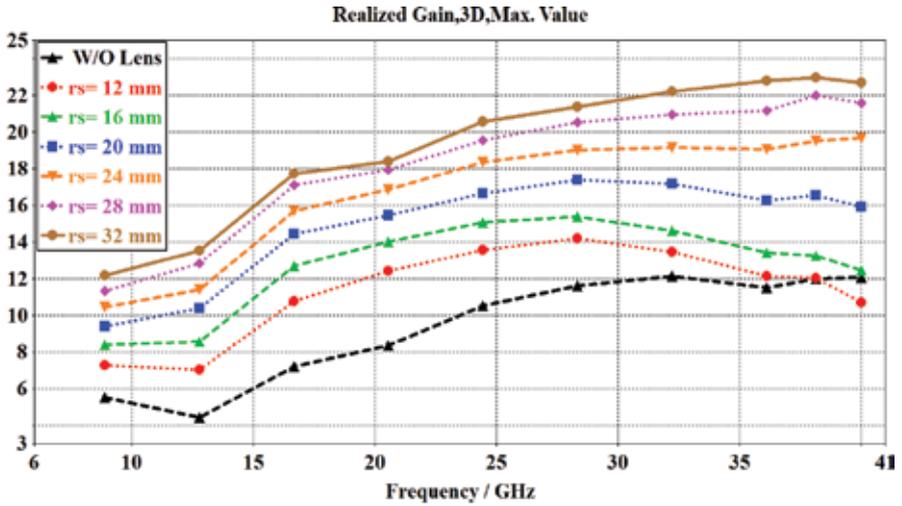

**Figure 10.** Gain versus frequency against different radii of dielectric lens loaded toward end-fire direction of ATSA [38].

|  | Peak gain (dBi) | Radiation efficiency (%) |
|---|---|---|
| Element 1 | 6.06 | 97 |
| Element 2 | 5.18 | 98 |
| Element 3 | 5.50 | 97 |

**Table 4.** Comparison of peak gain and radiation efficiency among the MIMO antenna systems.

### 5.3. Four-element beam-switched MIMO ATSA system

The geometry of the proposed conformal MIMO antenna system is shown in **Figure 11**. The four ATSA elements are placed along the ±*x*, *y* co-ordinate axis of a 12-mm-radius cylinder. The physical separation between the subsequent antenna elements is dc = 17 mm, which is 1.58 $\lambda$_0 at 28 GHz. The dimensions of conformal MIMO ATSAs are the same as mentioned previously. Antenna elements 1 and 2 are approximately perpendicular to each other, placed parallel to the *yz*-plane and *xz*-plane, respectively. Similarly, antenna elements 3 and 4 are placed opposite to antenna elements 1 and 2, respectively.

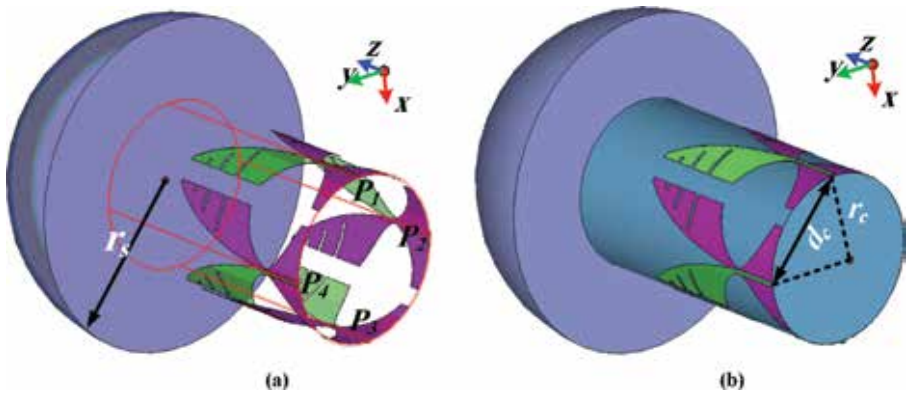

**Figure 11.** Geometry of four-element switched-beam MIMO antenna system, (a) transparent view and (b) solid view [38].

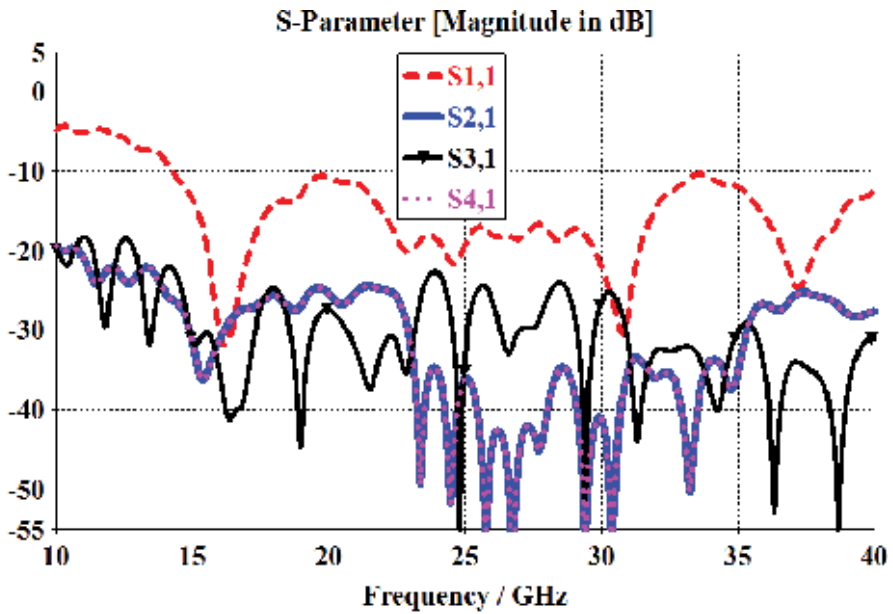

**Figure 12.** S-parameters of four-element beam-switched antenna system [38].

The four-element conformal ATSAs loaded with a dielectric lens are simulated to calculate and analyze S-parameters using the full-wave simulation program CST Microwave Studio [20]. The proposed conformal configuration of ATSAs does not affect the impedance band-width of the original design, since the mutual coupling between the antenna elements is below −20 dB over the complete spectrum, as shown in **Figure 12**. The minimum values of the reflection coefficient and mutual coupling values are below −10 and −20 dB, respectively, from 14.8 to 40 GHz. Moreover, better isolations are achieved between antenna elements 1 and 3. Considering the frequencies 28 and 38 GHz proposed for 5G wireless communications, the designed switched-beam ATSAs exhibit excellent S-parameter performance at those particular frequencies. Due to the symmetry of the designed configuration, only the S-parameter results for ATSA 1 are presented. The radiation performance of the proposed antenna configuration is calculated by exciting the particular element and terminating the other element with 50-Ω matched loads. Consider the proposed conformal configuration (see **Figure 11**) where antenna 1 and antenna 3 are placed opposite to each other at the −ve and +ve *x*-axis, respectively. Similarly, antenna 1 and antenna 3 are placed at the −ve and +ve y-axis, respectively. Since the dielectric lens is placed off-center with respect to each antenna element, exciting antenna 1 enables the focusing of electromagnetic energy toward the +ve $x$-axis and vice versa. The same phenomenon is observed between antenna 2 and antenna 4. The calculated H-plane and E-plane radiation patterns with their respective excitations at different ports are shown in **Figure 13**. For the excitation of antenna 2 and antenna 4, the H-plane pattern is calculated by taking theta ($\theta$) cut at the $\varphi = [\![90]\!]^\circ$o plane. Antenna 2 and antenna 4 find beams switched at $\theta = [\![12]\!]^\circ$o and $\theta = [\![-12]\!]^\circ$o, respectively. Similarly, the same radiation pattern results for antenna 1 and antenna 3 in the H-plane are observed at $\varphi = 0^\circ$o cut. The E-plane radiation pattern is calculated by taking phi ($\varphi$) cut at $\theta = [\![12]\!]^\circ$. It is worth noticing that the consecutive excitation of each individual port can result in four orthogonal switched beams placed 90° apart from each other.

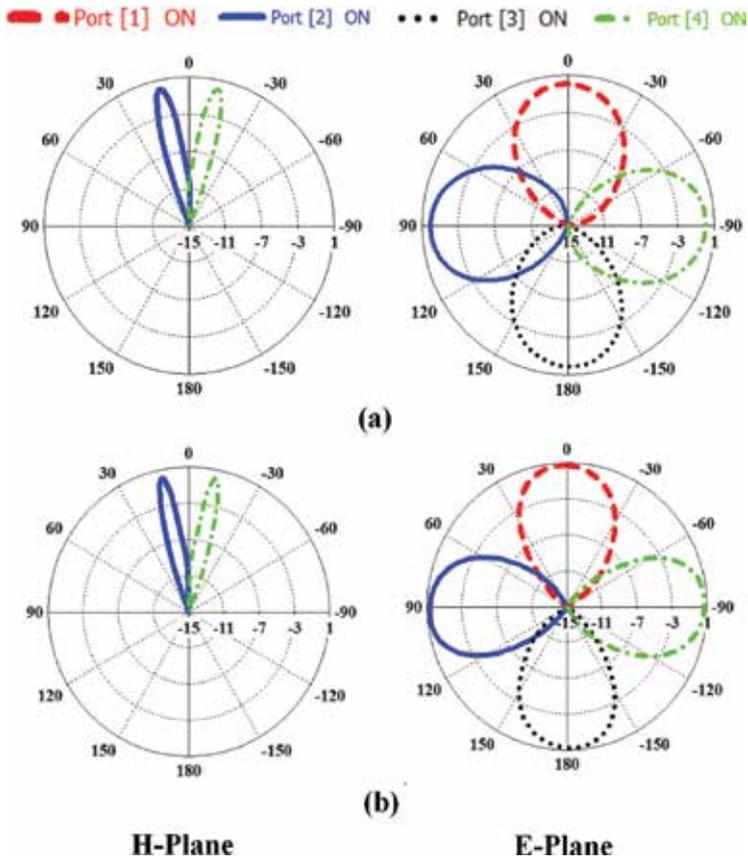

**Figure 13.** H-plane (antenna 2 and antenna 4) and E-plane (all antennas) switched beam radiation patterns at (a) 28 GHz and (b) 38 GHz. Beam switches at each co-ordinate axis (±*x*,*y*) displaced by 12° in H-plane [38].

## 6. 60-GHz radio or MMW antenna array for cloud computing

In this section, a simulated design of an MMW and array or 60-GHz radio band is presented for a cloud computing architecture. The design achieves the minimum requirements for the 60-GHz radio in terms of wide bandwidth (7 GHz at least, i.e., 57–64 GHz) and high gains (~8 dBi). The proposed antenna design is based on the aperture coupling technique [17, 38] that alleviates the problem of feedline and conductor losses while working at higher frequencies. The design consists of a multilayer structure with an aperture-coupled microstrip patch and a surface-mounted horn integrated on an FR4 substrate. The proposed antenna contributes an impedance bandwidth of 10.58% (58.9–65.25 GHz). The overall antenna gain and directivity are about 11.78 and 12.51 dBi, respectively. The antenna occupies an area of 7.14 mm × 7.14 mm × 4 mm with an estimated efficiency of 82%. In order to make the antenna more directive and to further increase the gain, a 2 × 2 and 4 × 4 array structure with a corporate feed network is introduced as well. The side lobe levels of the array designs are minimized, and the back radiations are reduced by utilizing a reflector at a $\lambda/4$ distance from the corporate feed network. The 2 × 2 array structure resulted in an improved gain of 15.3 dB with an efficiency of 83%, while the 4 × 4 array structure provided further gain improvement of 18.07 dB with 68.3% efficiency. The proposed design is modeled in CST Microwave Studio, and its results are verified using HFSS.

### 6.1. Wideband and high-gain aperture-coupled microstrip patch antenna (ACMPA) with mounted horn for MMW communication

The geometry of the single-element multilayer ACMPA integrated with a mounted horn antenna on an FR4 substrate is shown in **Figure 14**. The 3D exploded view shows the entire multilayer structure with relevant parameters. For the first and second layers, RT/duroid® 5880 Laminate having dielectric constant $\varepsilon_r$ = 2.2 and loss tangent 0.003 is used, while the third layer has an FR4 substrate with a dielectric constant of 4.3. A Rohacell foam is also placed on top of the FR4 substrate to assist the mounted horn antenna. The optimized dimensions of the horn antenna can be obtained from the guidelines listed in Ref. [40]. The conducting materials for the substrate have copper as an element with a thickness of $t$ = 0.0175 mm. The optimized dimensions of the proposed ACMPA are listed in **Table 5**.

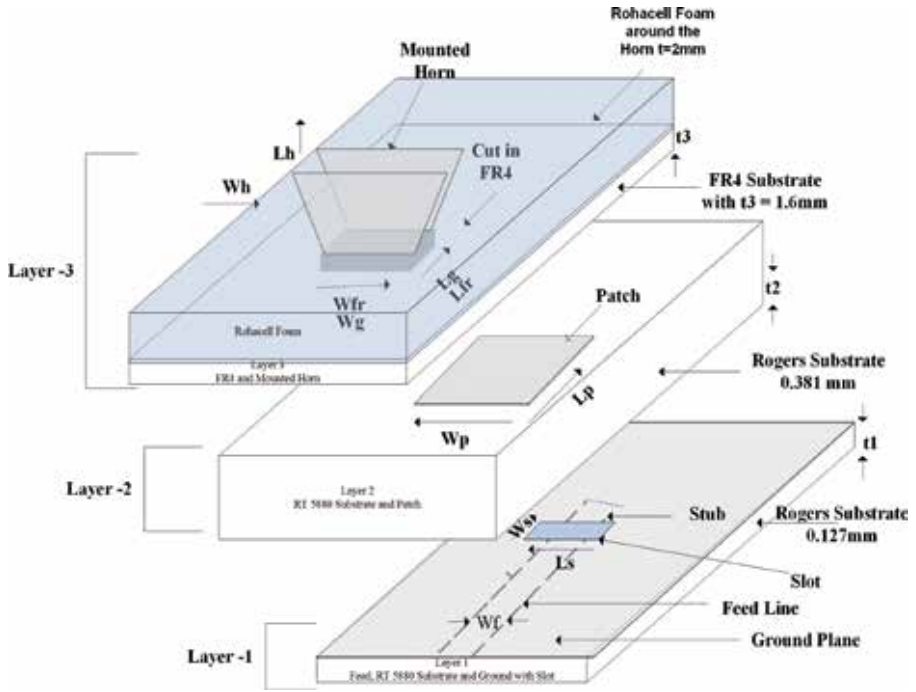

**Figure 14.** 3D exploded view of proposed ACMPA with relevant dimensions [39].

| Design | Antenna element | Dimensions/parameters (mm) |
|---|---|---|
| Layer I | Microstrip feed | Feed width, $W_f$ = 0.386 |
| | | Thickness, $t$ = 0.0175 |
| | | Stub length, $L_{fs}$ = 0.45 |
| | Substrate | RT Duroid 5880 |
| | | Length, $L$ = 30 |
| | | Width, $W$ = 30 |
| | Ground | Thickness, $t_1$ = 0.127 |
| | | Length $L$ = 30 |
| | | Width $W$ = 30 |
| | Rectangular slot | Slot length, $L_s$ = 1 |
| | | Slot width, $W_s$ = 0.2 |
| Layer II | Substrate | RT Duroid 5880 |
| | | Thickness, $t_2$ = 0.381 |
| | | Length, $L$ = 30 |

| Design | Antenna element | Dimensions/parameters (mm) |
|---|---|---|
| | | Width, $W$ = 30 |
| | Patch | Length, $L_p$ = 1.2 |
| | | Width, $W_p$ = 1.2 |
| Layer III | Substrate | FR 4 |
| | | Thickness, $t_3$ = 1.6 |
| | Cut in FR-4 | Length, $L_{fr}$ = 3 |
| | | Width, $W_{fr}$ = 4.25 |
| Horn | Horn dimensions | Horn length, $L_h$ = 7.14 |
| | | Horn width, $W_h$ = 7.14 |
| | | Waveguide length, $L_g$ = 3 |
| | | Waveguide width, $W_g$ = 4.25 |
| | | Thickness of metal horn, $t$ = 2 |
| Full structure | Total height | 4 |

**Table 5.** Optimized dimensional parameters of the proposed ACMPA.

## 6.2. Results and discussion

Antenna design simulation tools were used to optimize and verify the proposed ACMPA design. The return loss S11 parameters below the −10 dB resonance and the gain of the antenna are shown in **Figure 15**. The antenna achieves an impedance of 10.58% (58.9–65.25 GHz) with a gain and efficiency of 11.78 dB and 88%, respectively. Substrate and metallic losses were taken into account during simulations. **Figure 16(a)** and **(b)** show the E-plane and H-plane radiation patterns, simulated in CST and HFSS, of the proposed antenna, for the frequencies at 59, 62, and 65 GHz, respectively. Thus, for the multilayer structure at 62 GHz, the E-plane has a side lobe of level −5 dB, half-power beamwidth of 31°, and back radiation of −18.3 dB. The H-plane radiation pattern at 62 GHz has a side lobe of −13.2, half-power beamwidth of 69.8°, back radiation of −17dB, and cross-polarization level of >−30 dB.

## 6.3. 2 × 2 and 4 × 4 ACMPA array design

MMWs or 60-GHz radio bands offer wide bandwidth and higher gains for short-range communications. In order to fulfill these requirements, especially that of higher gain, the proposed ACMPA was optimized in terms of arrays (i.e., 2 × 2 and 4 × 4). Two factors are important when designing arrays: (1) array factor and (2) feeding network impedances. The theory behind the antenna array factor was utilized as explained in Ref. [41], where each antenna element is treated as an individual isotropic point source. Energy contributions from each point source are derived in the far field expressed as array factor (AF). For the feed network, one can select either single feed or parallel/corporate feed depending on the design requirements. For our proposed design, since we are working at higher (i.e., 60 GHz) bands, we opted for the corpo-rate feed network, as it would suppress further losses encountered during analysis. A general 2 × 2 and 4 × 4 corporate feed network is shown in **Figure 17(a)** and **(b)** with relevant imped-ances. Corporate feed networks are in general very versatile, as they offer power splits of $2n$ (i.e., $n = 2, 4, 8, 16, 32$, etc.) and control to the designer in terms of amplitude and phase selection of the individual feed element and its power division among the transmission lines. It is ideal for scanning phased arrays, shaped-beam arrays, and multibeam arrays [41]. The length and width of the transmission lines can be varied as per the requirement of the power division. The feed network consists of a 50-Ω transmission line and a 70.7-Ω quarter-wavelength transformer matched to a primary 50-Ω feeding line. For the 2 × 2 and 4 × 4 arrays with reflectors, the corporate feed network's impedance values can be retrieved from the guidelines provided in Ref. [42–44]. **Figure 18(a)** and **(b)** show the exploded view of the proposed array designs.

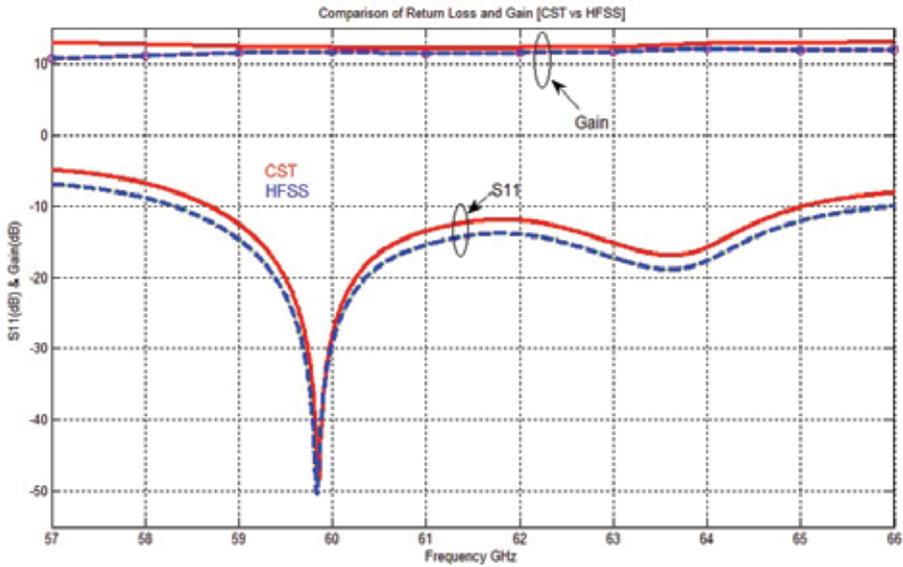

**Figure 15.** S-parameters and gain of proposed ACMPA.

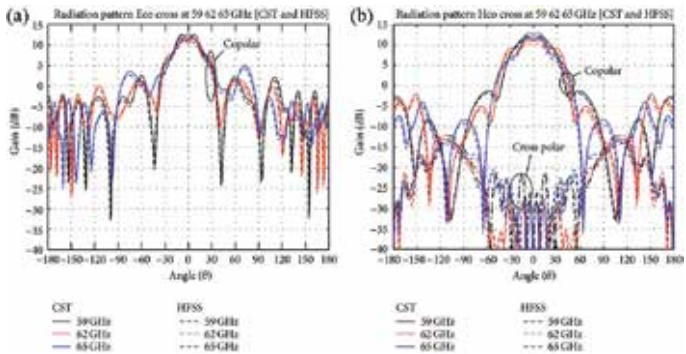

**Figure 16.** (a) Simulated E-plane radiation pattern at 59, 62, and 65 GHz and (b) simulated H-plane radiation pattern at 59, 62, and 65 GHz [39].

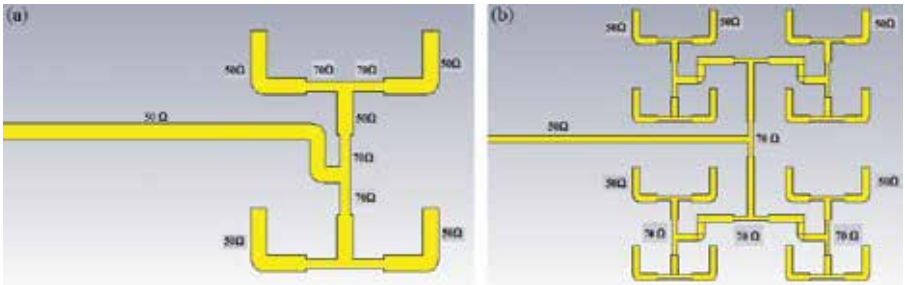

**Figure 17.** Corporate feed network (a) 2 × 2 array and (b) 4 × 4 array.

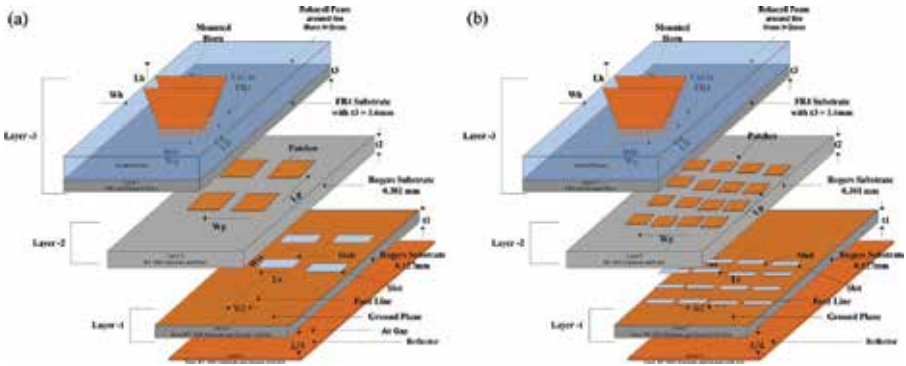

**Figure 18.** Exploded view (a) 2 × 2 array and (b) 4 × 4 array [39].

The E-plane and H-Plane radiation patterns for the 2 × 2 array and 4 × 4 arrays are shown in **Figures 19** and **20**. For the 2 × 2 array, it is observed that the E-plane at 62 GHz has a side lobe of level −13.7 dB, half-power beamwidth of 22.1°, and back radiation of −25.3 dB. The H-plane radiation pattern at 62 GHz has a side lobe of −9.1 dB, half-power beamwidth of 22.2°, back radiation of −21.8 dB, and cross-polarization level of >−30 dB. For the 4 × 4 array, the E-plane at 62 GHz has a side lobe of level −11.8 dB, half-power beamwidth of 13.6°, and back radiation of −23.07 dB. The H-plane radiation pattern at 62 GHz has a side lobe of −12.4, half-power beamwidth of 16.1°, and back radiation of −23.07 dB. **Table 6** shows the comparison of the improved gain from a single element to 2 × 2 and 4 × 4 arrays.

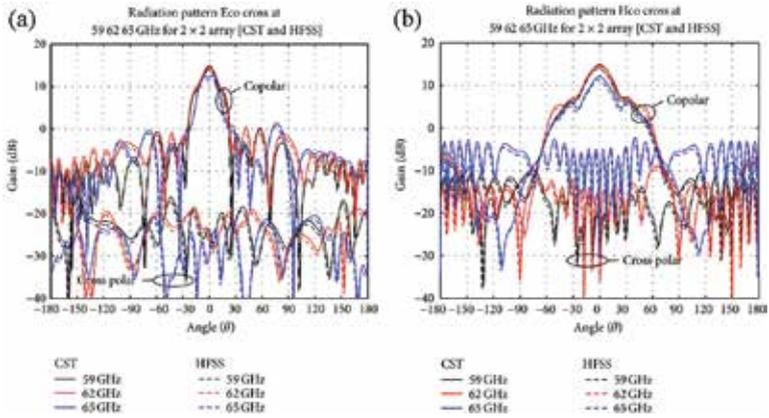

**Figure 19.** (a) Simulated E-plane radiation pattern at 59, 62, and 65 GHz and (b) simulated H-plane radiation pattern at 59, 62, and 65 GHz [39].

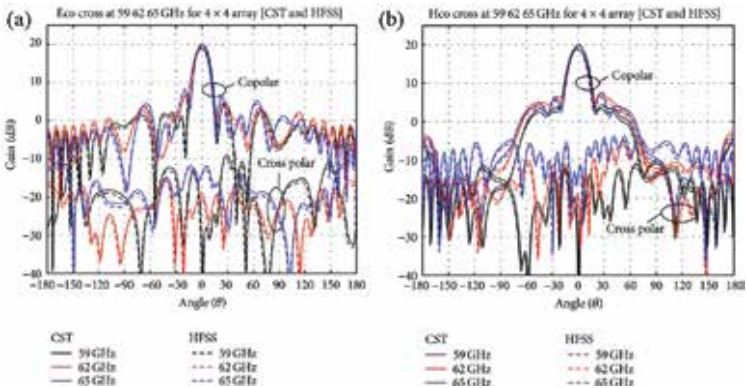

**Figure 20.** (a) Simulated E-plane radiation pattern at 59, 62, and 65 GHz and (b) simulated H-plane radiation pattern at 59, 62, and 65 GHz [39].

| Array/parameters | Single element | 2 × 2 array | 4 × 4 array |
|---|---|---|---|
| Bandwidth (%) | 10.58 | 10.55 | 10.51 |
| Gain (dB) | 11.78 | 15.3 | 18.07 |
| Efficiency (%) | 88 | 83 | 68.3 |

**Table 6.** Simulated results of single element, 2 × 2 and 4 × 4 array.

## Acknowledgements

This research is supported by King Abdul Aziz City for Science and Technology and Lockheed Martin (KACST-LM) University funding program.

## Author details

Waleed Tariq Sethi[1], Abdullah Alfakhri[2], Muhammad Ahmad Ashraf[1]*, Amr G. Alasaad[2] and Saleh Alshebeili[1]

*Address all correspondence to: mashraf@ksu.edu.sa

1 KACST Technology Innovation Center in Radio Frequency and Photonics for the e-Society (RFTONICS), King Saud University, Riyadh, Saudi Arabia

2 Center of Excellence in Future Telecommunication Applications, KACST, Riyadh, Saudi Arabia

# Cloud Computing for Next-Generation Sequencing Data Analysis

Shanrong Zhao, Kirk Watrous, Chi Zhang and Baohong Zhang

Additional information is available at the end of the chapter

**Abstract**

High-throughput next-generation sequencing (NGS) technologies have evolved rapidly and are reshaping the scope of genomics research. The substantial decrease in the cost of NGS techniques in the past decade has led to its rapid adoption in biological research and drug development. Genomics studies of large populations are producing a huge amount of data, giving rise to computational issues around the storage, transfer, and analysis of the data. Fortunately, cloud computing has recently emerged as a viable option to quickly and easily acquire the computational resources for large-scale NGS data analyses. Some cloud-based applications and resources have been developed specifically to address the computational challenges of working with very large volumes of data generated by NGS technology. In this chapter, we will review some cloud-based systems and solutions for NGS data analysis, discuss the practical hurdles and limitations in cloud computing, including data transfer and security, and share the lessons we learned from the implementation of Rainbow, a cloud-based tool for large-scale genome sequencing data analysis.

**Keywords:** next-generation sequencing, cloud computing, data analysis, workflow, pipeline

## 1. Introduction

High-throughput next-generation sequencing (NGS) technologies have evolved rapidly and are reshaping the scope of genomics research [1, 2] and drug development [3, 4]. The significant advances in NGS technologies, and consequently, the exponential expansion of biological data have created a huge gap between the computer capabilities and sequencing throughput [5, 6]. Technical improvements have greatly decreased the sequencing costs and, as a result, the size and number of datasets generated by large sequencing centers have increased dramatically. The lower cost also made the sequencing data more affordable to small and midsize research groups. As always, digging out the "treasure" from NGS data is the primary challenge in bioinformatics, which places unprecedented demands on big data storage and analysis. It is becoming increasingly daunting for small laboratories or even large institutions to establish and maintain their own computational infrastructures for large-scale NGS data analysis.

A promising solution to address this computational challenge is cloud computing [7–10], where CPU, memory, and storage are accessible in the form of virtual machines (VMs). In recent years, cloud computing has spread very rapidly for the supply of IT resources (hardware and software) of different nature, and is emerging as a viable option to quickly and easily acquire the computational resources for large-scale NGS data analyses. Cloud computing offers a wide selection of VMs with different hardware specifications and users can choose and configure these VMs to meet their computational demands. With the massive scale of users, cloud computing providers, such as Amazon, are continuously driving costs down, which in turn has led to the use of cloud computing for NGS data analyses attractive within the bioinformatics community. Despite the apparent benefits associated with cloud computing, there are also issues to be addressed. Data privacy and security are particularly important when managing sensitive data, such as the patients' information from clinical genomics studies [11].

The aim of this chapter is to describe the application of cloud computing in large-scale NGS data analysis and to help scientists to understand advantages and disadvantages of cloud computing, and to make an informed-choice on whether to perform NGS analysis on cloud services or to build the infrastructure themselves. It is organized as follows. First, we give a brief introduction to NGS technology, including DNA sequencing, RNA sequencing, and ChIP-sequencing. Secondly, we briefly introduce cloud computing and its services. Thirdly, we summarize and review publicly available cloud-based NGS tools and systems, with some particular emphasis on "Rainbow" [12], a cloud-based tool for large-scale whole-genome sequencing. Finally, we will discuss the challenges and remaining problems related to the full adoption of cloud computing in the NGS data analysis.

## 2. Next-generation sequencing

Next-generation sequencing [13] platforms allow researchers to ask virtually any question related to the genome, transcriptome, or epigenome of any organism. It has already profoundly changed the nature and scope of genomic research in the past few years. Sequencing methods differ primarily by how the DNA or RNA samples are obtained (e.g., organism, tissue type, normal vs. affected, experimental conditions) and by the data analysis options used. After the sequencing libraries are prepared, the actual sequencing processes are similar regardless of the method. There are a number of standard library preparation kits from different vendors that offer solutions for whole-genome sequencing (WGS), RNA sequencing (RNA-seq), targeted sequencing (such as exome sequencing, targeted RNA-seq or 16S sequencing), and detection of DNA methylation and protein-DNA interactions. As the number of NGS methods is constantly growing, a brief overview covering the most common methods is presented below.

## 2.1. Genomics

A breakthrough in NGS in the last decade has provided an unprecedented opportunity to investigate the contribution of genetic variation to health and disease [14]. WGS and whole-exome capture sequencing (WES) have emerged as compelling paradigms for routine clinical diagnosis, genetic risk prediction, and rare diseases [15–18]. WGS of tumors [19] is an unbiased approach that provides extensive genomic information about a tumor at the single nucleotide level as well as structural variations such as large insertions, genomic rearrangements, gross deletions, and duplications. Using low-coverage WGS of many individuals from diverse human populations, the 1000 Genomes Project [20] has characterized common variations and a considerable proportion of rare variations present in human genomes. With falling costs, it is now possible to sequence genomes of many individuals for association studies and other genomic analyses [21].

The WGS workflow is depicted in **Figure 1**. A human genome is fragmented into many short pieces that are sequenced by a sequencer. The sequencing step typically generates billions of short reads. All short reads are mapped to a reference genome, and genetic and structural variants can be identified with respect to the reference genome sequence. Human DNA is comprised of approximately 3 billion base pairs. 30× coverage sequencing of a personal genome will produce approximately 100 gigabytes (GB) of nucleotide bases, and its corresponding FASTQ file will be about 250 GB. For a WGS project consisting of 400 subjects, 100 terabytes of disk space is required to store the raw reads alone. Additional space is required for storing intermediate files generated during data analyses. Transferring and processing a dataset of such size would be extremely time-consuming and heavily computation-intensive and thus they pose huge practical challenges in data analyses.

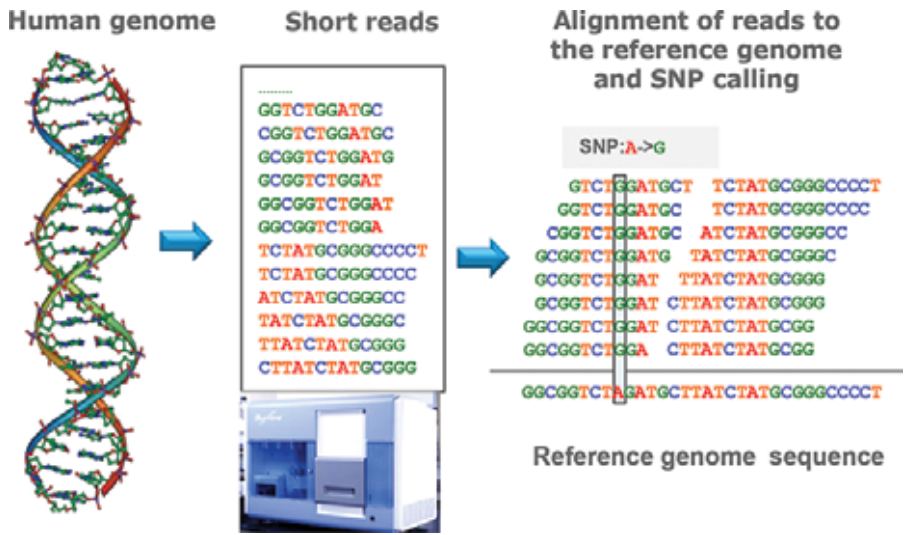

**Figure 1.** WGS workflow. A human genome is fragmented and sequenced, and billions of short reads are generated by a sequencer. All short reads are aligned to the reference genome and genetic variants are identified accordingly.

## 2.2. Transcriptomics

RNA sequencing (RNA-seq) has emerged as a powerful technology for transcriptome profiling [22–25]. It allows both quantification of known or predefined RNA transcripts and the capability to detect and quantify rare and novel transcripts within a sample. Compared to microarray, RNA-seq has a broader dynamic range, which allows for the detection of more differentially expressed genes with higher fold-change [26]. It is also superior in detecting low abundance transcripts, differentiating biologically critical isoforms, and allowing the identification of genetic variants. Not only RNA-seq can detect underlying genomic alterations at single-nucleotide resolution within expressed regions of the genome, but also it can quantify expression levels and capture variation not detected at the genomic level, including the expression of alternative transcripts. In the past decade, RNA-seq has become one of the most versatile applications of NGS technology and has revolutionized the researches on transcriptome [27]. As in WGS, RNA-seq generates vast number of short reads that must be computationally aligned or assembled to quantify expression of hundreds of thousands of RNA transcripts. Similar to DNA sequencing, the enormous data from large-scale RNA-seq studies poses a fundamental challenge for data management and analysis in a local environment [28–30]. Consequently, limited access to computational infrastructure and high-quality bioinformatics tools, and the demand for personnel skilled in data analysis and interpretation, remains a serious bottleneck for most researchers.

## 2.3. Epigenomics and protein-DNA interactions

While genomics involves the study of heritable or acquired alterations in the DNA sequence, epigenetics is the study of heritable changes in gene activity caused by mechanisms other than DNA sequence changes [31, 32]. Mechanisms of epigenetic activity include DNA methylation, histone modification and more. A focus in epigenetics is the study of cytosine methylation (5-mC) states across specific areas of regulation such as promotors or heterochromatin. Cytosine methylation can significantly modify temporal and spatial gene expression and chromatin remodelling. Two methylation sequencing methods are widely used: whole-genome bisulfite sequencing (WGBS) and reduced representation bisulfite sequencing (RRBS). With WGBS-seq, sodium bisulfite chemistry converts nonmethylated cytosines to uracils, which are then converted to thymines in the sequence reads. In RRBS-seq, DNA is digested with MspI—a restriction enzyme unaffected by methylation status. Fragments in the 100–150 bp size range are isolated to enrich CpG and promotor containing DNA regions. Sequencing libraries are then constructed using the standard NGS protocols.

ChIP-sequencing, also known as ChIP-seq [33, 34], is a method used to analyze protein interactions with DNA. ChIP-seq combines chromatin immunoprecipitation (ChIP) with massively parallel DNA sequencing to identify the binding sites of DNA-associated proteins. It can be used for genome-wide mapping of transcription factor binding sites. Protein-DNA interactions have a significant impact on many biological processes and disease states. The sequence reads generated by ChIP-seq are massive and need to be aligned to reference genome first, and then the locations of protein-DNA interactions are inferred based on enrichment of sequence reads along the genome.

## 3. Cloud computing

"Cloud Computing," by definition, refers to the on-demand delivery of IT resources and applications via the Internet with pay-as-you-go pricing. Cloud computing is a model for enabling ubiquitous, on-demand access to a shared pool of configurable computing resources (e.g., networks, servers, storage, applications, and services), which can be rapidly provisioned and released with minimal management effort. With cloud computing, you do not need to make large upfront investments in hardware and spend a lot of time on the heavy lifting of managing hardware. Instead, cloud computing providers such as Amazon Web Services own and maintain the network-connected hardware, and you can provision exactly the right type and size of computing resources you need. You can access as many resources as you need, almost instantly, and only pay for what you request and own. These computing resources include networks, servers, storage, applications, and services. There are several essential characteristics of the cloud computing model.

a. Rapid elasticity: you only allocate resources when you need them, and you are able to dynamically scale-up and -down your allocated resources as your needs change over time.

**b.** Pay-as-you-go: you only pay when you consume computing resources, and only pay for how much you consume.

**c.** On-demand self-service: the user can request and manage the computing resources without help from the service providers.

**d.** Cost-effective: classical computational infrastructure for data processing has become ineffective and difficult to easily scale-up and -down, and cloud computing is a viable and even a cheaper technology that enables large-scale data analysis.

Existing cloud-based services can be classified into four categories or layers (see **Figure 2**). The first one is Infrastructure as a Service (IaaS). This service model is offered in a computing infrastructure that includes servers (typically virtualized) with specific computational capability and/or storage. The user has full control on the operating system and applications that are deployed to, but with limited control, over the network settings. A good example is Amazon elastic compute cloud (EC2), which allows the user to request and manage virtual machines, and Amazon simple storage service (S3), which allows storing and accessing data. The second category of service is Platform as a Service (PaaS) in which the provider offers the customer the authority to create applications using developing tools supported by the provider. PaaS features rapid application development and good scalability, presenting usefulness in developing specific applications for big biological data analysis. Typically, the environment delivered by PaaS includes programming language environments, web servers, and databases. The Amazon Web Services (AWS) software development kit (AWS SDK) and Google App Engine are good examples of this service.

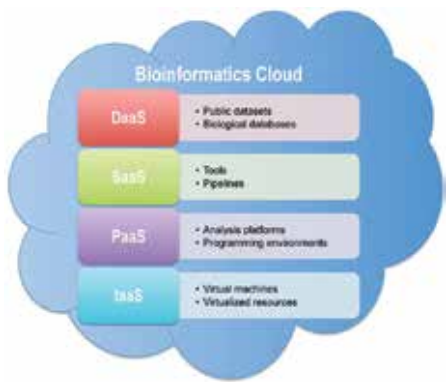

**Figure 2.** Illustration of cloud services [8]. Cloud-based services in are grouped into Data as a Service (DaaS), Software as a Service (SaaS), Platform as a Service (PaaS), and Infrastructure as a Service (IaaS).

The third service is Software as a Service (SaaS). SaaS eliminates the need for local installation and eases software maintenances and updates, providing up-to-date cloud-based services for data analysis. Customers do not manage the cloud infrastructure or network components, servers, operating systems, or storage, and can use the applications provided by the cloud provider. Most bioinformatics applications are open-source projects, and difficult to build, configure and maintain, primarily because they lack good documentation and have complex library dependencies. However, as all software applications are installed and configured within the VM, SaaS provides the perfect solution. The fourth layer is Data as a Service (DaaS). Bioinformatics clouds are heavily dependent on data, as data are fundamentally crucial for downstream analyses and knowledge discovery. Due to such unprecedented growth in biological data, delivering Data as a Service (DaaS) via the Internet is of utmost importance. DaaS enables dynamic data access and provides up-to-date data that are accessible by a wide range of devices that are connected over the web. AWS provides a centralized repository of public data sets, including GenBank [35], 1000 Genomes [20], encyclopedia of DNA elements [36], etc., and all public datasets are delivered as services in AWS and thus can be seamlessly integrated into cloud-based applications.

## 4. NGS data analysis on cloud computing

In recent years, cloud computing offers an alternative approach to quickly and easily acquire computational resources for large-scale NGS data analysis. As a result, many cloud-based services and bioinformatics platforms (see **Table 1**), applications (see **Table 2**), and resources have been developed to address the specific challenges of working with the large volumes of data generated by NGS technology. Cloud computing has created new possibilities to analyze NGS data at reasonable costs, especially for laboratories lacking a dedicated bioinformatics infrastructure. From the perspective of end users, there are three options to analyze NGS data on cloud computing (**Tables 1** and **2**). First, commercial systems such as DNAnexus and Seven Bridges can be used out of box to carry out the entire NGS data analysis. Second, commercial or open bioinformatics platforms are further customized to meet users' computational needs. Third, open-source tools (**Table 2**) can be deployed into the cloud for any customized data analysis.

## 4.1. Commercial services

Commercial services provide the users with well-established pipelines, user interfaces, and even application programming interfaces (APIs), and can reduce the time and effort required for setting up pipelines for NGS data analysis. For instance, DNAnexus and Seven Bridges offer various customizable NGS data analysis pipelines. In addition, DNAnexus also provides software that can directly upload the sequencing data produced. BaseSpace, launched by Illumina in collaboration with Amazon, is a genomics cloud computing platform that provides NGS data analysis services, such as mapping, de novo assembling, small RNA analysis, library quality control (QC), metagenomics analysis, and data storage. It is designed to bring simplified data management and analytical sequencing tools directly to researchers in a user-friendly manner. BaseSpace provides flexibility and convenience with an array of tools, significantly simplifying the process of yielding meaningful results from NGS data. Bina Technologies offers a service that is composed of a specialized hardware called Bina Box and a cloud service. Bina Box can employ accelerated BWA [54] and GATK [55] for data analyses.

Some commercial services also provide APIs with which the users can manage their jobs or build their own applications. Variant calling on datasets of hundreds or thousands of genomes is time-consuming, expensive, and not easily reproducible given the myriad components of a variant calling pipeline. To address these challenges, the Mercury [47] analysis pipeline was developed on top of the DNAnexus platform. It integrates multiple sequence analysis components across various computational steps, from obtaining patient samples to providing a fully annotated list of variant sites for clinical applications. Mercury is an automated, flexible, and extensible analysis workflow that provides accurate and reproducible genomic results at scales ranging from individuals to large cohorts.

| Name | URL | Description |
|------|-----|-------------|
| BaseSpace | http://basespace.illumina.com | Commercial services |
| Bina | http://www.bina.com/ | Commercial services |
| DNAnexus | http://www.dnanexus.com | Commercial services |
| SevenBridges | http://www.sbgenomics.com | Commercial services |
| Eoulsan | http://transcriptome.ens.fr/eoulsan | Cloud-based platform |
| CloVR | http://clovr.org | Automated sequence analysis |
| Cloud BioLinux | http://cloudbiolinux.org | Virtual machine for bioinformatics cloud computing |
| CloudMan | https://wiki.galaxyproject.org/CloudMan | Cloud-scale Galaxy |
| Globus Genomics | https://www.globus.org/genomics | Cloud-based bioinformatics workflow for NGS analyses |
| GenomeCloud | http://www.genome-cloud.com/ | Analyze genome data |
| COSMOS | http://cosmos.hms.harvard.edu/ | Workflow management system |

**Table 1.** Cloud computing services and platforms.

| Software | URL | Description | Reference |
|---|---|---|---|
| Atlas2 | http://atlas2cloud.sourceforge.net | Genome analysis | [37] |
| CloudAligner | http://cloudaligner.sourceforge.net | Reads mapping | [38] |
| CloudBurst | http://cloudburst-bio.sourceforge.net | Reads mapping | [39] |
| Crossbow | http://bowtie-bio.sourceforge.net/crossbow | Read mapping/SNP call | [40, 41] |
| FX | http://fx.gmi.ac.kr | RNA-seq | [42] |
| Myrna | http://bowtie-bio.sourceforge.net/myrna | RNA-seq | [43] |
| Stormbow | http://s3.amazonaws.com/jnj_stormbow/index.html | RNA-seq | [44] |
| STORMSeq | http://www.stormseq.org/ | Read mapping | [45] |
| GenomeKey | https://github.com/LPM-HMS/GenomeKey | Whole genome analysis | [46] |
| Mercury | https://www.hgsc.bcm.edu/software/mercury | Workflow for genomic analysis | [47] |
| Rainbow | http://s3.amazonaws.com/jnj_rainbow/index.html | Whole genome analysis | [12] |
| PaekRanger | http://ranger.sourceforge.net | ChIP-seq | [48] |
| VAT | http://vat.gersteinlab.org | Variant annotation | [49] |
| YunBe | http://tinyurl.com/yunbedownload | Gene set analysis | [50] |
| BioVLAB-MMIA-NGS | https://sites.google.com/site/biovlab/ | microRNA-mRNA integrated analysis | [51, 52] |
| SURPI | http://chiulab.ucsf.edu/surpi/ | Pathogen identification | [53] |

**Table 2.** Open-source tools for cloud computing.

Although a number of cloud-based pipelines are available for analyses of sequencing data in massively parallel DNA sequencing, the majority of them can only identify variants within a single sample. While this approach has enough power for detecting variants in high-coverage sequencing, it performs worse than multiple-sample calling when applied to low-coverage sequencing data. To this end, another scalable DNAnexus-based pipeline for joint variant calling in large samples was developed and deployed to the Amazon cloud. Using this pipeline, Shringarpure et al. [21] identified 68.3 million variants in 2535 samples from Phase 3 of the 1000 Genomes Project. By performing the variant calling in a parallel manner, the data was processed within 5 days at a compute cost of just $7.33 per sample (a total cost of $18,590 for completed jobs and $21,805 for all jobs).

Despite their merits, these commercial services also have several disadvantages. First, the use of a commercial service requires extra expenses for the convenience of NGS data analysis and user-friendly interfaces. Second, compared to open-source tools on the cloud, the commercial services are less customizable with respect to the use of the services and access to the cloud service. Although DNAnexus and Seven Bridges provide APIs to access and control their cloud services, their functionalities are restricted and the users have to request the service provider to set up new application software on their cloud services.

### 4.2. Bioinformatics platforms

CloudBioLinux [56] is a publicly accessible virtual machine (VM) that is based on an Ubuntu Linux distribution and is available to all Amazon EC2 users for free. It comes with a user-friendly graphical user interface (GUI), along with over 135 preinstalled bioinformatics packages. CloudBioLinux instances provide an excellent environment for users to become familiar with BioLinux and cloud computing. Galaxy is an open, web-based platform for data-intensive biomedical research. Whether on the free public server or your own instance, you can perform, reproduce, and share the entire data analyses. Galaxy Cloud [57], a cloud-based Galaxy platform for the analysis of data at a large scale, is the most used platform for bioinformatics. Unlike commercial software service solutions, users can customize their deployment and have complete control over their instances and associated data. Currently, a public Galaxy Cloud deployment, called CloudMan [57], is provided on the AWS cloud, enables bioinformatics researchers to easily deploy, customize, and share their cloud analysis environment, including data, tools, and configurations. By combining three platforms, CloudBioLinux, CloudMan, and Galaxy, into a cohesive unit, researchers can gain access to more than 135 preconfigured bioinformatics tools and gigabytes of reference genomes on top of the flexible cloud computing infrastructure [56].

Although Galaxy cloud provides a convenient platform for researchers, challenges remain in moving large amounts of data reliably and efficiently and in adding domain-specific tools for specific analyses. To address these challenges, Globus Genomics [58, 59] was developed at the Computation Institute (CI), a joint institute between the University of Chicago and Argonne National Laboratory. Globus Genomics is a cloud-based integrated solution for NGS data analysis. It extends the existing Galaxy workflow system by adding data management capabilities for transferring large quantities of data efficiently and reliably (via Globus Transfer), domain-specific analyses tools preconfigured for immediate use by researchers (via user-specific tools integration), automatic deployment on cloud for on-demand resource allocation and pay-as-you-go pricing (via Globus Provision), and a cloud provisioning tool for auto-scaling (via HTCondor scheduler). Genome sequencing is notoriously data-intensive, and Globus Transfer [59] is designed for fast and secure movement of large amounts of data. Setting up a production instance of Galaxy is a nontrivial task that involves a number of manual installation and configuration steps for both the platform and any dependent software packages—steps that can be both error-prone and time-consuming. Globus Provision addresses the above issues by providing on-demand cluster reconfiguration, user-specific node provisioning, and automatic instance deployment on Amazon EC2.

GenomeCloud (http://www.genome-cloud.com/) is another on-demand Galaxy cloud. It was built upon Galaxy, and is composed of g-Analysis, g-Cluster, g-Storage, and g-Insight services, providing convenient services to the researchers and other users. GenomeCloud is a complete and integrated platform for analyzing genome data to the interpretation of analysis results. It combines the idea of cloud computing with bioinformatics to generate an integrated solution for data storage and sharing, database management, continuously updated computing and analysis tools, and security. GenomeCloud is designed to help researchers perform bioinformatics tasks more easily, as well as to support laboratories without the computational resources to conduct research without hurdles.

### 4.3. Open-source tools

The development of tools supporting NGS data analysis with cloud computing has recently become popular in the open-source community [8]. Currently, there are many pipelines and workflows that support cloud computing (**Table 2**). Despite their advantages in cost and flexibility, open-source tools on the cloud also have substantial drawbacks. The users are responsible for designing/setting up the entire analysis pipeline, the data management and hardware configuration, such as CPUs, memory, storage, and security. Quite often, the users have to overcome a laborious series of trial and error before setting up the proper configuration. Although several tools have been developed to date, in most cases, their cloud computing support is incomplete and their functionality is under-developed. Here, we will briefly report some existing bioinformatics tools and then describe Rainbow, a cloud-based tool for large-scale WGS data analysis, in detail in the next section.

CloudAligner [38] and CloudBurst [39] are parallel read mapping algorithms optimized for mapping short reads to human and other reference genomes and can produce alignments for a variety of downstream biological analyses including SNP discovery, genotyping, and personal genomics. Crossbow is a Hadoop- [60] based tool that combines the speed of the short read aligner bowtie [61], with the accuracy of the SNP caller SOAPsnp [62] to perform alignment and SNP detection from WGS data in parallel. Scalable tools for open-source read mapping (STORMseq) [45] is a graphical interface cloud computing solution that performs read mapping, read cleaning, variant calling, and annotation using personal genome data. Variant annotation tool (VAT) [49] has been developed to annotate variants from multiple personal genomes at the transcript level as well as to obtain summary statistics across multiple genes and individuals.

FX [42] is an RNA-seq analysis tool, which runs in parallel on cloud computing infrastructure, for the estimation of gene expression levels and genomic variant calling. Another cloud computing pipeline for calculating differential gene expression in large RNA-seq datasets is Myrna [43]. Myrna uses bowtie [61] for short read alignment and R/bioconductor for quantification, normalization, and statistical testing. These tools are combined in an automatic, parallel pipeline that runs in the cloud, exploiting the availability of multiple computers and processors wherever possible. Stormbow [44] is a scalable, cost-effective, and open-source-based tool for large-scale RNA-seq data analysis. Its performance has been tested by applying it to analyze 178 RNA-seq samples in the cloud. In the test, it took 6–8 h to process each RNA-seq sample with 100 million pair-ended reads in the m1.xlarge instance, and the average cost was only $3.50 per sample. BioVLAB-MMIA-NGS [51, 52] offers the integrated miRNA-mRNA analysis and can be used to identify the "many-to-many" relationship between miRNAs and target genes with high accuracy. PeakRanger [48]

is a software package for the analysis of ChIP-seq data. It can be run in a parallel cloud computing environment to obtain extremely high performance on large data sets. Unbiased NGS approaches enable comprehensive pathogen detection in the clinical microbiology laboratory [63] and have numerous applications for public health surveillance, outbreak investigation, and the diagnosis of infectious diseases. Sequence-based ultra rapid pathogen identification (SURPI™) [53] is a computational pipeline for pathogen identification from complex metagenomic NGS data generated.

### 4.4. Rainbow

Crossbow is a software tool that can detect SNPs in WGS data from a single subject; however, it has a number of limitations when applied to large-scale WGS projects. Rainbow [12] is a cloud-based software package that can assist in the automation of large-scale WGS data analyses. Rainbow was built upon Crossbow. By hiding the complexity of the Crossbow command-line options, Rainbow facilitates the application of Crossbow for large-scale WGS analysis in the cloud. Compared with Crossbow, the main improvements incorporated into Rainbow include the ability: (1) to handle BAM as well as FASTQ input files, (2) to split large sequence files for better load balance in downstream clusters, (3) to collect and track the running metrics of data processing and monitoring multiple Amazon EC2 instances, and (4) to merge SOAPsnp outputs from multiple individuals into a single file to facilitate downstream genome-wide association studies.

The workflow of Rainbow is shown in **Figure 3**. Multiple data drives are shipped to Amazon. After the BAM or FASTQ files have been uploaded to S3, large FASTQ files are split into smaller files in parallel. Then multiple clusters are launched in the cloud, with each cluster processing a single sample. Crossbow is responsible for mapping reads to the reference sequence and for SNP calling. The SNPs for all samples are then combined by a Perl script. When the analysis is complete, the results can either be downloaded directly or exported via Amazon Export. We applied Rainbow to analyze the 44 subjects, with 0.55–1 billion paired-ended 100 bp short reads per sample. The running environments were as follows. For step #1 in **Figure 3**, we chose the Amazon m1.large instance, which has two CPUs, 7.5 GB memory, and two 420 GB instance drives. For Crossbow run, each compute cluster has 40 c1.xlarge nodes as recommended by the Crossbow developers. Each c1.xlarge node has 8 CPUs, 7 GB memory, and 1690 GB instance storage.

The performance of Rainbow is summarized in **Figure 4**. In a 320-CPU (=40 × 8) cluster, the alignment of billions of reads takes between 0.8 and 1.6 h. The linear relationship shown in **Figure 4** indicates that the sequence data blocks in the Hadoop distributed file system (HDFS) are physically local to the nodes that processed them, which reduces virtual I/O delays. The SOAPsnp running time ranges from 1 to 1.8 h, which takes a little longer than the alignment. All EC2 instances and clusters are terminated immediately after the jobs are finished. On average, it costs less than 120 US dollars to analyze each subject, and the total cost for analyzing those 44 subjects was 5800 US dollars, including data import. More important than the cost is the ability to scale Rainbow up or down, so that the analyses can be accomplished in a reasonably short amount time, regardless of sample size. No upfront investment in infrastructure is required and there is no additional administrative costs involved using Amazon cloud. Rainbow is a scalable, cost-effective, and open-source tool for large-scale WGS data analysis. It is available for third-party implementation and use, and can be downloaded from the Rainbow website.

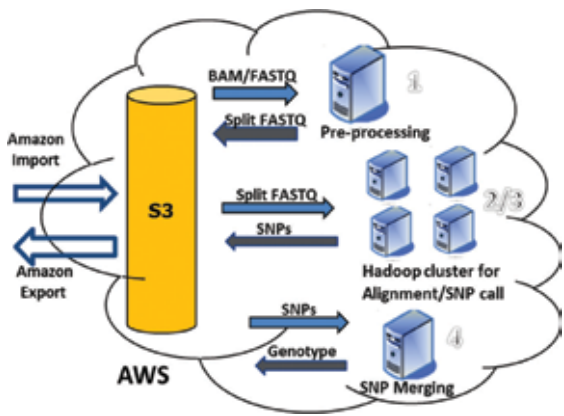

**Figure 3.** The Rainbow pipeline. S3 centralizes data storage, including inputs, intermediate results, and outputs. Alignment and SNP call are performed by Crossbow in a cluster with multiple nodes.

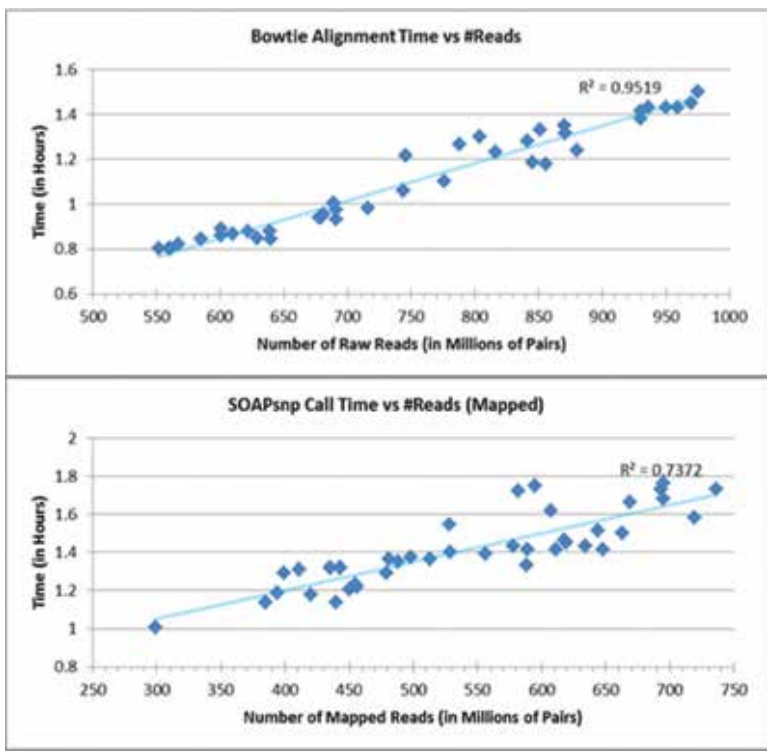

**Figure 4.** Top panel: running time of bowtie alignment vs. the number of paired sequence reads. Bottom panel: running time of SOAPsnp vs. the number of paired mapped reads. Note: each cluster consists of 40 c1.xlarge instances (8 CPUs per c1.xlarge instance).

In order to access the Rainbow cloud pipeline, the user must first set up an AWS account (http://aws.amazon.com/). Once registered, the user needs to sign up for Amazon EC2, S3, EMR, and SES services. The user can then start an instance based on the public AMI: ami-0f1f9866 in US-East (N. Virginia); or ami-b6bc89f3 in US-West (N. California). All required software is already preinstalled and configured in the AMI. Then, the user can connect to the instance and configure EC2, EMR, and S3cmd command-line tools. After a successful connection to the instance has been established, the user needs to prepare a sample manifest file in order to run Rainbow. A master manifest file is a plain text file to describe all subjects in a WGS project. Each subject has a corresponding entry in the manifest file and each entry consists of three fields separated by spaces or tabs: (1) a unique identifier; (2) locations of the raw reads in S3; and (3) an output folder in S3. Each individual step in the Rainbow workflow uses this same manifest file as input, thus all output files are named and stored consistently. After the creation of the manifest file, the user just needs to run a couple of command lines and all the analyses will be done automatically in the cloud.

Analyzing large datasets in the cloud is different from performing the same analysis in a local environment. When developing Rainbow, we have learned the following lessons.

- It is not trivial to move large datasets around within the cloud. Users should be prepared to handle network congestion or failures, never assuming data transfer is always successful. If data move fails, it is a good practice to wait a couple of minutes and then try to transfer data again. Repeat the "wait-and-transfer" cycle till data transfer is successful.

- Boot time should be taken into account when new resources are starting up. It is recommended to give cloud providers 10–15 min grace period (waiting time) before attempting to use a newly requested resource, for instance, elastic block store (EBS).

- Cloud providers typically offer a variety of compute instances to choose from. To choose the best option, it is important to understand the bottleneck (CPU, I/O, or network) for the algorithm that is to be run.

- When large amounts of data are moved between cloud resources, it is essential to ensure that they are in the same location or data center.

- It is difficult to debug workflows in the cloud without heavy logging.

## 5. Cloud computing hurdles

Albeit relatively new, cloud computing holds great promise in effectively addressing big data storage and analysis problems in NGS data analysis. Despite the potential gains achieved, there are also several important issues that need to be addressed. Below, we present the main hurdles on the adoption of cloud computing.

### 5.1. Big data transfer

To analyze the NGS data in the cloud, data have to be transferred across the wired network and uploaded onto AWS. The volume and complexity of NGS data have exponentially increased, giving rise to issues related to data analysis, management, and transfer to the cloud [64–66]. For example, WGS of 400 subjects at 30× coverage will generate approximately 100 TB raw sequence reads in FASTQ format. In the future, more and more sequencing projects would generate ultra-large volumes of biological data and thus require bioinformatics clouds for big data storage, sharing, and analysis. One of the most challenging issues of cloud computing is data transfer. Transferring vast amounts of biological data to the cloud is a significant bottleneck in cloud computing.

The speed of data transfer is usually slow and at present there are not many solutions available for moving the huge amount of data to cloud. Therefore, we need more efficient data transfer technologies in cloud computing. According to CloudHarmoy's report on download speed relative to the year 2010 (http://blog.cloudharmony.com/2010/02/cloud-speed-test-results.html), the download speed from Amazon AWS EC2 in North Virginia (U.S.) was 2.95 Mb/s, which corresponds to downloading a 10 GB file in 29,116 s ('8 h). Therefore, data transfer is a serious bottleneck in NGS data analysis on cloud service. To deal with the data transfer issue, Aspera (http://www.asperasoft.com/) has developed the fast and secure protocol (FASP) for data transfer with a speed of up to 5 GB/s. Ideally, using FASP, the user can download a 10 GB file in 17.2 s, which is a revolutionary improvement. But still it cannot transfer data at the TB scale. Alternatively, sequencing service providers such as BGI and Illumina offer a service in which they deliver a hard disk drive (HDD) containing the sequencing data. However, the shipping has limitations aside from the time taken up by travel, and once a package (of hard drives) is stamped and sealed, researchers cannot control when (and sometimes if) the parcel arrived or what condition it arrived. This service cannot guarantee safe transfer of data, since the HDD can get lost, stolen, or physical errors can occur.

### 5.2. Most bioinformatics tools are not cloud-aware

Most bioinformatics software tools are written for desktop (rather than cloud) applications and are therefore not provided as cloud-based web services accessible via the Web, making it infeasible to perform complex bioinformatics tasks in the cloud. For instance, bowtie [61] is one of the most popular mapping algorithms, but it requires that input files are stored on local disk when mapping reads and is not compatible with Amazon S3. Even if you run bowtie in an EC2 instance, the raw FASTQ files have to be fetched to an elastic block store (EBS) volume that is attached to the EC2 instance. In other words, bowtie does not have built-in support for S3. Spliced transcripts alignment to a reference (STAR) [67–69] is a popular RNA-seq mapper that performs highly accurate spliced sequence alignment at an ultrafast speed. However, it is not cloud-friendly either. Like bowtie, STAR does not take advantage of AWS cloud services, and cannot work with S3 either. Unfortunately, the majority of bioinformatics tools are developed without native support for cloud computing.

MapReduce [70, 71], developed by Google, is an easy-to-use and general-purpose parallel programming model that is suitable for large data set analysis on a commodity hardware cluster. MapReduce is a software framework, written in Java, designed to run over a cluster of machines in a distributed way. A MapReduce program is composed of a user-defined map function and a reduce function. When a program that is implemented with the map and reduce functions has been launched, the map function processes each key/value pair and produces a list of intermediate key/value pairs, while the reduce function aggregates all the intermediate values with the same keys. MapReduce is an important advancement in cloud computing because it can process huge data sets quickly and safely using commodity hardware.

Hadoop, comprised of MapReduce and the Hadoop distributed file system (HDFS), is based on a strategy of colocating data and processing to significantly accelerate computing performance [60]. Hadoop allows for the distributed processing of large datasets across multiple computer nodes, supports big data scaling, and enables fault-tolerant parallel analysis. The Hadoop framework has been recently deemed as the most suitable method for handling bioinformatics data [70]. Unfortunately, many traditional bioinformatics tools and algorithms have to be redesigned and implemented in order to support and benefit from Hadoop MapReduce infrastructure. Even with the help of the corresponding developers, it will take a while for most bioinformatics tools currently available to add this feature.

Apache Spark™ (https://spark.apache.org/) is a fast and general engine for large-scale data processing, natively supported in Amazon EMR. Apache Spark supports a variety of languages, including Java, Scala, and Python, for developers to build applications. Hadoop and Apache Spark are both big data frameworks, but they do not really serve the same purposes. Hadoop is essentially a distributed data infrastructure. It distributes massive data collections across multiple nodes within a cluster of commodity servers, Spark, on the other hand, is a data-processing tool that operates on those distributed data collections; it does not do distributed storage. To study the utility of Apache Spark in the genomic context, SparkSeq was created [72]. It is a general-purpose, flexible, and easily extendable library for genomic cloud computing, and can be used to build genomic analysis pipelines in Scala and run them in an interactive way. Recently, SparkBWA [73] was introduced; a new tool that exploits Spark to boost the performance of one of the most widely adopted sequence aligner, the Burrows-Wheeler Aligner (BWA). It is hoped more Apache Spark-based bioinformatics algorithms will be developed for large-scale genomic data analysis in the future.

## 5.3. Open clouds for bioinformatics

Currently, the largest cloud computing provider is Amazon, which provides commercial clouds for processing big data. Additionally, Google also provides a cloud platform to allow users to develop and host applications, and to store and analyze data. However, commercial clouds are not yet able to provide ample data and software for bioinformatics analysis. By placing public biological database and software into the cloud and delivering them as services, data and software can be seamlessly and easily integrated into the cloud. AWS hosts a variety of public data sets for free access (https://aws.amazon.com/public-data-sets/). All public datasets in AWS are delivered as services. Previously, large data sets, such as the mapping of the human genome, required hours or days to locate, download, customize, and analyze.

Now, anyone can access these data sets via the AWS centralized data repository from any Amazon EC2 instance or Amazon EMR cluster. Google Genomics also helps the life science community organize the world's genomic data and make them accessible and useful.

In the era of big data, however, only a tiny amount of biological data is accessible in the cloud at present (only AWS, including GenBank [35], Ensembl [74], 1000 Genomes [20], etc.) and the vast majority of data are still deposited in conventional biological databases. It is difficult for commercial clouds to keep pace with the emerging needs from academic research, opening up the demand for specific open clouds for bioinformatics studies. Needless to say, open access and public availability of data and software are of great significance to biological science. To satisfy the need for big data storage, sharing, and analysis with lower cost and higher efficiency, it is essential that a large number of biological data as well as a wide variety of bioinformatics tools should be publicly accessible in the cloud and delivered as services. Therefore, future efforts should be devoted to building open bioinformatics clouds for the bioinformatics community. GenomeSpace [75] is a cloud-based, cooperative community resource that currently supports the streamlined interaction of 20 bioinformatics tools and data resources. To facilitate integrative analysis by nonprogrammers, it offers a growing set of 'recipes', short workflows to guide investigators through high-utility analysis tasks. The potential benefits of open bioinformatics clouds include maximizing the scope for data sharing, easing large-scale data integration, and harnessing collective intelligence for knowledge discovery.

### 5.4. Security and privacy

The many characteristics of cloud computing have made the long-dreamed vision of "computing as a utility" a reality. The cloud computing offers scalable and competitively priced computing resources for the analysis and storage of data from large-scale genomics studies, but it must also ensure that genetic data coming from human subjects are hosted in a context that is both secure and compliant with regulations [76]. When deciding whether to move the analyses into the cloud or not, potential cloud users need to weigh all the factors including system performance, service availability, cost, and most importantly, data security. Genomics data extracted from clinical samples are sensitive data and present unprecedented requirements of privacy and security [11, 77]. In general, there are concerns that genomics and clinical data managed through a cloud are susceptible to loss, leakage, theft, unauthorized access, and attacks. The centralized storage and shared tenancy of physical storage space means the cloud users are at higher risk of disclosure of their sensitive data to unwanted parties. A secure protection scheme will be necessary to protect the sensitive information from medical records. There is considerable amount of work to enforce data protection against security attacks.

However, the question of security in cloud computing is intrinsically complicated. Cloud computing is built on the top of existing architectures and techniques such as SaaS and distributed computing. When combining all the benefits of these architectures and techniques, cloud computing also inherits almost all of their security issues at various levels of the system stack. When cloud users move their applications from within their enterprise/organization boundary into the open cloud, they will lose physical control over their data, and traditional security protection mechanisms such as firewalls are no longer applicable to cloud applications. As a result, cloud users have to heavily rely upon the service providers for data privacy and security protection. In cloud computing, the data and applications from different customers reside on the same physical computing resources. This fact will inevitably bring forth more security risks in the sense that any intentional or inadvertent misbehavior by one cloud user would make other coresidences victims.

## 6. Conclusion

Pharmacogenomics is an important branch of genomics that studies the impact of SNP on drug response in patients, the toxicity or efficacy of a drug, and the development of diseases [78, 79]. Pharmacogenomics aims to improve drug therapy with respect to the patients' genotype, to ensure maximum efficacy with minimal adverse effects, and is at the basis of the idea of "precision medicine" where drugs are chosen or optimized to meet the genetic profile of each patient. Thus, the presence (or the absence) of specific SNPs may be used as a clinical marker to predict drug effectiveness, and to foresee the drug responses of individuals with different SNPs. Precision medicine also opens up the possibility to treat diseases based on the genetic makeup of the patient. Identification of the genetic defect underlying early-onset diabetes is important for determining the specific diabetes subtype, providing personalized treatment. The study by Artuso et al. [14] has revealed that NGS provides a highly sensitive method for identification of variants in a new-set of "driver genes" causing diabetes. NGS can be used to analyze the comprehensive landscape of genetic alterations, including known disease-causing gene fusions in transcripts, which brings new insights to study diseases with a highly complex and heterogeneous genetic composition such as cancer [19]. Therefore, NGS facilitates precision medicine and changes the paradigm of cancer therapy, and holds expanded promise for its diagnostic, prognostic, and therapeutic applicability in various diseases.

The substantial decrease in the cost of NGS techniques in the past decade has dramatically reshaped the biomedical research and has led to its rapid adoption in biological research and drug development [3, 4]. Nowadays, massive amount of data, targeting a variety of biological questions, can be generated quickly using NGS platforms. These data range from the function and regulation of genes, the clinical diagnosis and treatment of diseases, to the omics profiling of individual patients for precision medicine. To better understand the association between SNPs and diseases, and to gain deeper insights into the relation between drug response and genetic variations, large-scale sequencing projects are continuously being initiated in research institutes and pharmaceutical companies. The availability of NGS and the genomics studies of large populations are producing an increasing amount of data. However, the storage, preprocessing, and analysis of NGS data are becoming the main bottleneck in the analysis pipeline. With the exponential increase in volume and complexity of NGS data, cluster or high performance computing (HPC) systems are essential for the analysis of large amounts of NGS data. But the associated costs with the infrastructure itself and the maintenance personnel will likely be prohibitive for small institutions or laboratories, and even too much to swallow for big institutions and pharmaceutical companies.

This chapter provides a useful perspective on cloud computing and helps researchers, who plan to analyze their NGS data on cloud services, to gain an understanding of the basic concept. As discussed earlier, cloud computing drives down infrastructure costs both up-front and on an on-going basis. It offers many operational advantages, such as fast completion of massive computational projects and infrastructures can be set up in minutes rather than months. There are growing demands for cloud-based NGS data analysis, which constitutes a good alternative for researchers with little interest in investing in a cluster system. Cloud computing is becoming a technology mature enough for its use in genomic researches. The use of large datasets, the demanding analysis algorithms, and the urgent need for computational resources, make large-scale sequencing projects an attractive test-case for cloud computing. It is likely that in the future, researchers will be able to analyze their NGS data on cloud services for a cost low enough to eliminate the need to introduce clusters into their laboratories. However, we must emphasize that there are a number of open issues and problems associated with the use of cloud, such as privacy and security, especially when managing patients' clinical data. Data transfer of large amounts of NGS data to the cloud remains a challenge and bottleneck in NGS data analysis. Furthermore, many traditional NGS tools are not designed to use in cloud environment, and thus cannot harness the benefits of cloud services. With the ever-increasing demand, commercial companies and bioinformaticians will develop more NGS analysis tools, using APIs offered by service providers, in the foreseeable future, making NGS data analysis in the cloud more efficient, friendly, and secure.

## Author details

Shanrong Zhao[1]*, Kirk Watrous[2], Chi Zhang[1] and Baohong Zhang[1]

*Address all correspondence to: shanrong.zhao@pfizer.com

1 Early Clinical Development, Pfizer Worldwide Research and Development, Cambridge, MA, USA

2 Business Technology, Pfizer Worldwide Research and Development, Groton, Connecticut, USA

# Green-Aware Virtual Machine Migration Strategy in Sustainable Cloud Computing Environments

Xiaoying Wang, Guojing Zhang, Mengqin Yang and
Lei Zhang

**Abstract**

As cloud computing develops rapidly, the energy consumption of large-scale datacenters becomes unneglectable, and thus renewable energy is considered as the extra supply for building sustainable cloud infrastructures. In this chapter, we present a green-aware virtual machine (VM) migration strategy in such datacenters powered by sustainable energy sources, considering the power consumption of both IT functional devices and cooling devices. We define an overall optimization problem from an energy-aware point of view and try to solve it using statistical searching approaches. The purpose is to utilize green energy sufficiently while guaranteeing the performance of applications hosted by the datacenter. Evaluation experiments are conducted under realistic workload traces and solar energy generation data in order to validate the feasibility. Results show that the green energy utilization increases remarkably, and more overall revenues could be achieved.

**Keywords:** virtual machine migration, resource management, power management, renewable energy aware

## 1. Introduction

Large-scale datacenters, as the key infrastructure of cloud environments, usually own massive computing and storage resources in order to provide online services for thousands of millions of customers simultaneously. This leads to significant energy consumption, and thus high carbon footprint will be produced. Recent reports estimate that the emissions brought by information and computing technologies grow from 2% in 2010 [1] to 8% in 2016 and will grow to 13% by 2027 [2]. Hence, considering the heavy emissions and increasing impact on climate change, governments, organizations, and also IT enterprises are trying to find cleaner ways to manage the datacenters, for example, exploiting renewable energy such as wind, solar, and tidal.

However, the intermittency and the instability of the renewable energy sources make it difficult to efficiently utilize them. Fortunately, we know that the datacenter workloads are usually variable, which give us opportunities to find ways to manage the resources and power together inside the datacenters to utilize renewable energy sources more efficiently. On the other hand, to provide guaranteed services for third-party applications, the datacenter is responsible of keeping the quality of service (QoS) at a certain level, subject to the service level agreements (SLAs) [3].

In modern datacenters, applications are often deployed in virtual machines (VMs). By virtualization mechanisms, VMs are flexible and easy to migrate across different servers in the datacenter. In this chapter, we attempt to conduct research on energy-aware virtual machine migration methods for power and resource management in hybrid energy-powered datacenters. Especially, we also employ thermal-aware ideas when designing VM migration approaches. The holistic framework is described, then the model is established, and heuristic and stochastic strategies are presented in detail. Experimental results show the effectivity and feasibility of the proposed strategies. We hope that this chapter would be helpful for researchers to study the features of VM workloads in the datacenter and find ways to utilize more green energy than traditional brown energy.

The remainder of this chapter is organized as follows. Section 2 introduces some relevant prior work in the field of energy-aware and thermal-aware resource and power management. Section 3 presents the entire system architecture we discuss in this chapter. Section 4 formulates the optimization problem corresponding to the issue we need to address. Section 5 describes the methods and strategies we designed to solve the problem. Section 6 illustrates the experimental results by comparing three different strategies, and finally conclusion is given out in Section 7, in which we also discuss about some of the possible future work.

## 2. Literature review

This section reviews the literature in the area of energy-aware resource management, thermal-aware power management, and green energy utilization in datacenters.

In the recent decade, many researchers started to focus on power-aware management methods to manage workload fluctuation and search trade-off between performance and power consumption. Sharma et al. [4] have developed adaptive algorithms using a feedback loop that regulates CPU frequency and voltage levels in order to minimize the power consumption. Tanelli et al. [5] controlled CPUs by dynamic voltage scaling techniques in Web servers, aiming at decreasing their power consumption. Berl et al. [6] reviewed the current best practice and progress of the energy efficient technology and summarized the remaining key challenges in the future. Urgaonkar et al. [7] employed queuing theory to make decision aiming at optimizing the application throughput and minimizing the overall energy costs. The above work attempts to reduce the power consumption while guaranteeing the system performance. On the basis of such ideas, we incorporate the usage of renewable energy into the optimization model, which might support performance improvement when the green energy is sufficient enough.

Besides, thermal-aware resource management approaches also attracted some interest of researchers recently. For example, Mukherjee et al. [8] developed two kinds of temperature-aware algorithms to minimize the maximum temperature in order to avoid hot spots. Tang et al. [9] proposed *XInt* which can schedule tasks to minimize the inlet temperatures and also to reduce the cooling energy costs. Pakbaznia et al. [10] combined chassis consolidation and efficient cooling together to save the power consumption while keeping the maximum temperature under a controlled level. Wang et al. [11] designed two kinds of thermal-aware algorithms aiming at lowering the temperatures and minimizing the cooling system power consumption. Islam et al. [12] proposed DREAM which can manage the resources to control allocate capacity to servers and distribute load considering temperature situations. Similarly, we consider the impact of temperature on two kinds of cooling devices in this chapter, which directly decide the cooling power consumption.

As renewable energy becomes more widely used in datacenters, corresponding research starts to put insights into green energy–oriented approaches for managing the resources and power. Deng et al. [13] treated carbon-heavy energy as a primary cost and designed some mechanisms to allocate resources on demand. Goiri et al. designed *GreenSlot* [14] aiming at scheduling batch workloads and *GreenHadoop* [15] which could deal with MapReduce-based tasks. Both of them tried to efficiently utilize green energy to improve the application performance. Li et al. [16] proposed *iSwitch*, which can switch the power supply between wind power and utility grid according to the renewable power variation. Arlitt et al. [17] defined the "Net-Zero energy" datacenter, which needs on-site renewable generators to offset the usage of power coming from the electricity grid. Deng et al. also conducted research on Datacenter Power Supply System (DPSS) and proposed an efficient, online control algorithm SmartDPSS [18] helping to make online decisions in order to fully leverage the available renewable energy and varying electricity prices from the grid markets, for minimum operational cost. Zhenhua et al. [19] presented a holistic approach that integrates renewable energy supply, dynamic pricing, cooling supply, and workload planning to improve the overall attainability of the datacenter.

Upon the basic concepts of these work, we exploit the possibility and of efficient VM migration management toward sufficiently utilizing renewable energy supply, incorporating the flexibility of transactional workloads, cooling power consumption, and the amount of available green energy.

## 3. Datacenter architecture

This section describes the datacenter architecture, including the hybrid power supply and virtualization infrastructure.

**Figure 1** shows the system architecture of the sustainable datacenter powered by both renewable energy and traditional energy supplies. The grid utility and renewable energy are combined together by the automatic transfer switch (ATS) in order to provide power supply for the datacenter. Both functional devices and cooling devices have to consume power, as shown in the bottom part of the figure.

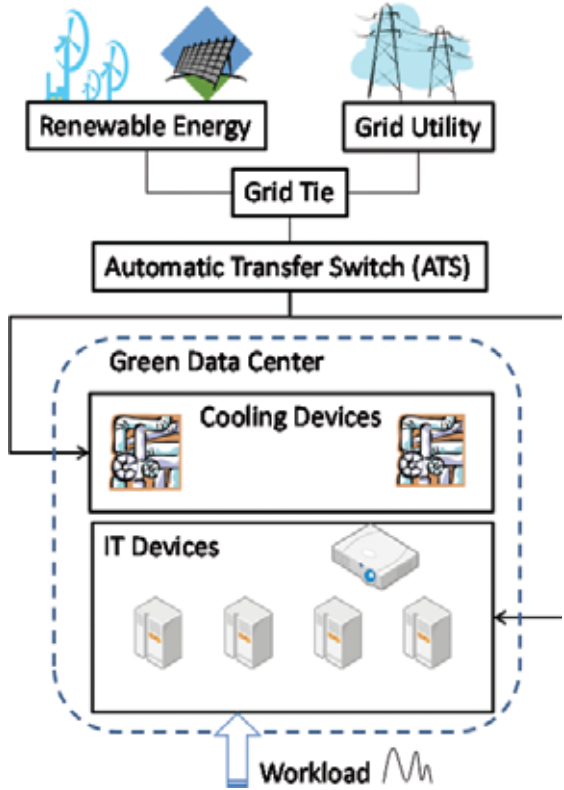

**Figure 1.** Architecture of the sustainable datacenter powered by both green energy and brown energy.

**Figure 2** illustrates the infrastructure of virtualized cloud datacenter. As shown, the underlying infrastructure of the datacenter is comprised of many physical machines (PMs), which are placed onto groups of racks. The utility grid bus and the renewable energy bus are connected together to supply power for the datacenter devices. Renewable sources will be used first, and the grid power will be leveraged as the supplementary energy supply.

As mentioned before, virtual machines (VMs) are running on the underlying infrastructure as used to host multiple applications, as shown in the virtualization layer in **Figure 2**. Different VMs on the same PM might serve for different applications. In this chapter, we mainly discuss about transactional applications which needs CPU resources mostly, other than other types of resources.

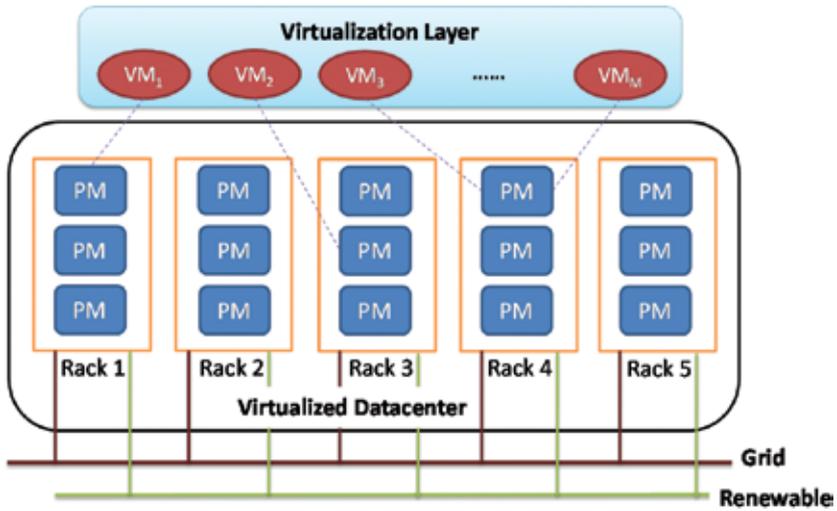

**Figure 2.** Virtualized datacenter.

## 4. Problem definition

This section defines necessary variables and also the problem we need to solve throughout this chapter.

### 4.1. Model of computing and service units

In the target problem, there are $N$ heterogeneous physical machines in the virtualized cloud environment, and the available CPU resource capacity of PM $i$ is denoted as $\Phi_i$. The entire environment is hosting $M$ kinds of different applications, deployed on $M$ different VMs. Denote the $j$th VM as $VM_j$. Then, denote $x_j$ as the index of the PM which is hosting $VM_j$. Denote $\phi_i$ as the allocated CPU capacity to $VM_j$ and $d_i$ as the demanded CPU capacity of application $j$ at the current time slot.

### 4.2. Power consumption model

According to the mechanisms of dynamic voltage and frequency scaling (DVFS) techniques, here we use a simple power model which assumes that the power consumption of other components in the PM correlate well with CPU [20]. Denote $p_i$ as the power consumption of PM $i$ in each time slot and $p_i^{MAX}$ as the maximum power consumption of PM $i$ (100% occupied by workloads). Then, the following equation can be used to compute the PM power consumption:

$$p_i = p_i^{\text{MAX}} \cdot \left( c + (1 - c) \cdot \theta_i \right) \tag{1}$$

where $c$ is a constant number representing the ratio of the idle-state power consumption of a PM compared to the full-utilized-state power consumption [21] and $\theta_i$ is the current CPU utilization of PM $i$.

Besides, we also consider the power cost spent on cooling devices when establishing the power model, which is usually much related to temperature. The cooling system we discuss here consists of both the traditional computer room air conditioning (CRAC) unit and the air economizer. According to relevant studies [10], the coefficient of performance (CoP) is often used to indicate the efficiency of a cooling system, which can be computed by

$$\text{CoP} = \begin{cases} 1/k\left(T_{\text{sup}} - T_{\text{out}}\right), & \text{when } T_{\text{out}} \leq T_{\text{sup}} \\ 0.0068\, T_{\text{sup}}^2 + 0.0008\, T_{\text{sup}} + 0.458, & \text{otherwise} \end{cases} \tag{2}$$

where $k$ is a factor reflecting the difference between outside air and target temperature, $T_{sup}$ is the target supply temperature, and $T_{out}$ is the outside temperature. As it can be observed, Eq. (2) contains two parts, corresponding to the situation whether the CRAC or the air economizer will be used for cooling, respectively.

Hence, the total power consumed by both functional devices and cooling devices can be calculated by

$$p_{DC} = \left(1 + \frac{1}{\text{CoP}}\right) \cdot \sum_{i=1}^{N} p_i \tag{3}$$

Furthermore, considering the impact of environmental temperature inside the datacenter, we also tried to exploit thermal-aware VM migration strategies. The power consumption of the servers will make the surrounding environmental temperature increase, due to the dissipated heat. Prior studies [11] provided ways to model the vector of inlet temperatures $T_{\text{in}}$ as

$$T_{\text{in}} = T_s + D \bullet p \tag{4}$$

where $D$ is the heat transferring matrix, $p$ is the power consumption vector, and $T_s$ is the supplied air temperature vector.

The thermal-aware strategy tries to reduce the cooling power by balancing the temperature over the servers. Accordingly, the workload on different PMs should also be maintained balanced. Denote $T_{safe}$ as the safe outlet temperature and $T_{server}$ as the outlet temperature of the hottest server. In order to lower the server temperature to the safe level, the output temperature of cooling devices should be adjusted by $T_{adj} = T_{safe} - T_{server}$. Then, the output temperature after adjusted will be $T_{new} = T_{sup} + T_{adj}$. Hereafter, the *CoP* value can be determined by $T_{new}$ and $T_{out}$ [22].

### 4.3. Modeling overhead and delay

To reduce the power consumption of the PM, it can be switched to sleeping state which can help save energy as much as possible. In addition, the operational costs also include the VM migration costs, since migrating VMs dynamically will definitely lead to some overhead. Denote $a_i$ as the flag recording whether PM $i$ is active or sleeping. Denote $c^A$ as the cost for activating a PM from sleeping state and $c^{MIG}$ as the cost for migrating a VM from one PM to another. Besides, the time delay is also considered and integrated into the experiments in Section 6 for waking up a PM and migrating a VM.

### 4.4. Optimization problem formulation

From the resource providers' point of view, the objective should be maximizing the total revenues by meeting the requirements of the hosted applications while minimizing the consumed power and other costs. Usually, the revenues from hosting the applications are related to service quality and the predefined level in the SLA. Assume here that the service quality is reflected by the CPU capacity scheduled to the target application. Denote $d_j$ as the demanded CPU capacity of APP $j$ and $\phi_j$ as the CPU capacity amount scheduled to APP $j$. Denote $\Omega^j(\bullet)$ as the profit model for APP $j$, which gives the actual revenue by serving APP $j$ at a certain quality level.

Since the dynamic action decisions are made during constant time periods, denote $\tau$ as the length of one time slot. Denote $t$ as the current time slot, and then in time slot $t+1$, the goal is to maximize the net revenue subject to various constraints. Denote $x_j$ as the index of PM currently hosting VM $j$, and then the VM placement vector $X$ can be denoted as

$$X = (x_1, x_2, ..., x_j, ..., x_M) \tag{5}$$

Hence, the optimizing objective of the defined problem can be expressed as

$$\max \begin{array}{l} \sum_{j=1}^{M} \Omega^j(d_j, \varphi_j) - c^P \cdot p_{DC} \\ -c^A \cdot \sum_{i=1}^{N} \max(0, a_i(t+1) - a_i(t)) \\ -c^{MIG} \cdot \sum_{j=1}^{M} (x_j(t+1) - x_j(t)) \end{array} \tag{6}$$

where the first term is the total revenue summarized over all of the hosted applications, the second term represents the power consumption costs of the entire datacenter, the third term is the PM wake-up cost, and the last term represents the VM migration cost.

With respect to the objective defined above, the constraints could be expressed as

$$\sum_{x_j=i} \varphi_j \le \Phi_i \cdot a_i \tag{7}$$

$$0 \le \varphi_j \le d_j, j = 1, 2, \ldots, M \tag{8}$$

$$a_i \in \{0, 1\}, x_j \in [1, N] \tag{9}$$

where Eq. (7) means that the allocated capacity cannot exceed the PM CPU capacity, Eq. (8) means that the CPU scheduled to a VM should be less than its demanded value, and Eq. (9) gives the validated ranges of the defined variables.

## 5. Methods and strategies

In this section, we design some heuristic methods and also the joint hybrid strategy, and describe the ideas in detail.

### 5.1. Dynamic load balancing *(DLB)*

The idea of the *DLB* strategy is to make the workload on different PMs balanced by dynamically placing VMs. To achieve the balancing effect, if one PM is detected to be more utilized than the specified upper threshold, some VMs on this PM will be chosen to migrate other-where. As a result, the PM utilization ratio will be controlled in a certain range, and there will be as few overloaded PMs as possible.

## 5.2. Dynamic VM consolidation (*DVMC*)

According to the features of virtualization techniques, VMs could be consolidated together onto a few PMs to make other PMs zero loaded. Hence, the main idea of the *DVMC* strategy is to consolidate VMs as much as possible aiming at saving more power. Both the upper threshold and the lower threshold of the PM utilization level are defined. If one PM is light loaded enough that its utilization is less than the lower threshold, the VM consolidation process will be triggered. After this process, VMs upon underutilized PM will be migrated onto other PMs. Finally, zero-loaded PMs could be turned into inactivate state in order to save more power.

## 5.3. Joint optimal planning (*JOP*)

The *JOP* strategy aims to optimize the VM placement scheme with the objective of sufficiently utilizing the renewable energy and reducing the total costs.

### 5.3.1. Renewable energy forecasting

Since renewable energy is used as one source of power supply, we have to forecast the input power value in the next time slot. Here the *k*-nearest neighbor (*k*-NN) algorithm is adopted. A distance weight function is designed to calculate the distance each solar radiation values, as follows:

$$w_i = (1/d^i)/(1/d^1 + 1/d^2 + \ldots + 1/d^k) \tag{10}$$

where $d^i$ is the distance between the $i$th neighbor and the current point.

**Figure 3** shows the forecasting effect on one day in October 2013. The data were measured and collected in Qinghai University, Xining, Qinghai Province of China. By analyzing the data points, the allowed absolute percentage errors (AAPE) of 97.01% data are less than 30%. The accuracy of the prediction method depends on the similar weather conditions in the recent past and may be affected by weather forecast data.

### 5.3.2. Stochastic search

In order to look for the best scheme of VM placement, we use stochastic search to do the optimization. Specifically, the genetic algorithm (GA) is modified and employed as follows:

For a typical genetic algorithm, there are two basic items as follows:

**1.** A genetic representation of solution space

Here, for this problem, the decision variable is the vector of VM placement, which can be denoted as $\mathbf{X = (x_1, x_2 \ldots, x_M)}$.

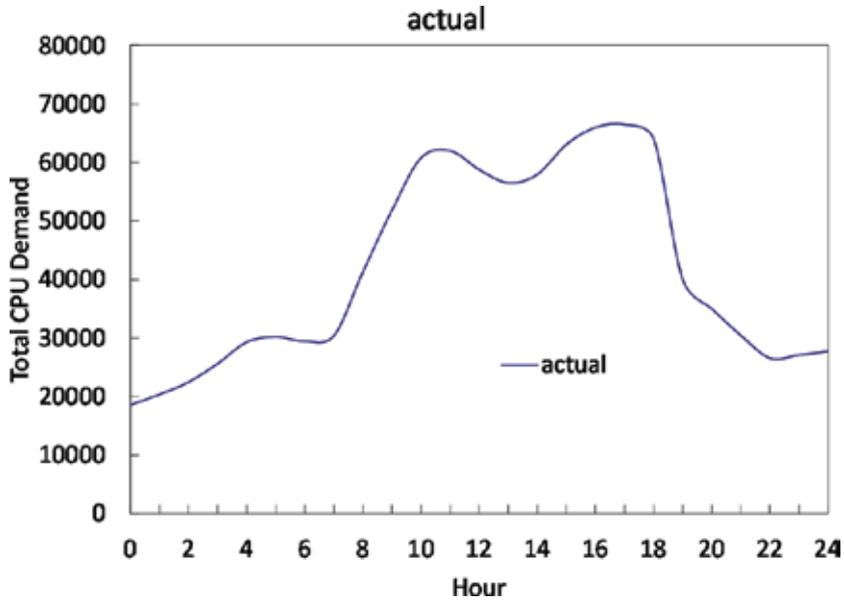

**Figure 3.** An example of renewable energy forecasting effect.

**2.** A fitness function to compute the value of each solution

As described, the objective function defined by Eq. (6) could be used as the fitness function. It is functional in measuring the quality of a certain solution. Hereafter, the fitness function will be denoted as $F(X)$.

The procedure of genetic algorithm can be divided into following steps:

i.  Initialization

First, we add the current configuration vector in the last time epoch into the initial generation. Besides, a fixed number (denoted as $n_g$) of individual solutions will be randomly generated. Specifically, a part of the elements of each solution will be generated randomly, in the range of 0~$N$−1.

ii.  Selection

After initialization, the generations will be produced successively. For each generation, $n_b$ best-ranking individuals from the current and past population will be selected to breed a new generation. Then, in order to keep the population constant, the remained individuals will either be removed or replicated based on its quality level. The selection procedure is conducted based on fitness, which means that solutions with higher fitness values are more prone to be selected.

According to such concepts, the probability to select an individual $X_i$ can be calculated as

$$P(X_i) = \frac{F(X_i)}{\sum_{i=1}^{n} F(X_i)} \tag{11}$$

In this way, less fit solutions will be less likely to be selected, and this helps to keep the diversity of the population and to keep away from premature convergences on poor solutions.

**1.** Reproduction

After selection, a second generation of population should be generated from those selected solutions through two kinds of genetic operators: crossover and mutation.

The crossover operator first selects two different individuals, denoted as $X^1 = (x_1^1, x_2^1, ..., x_M^1)$ and $X^2 = (x_1^2, x_2^2, ..., x_M^2)$. Then, a cutoff point $k$ is set from the range 1~$M$. Both $X^1$ and $X^2$ are divided into two halves, and the second half of them will be swapped and then $X^{1'} = (x_1^1, x_2^1, ...x_k^1, x_{k+1}^2, ..., x_M^2)$ and $X^{2'} = (x_1^2, x_2^2, ..., x_k^2, x_{k+1}^1, ..., x_M^1)$. As a result, two new individuals will come out, which is perhaps already in the current population or not.

After crossover, the mutation operator will mutate each individual with a certain probability. The mutation process starts by randomly choosing an element in the vector and then changing its value, and then converts an individual into another.

**2.** Termination

This production process will repeat again and again until the number of generations reaches to a predefined level.

# 6. Evaluation results

This section shows our experiments comparing different strategies, and then the results and some details will be discussed.

### 6.1. Parameter settings

For the following experiments, we used C#.NET to develop the simulation environment and set up the prototype test bed. Specifically, a virtualized datacenter is established, comprised of 40 PMs with CPU capacity of 1500 MIPS each. For the power model, $p_i^{MAX}$ is set to 259W according to Ref. [23] and, $c$ is set to 66% according to Ref. [21]. Then, 100 VMs hosting different applications were simulated and put on the PMs. The workload on each VM fluctuates with time, with the value randomly generated under the uniform distribution.

**Table 1** shows all of the parameter settings in detail, and **Figure 4** shows variation of the total CPU demand summarized over all of the workloads, from which it can be seen that there are two peaks in the 24-h period.

We defined a nonlinear revenue function for each application, as mentioned in Section 4. **Figure 5** shows some three typical examples. It can be seen that the revenue of every application changes elastically in a certain range.

|  | APP 1 | APP 2 | APP 3 |
|---|---|---|---|
| Lower bound | 50 | 40 | 30 |
| Upper bound | 90 | 60 | 70 |
| $U_j^{max}$ | 100 | 60 | 80 |

**Table 1.** Parameter settings for example applications.

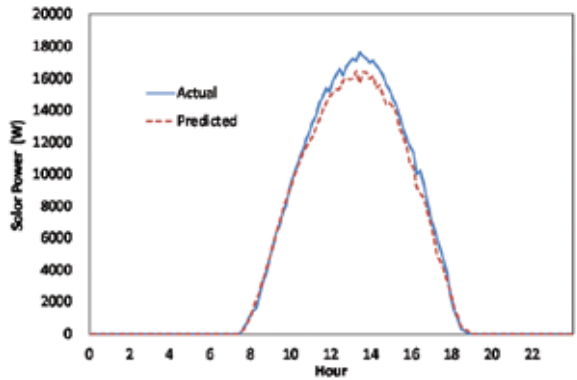

**Figure 4.** The workload variation.

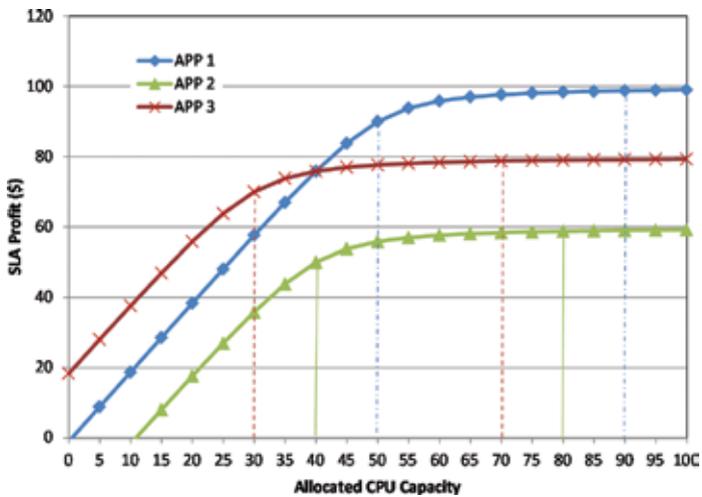

**Figure 5.** The revenue functions of three example applications.

The control interval for reconfiguration actions in the experiment is set to 60 minutes. According to Refs. [24–26], we set $c^P$ to \$0.08, set $c^A$ to \$0.00024, and set $c^{MIG}$ to \$0.00012. The VM migration delay is set to 5s, and the PM wakeup delay is set to 15s. The total experiment time is set to 1440 minutes. The temperature data used in the experiments come from the realistic data measured on 4 October 2013, recorded in the campus of the Qinghai University, Xining, Qinghai Province, China, as shown in **Figure 6**.

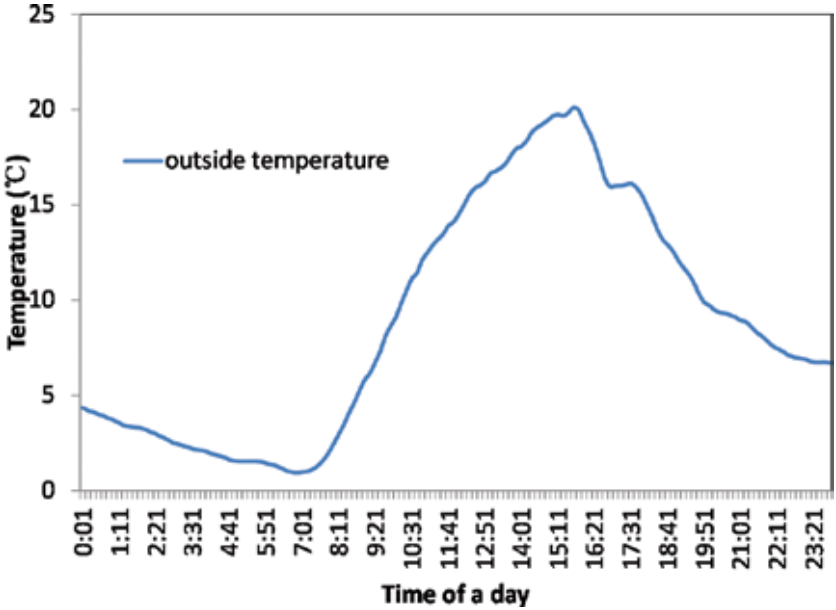

**Figure 6.** Variation of the outside temperature in 1440 min.

## 6.2. Results

In order to investigate the effectiveness of the proposed strategy, we will compare the performance among three different strategies – *DLB*, *DVMC*, and *JOP*, as stated in Section 5.

### 6.2.1. Revenues

As described in Section 4, the net revenue is a main optimizing objective in our problem. **Figure 7** shows the total accumulated net revenues throughout the 1440-min experiment time. It can be observed that the *JOP* strategy can keep the net revenue relatively higher than other ones. Moreover, the *DVMC* approach behaves relatively better than *DLB* since it can save more power by VM consolidation. By examining the detailed data, we found that *JOP* could make the gains 38.2 and 24.2% higher than *DLB* and *DVMC*, respectively, with respect to the accumulated revenue.

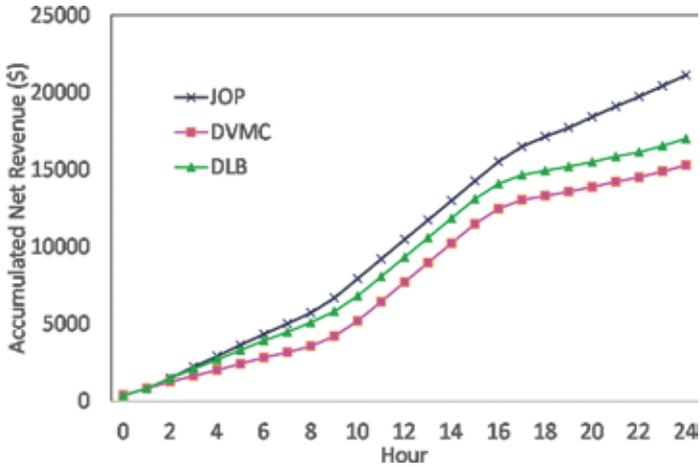

**Figure 7.** Accumulated net revenue variation.

### 6.2.2. Power consumption

Now we intend to investigate the power consumption in detail when using *JOP*, as **Figure 8** illustrates. It can be observed from the figure that *JOP* is able to follow the solar energy variation quite well. When the solar power drops to insufficient level, *JOP* is prone to degrade the application performance to save more power. On the contrary, when the solar power arises, *JOP* allows both functional and cooling devices to consume more power, under the constraints of the input power. Interestingly, we can see that the temperature varies more or less in coincidence with solar energy generation, which implies that thermal-aware coscheduling of energy supply and consumption might be promising, since the temperature also affects energy consumption to some extent.

### 6.2.3. PM Management

**Figure 9** shows the number of active servers when using the three different strategies. We can see that *JOP* can increase or decrease the number of active servers according to the variation of the solar power generation amount. Under the *DLB* strategy, all PMs are kept active so that the system-wide workload could be balanced. Comparatively, *DVMC* uses much fewer active PMs than *DLB* due to VM consolidation. However, it still uses more PMs at night time because it cannot effectively deal with the relationship of revenues and costs. Overall, *JOP* tries to manage PMs dynamically toward the optimization objective and thus can keep the number of active PMs as needed.

### 6.2.4. Energy for cooling

The cooling energy consumption is also investigated when using the three different strategies, as shown in **Figure 10**. As illustrated, *JOP* allows cooling devices to consume more power until after 18:00, showing its capability of catching the solar energy variation. By forecasting the solar power generation amount, *JOP* is able to make better decisions for migrating VMs according to the optimized scheme.

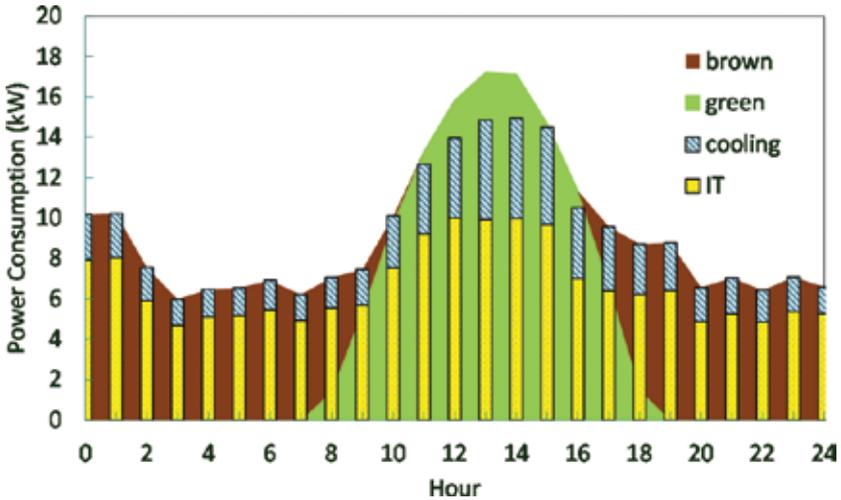

**Figure 8.** Power consumption details under *JOP* strategy.

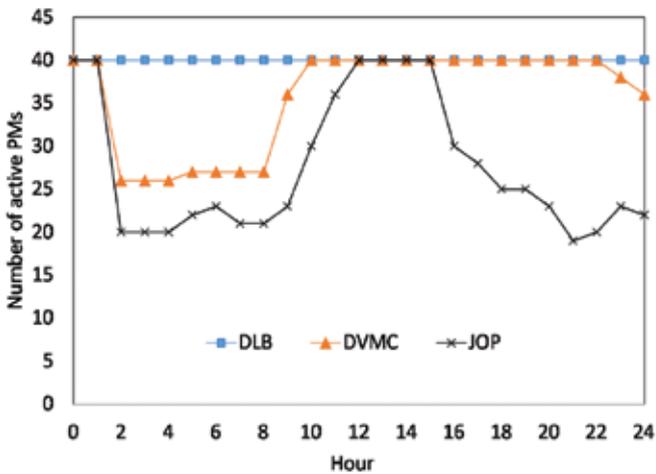

**Figure 9.** Variation of the number of activated PMs.

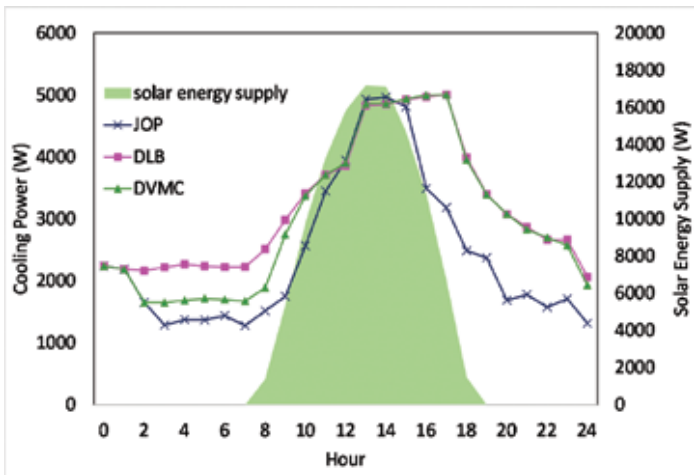

**Figure 10.** Cooling energy consumption under three different strategies.

## 7. Conclusion and future work

As the energy consumption of large-scale datacenters becomes significant and attracts more attentions, renewable energy is being exploited by more enterprises and cloud providers to be used as a supplement of traditional brown energy. In this chapter, we introduced the target system environment using hybrid energy supply mixed with both grid energy and renewables. From the datacenter's own point of view, the optimization problem was defined aiming at maximizing net revenues. Accordingly, three different strategies were designed to migrate VMs across different PMs dynamically, among which the *JOP* strategy could leverage stochastic search to help the optimization process. Results illustrate the feasibility and effectiveness of the proposed strategy and further investigation about the accumulated revenues, PM states, and cooling power consumption helps us to see more details of the working mechanisms of the proposed strategy.

As datacenters become larger and larger and thus enormous amount of energy is still needed to power these datacenters, it can be expected that green sources of energy will attract more insights to provide power supplies instead of traditional brown energy. Our work tries to explore some strategies to migrate VMs inside a datacenter in a green-aware way. Nevertheless, there are still a lot of challenges in the field of leveraging sustainable energy to power the datacenters. On one hand, more kinds of clean energy sources besides wind and solar could be exploited, such as hydrogen and fuel cell, and their features should be studied and developed. On the other hand, how to synthetically utilize the battery, utility grid, and datacenter loads to solve the intermittency and fluctuation problems of the energy sources remains a difficult problem for system designers. In addition, it is also necessary and interesting to conduct some research on the air flow characteristics among racks and server nodes inside the datacenter room and develop some thermal-aware scheduling approaches correspondingly.

## Acknowledgements

This work is partially supported in part by National Natural Science Foundation of China (No. 61363019, No. 61563044, and No. 61640206) and National Natural Science Foundation of Qinghai Province (No. 2014-ZJ-718, No. 2015-ZJ-725).

## Author details

Xiaoying Wang[1]*, Guojing Zhang[1], Mengqin Yang[1] and Lei Zhang[1,2]

*Address all correspondence to: xy.wang@foxmail.com

1 Department of Computer Technology and Application, State Key Laboratory of Plateau Ecology and Agriculture, Qinghai University, Xining, China

2 College of Computer Science, Sichuan University, Chengdu, China

# M-ary Optical Computing

Jian Wang and Yun Long

**Abstract**

The era of cloud computing has fuelled the increasing demand on data centers for high-performance, high-speed data storage and computing. Digital signal processing may find applications in future cloud computing networks containing a large sum of data centers. Addition and subtraction are considered to be fundamental building blocks of digital signal processing which are ubiquitous in microprocessors for arithmetic operations. However, the processing speed is limited by the electronic bottleneck. It might be valuable to implement high-speed arithmetic operations of addition and subtraction in the optical domain. In this chapter, recent results of M-ary optical arithmetic operations for high base numbers are presented. By exploiting degenerate and nondegenerate four-wave mixing (FWM) in highly nonlinear fibers (HNLFs), graphene-assisted optical devices, and silicon waveguide devices, various types of two-/three-input high-speed quaternary/octal/decimal/hexadecimal optical computing operations have been demonstrated. Operation speed up to 50 Gbaud of this computing approach is experimentally examined. The demonstrated M-ary optical computing using high base numbers may facilitate advanced data management and superior network performance.

**Keywords:** high-base optical signal processing, multilevel modulation format, four-wave mixing, wavelength conversion, optical computing

## 1. Introduction

The great progress of fiber-optic communication has driven the success in transmitting/receiving very high-speed data signals in optical fiber links [1–5]. Recently, the era of cloud computing has fuelled the increasing demand on data centers for high-performance, high-speed data storage and computing. Optical interconnection is considered to be a promising technology for data interconnection in data centers. In future cloud computing networks containing a large sum of data centers, optical technology will play an important part [6–8]. For inter-data center communication, modern optical communication links will be used. Advanced modulation formats and wavelength division multiplex (WDM) can be used to enhance the transmission capacity of inter-data center links. And for intra-data center links, low-cost short-reach optical interconnection technologies, such as vertical-cavity surface-emitting laser (VCSEL) and multimode fiber, will be adopted. The rapid development of optical interconnection in data centers has also promoted increasing interest for digital signal processing used in data centers for wavelength management or routing. Among various digital signal processing operations, two important arithmetic modules, i.e., addition and subtraction, are considered to be fundamental building blocks of digital signal processing which are ubiquitous in microprocessors for arithmetic operations. However, the processing speed is limited by the electronic bottleneck. It might be valuable to implement high-speed arithmetic operations of addition and subtraction in the optical domain.

Remarkably, nonlinear optics has offered great potential to develop high-speed optical signal processing using optical nonlinearities [9–21]. Multitudinous optical signal processing functionalities have been demonstrated. Commonly used optical signal processing functionalities include wavelength (de)multiplexing, wavelength conversion, data exchange, optical addressing, optical switching, optical logic gate and computing, optical format conversion, optical equalization, tunable optical delay, optical regeneration, optical coding/decoding, and more [22–54]. As depicted in **Figure 1**, the material platforms for nonlinear optical signal processing mainly include highly nonlinear fiber (HNLF) [51, 55–57], semiconductor optical amplifier (SOA) [58–60], periodically poled lithium niobate (PPLN) waveguide [32, 35, 36, 61, 62], chalcogenide ($As_2S_3$) waveguide [63], silicon waveguide [64–66], and graphene-assisted device [67]. Previously, optical arithmetic or optical logic operations have been reported in these material systems. It is noted that most of previous research efforts are dedicated to optical computing for binary modulation formats such as on-off keying (OOK) and binary phase-shift keying (BPSK). Despite favorable operation performance achieved for binary operation, it suffers the limited bit rate and low spectral efficiency because each symbol for binary modulation formats only carries single-bit information.

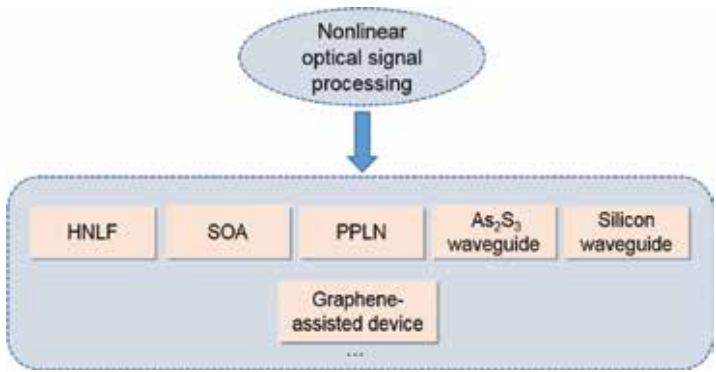

**Figure 1.** Material classification for nonlinear optical signal processing.

The use of M-ary phase-shift keying (m-PSK) and M-ary quadrature amplitude modulation (m-QAM) in coherent systems has become a key technique for efficient increase of the transmission capacity and spectral efficiency of optical communication systems. For instance, quadrature phase-shift keying (QPSK) with 2-bit information in one symbol has been extensively used in high-speed optical fiber transmission systems [68, 69]. Multilevel modulation format containing multiple constellation points in the constellation diagram can also be used to represent M-ary numbers. Taking QPSK as an example, four constellation points (i.e., four-phase levels) in the constellation diagram of QPSK signal can donate a quaternary base number (i.e., 0, 1, 2, 3), as shown in **Figure 2**. Similarly, 8 PSK (16 PSK) signal which has 8 (16) points in its constellation plane can represent an octal (hexadecimal) base number. The related optical signal processing functions to multilevel modulation formats could be addition and subtraction of high base numbers. In this scenario, a laudable goal would be to perform addition and subtraction of high base numbers because (i) high capacities might be achievable, (ii) optical spectra might be utilized efficiently, and (iii) processing throughput might be improved.

In this chapter, we tend to provide a comprehensive report of our recent research works on M-ary optical computing for multilevel modulation formats by exploiting optical nonlinearities [70–75]. Various material platforms, including HNLFs, graphene-assisted optical devices, and silicon waveguide devices, are adopted to performing high-speed M-ary addition and subtraction. First, we report the experimental results of optical addition and subtraction using HNLFs. Functionalities of quaternary addition/subtraction are examined. Second, we show the graphene-enhanced optical nonlinearities in graphene-assisted optical devices and its application in optical computing. Finally, we present the latest results of high-speed optical computing using ultracompact on-chip silicon waveguides. Quaternary/hexadecimal hybrid optical computing is suc-cessfully demonstrated in a complementary metal oxide semiconductor (CMOS)-compatible platform, which can be potentially integrated with standard CMOS large-scale integrated circuit.

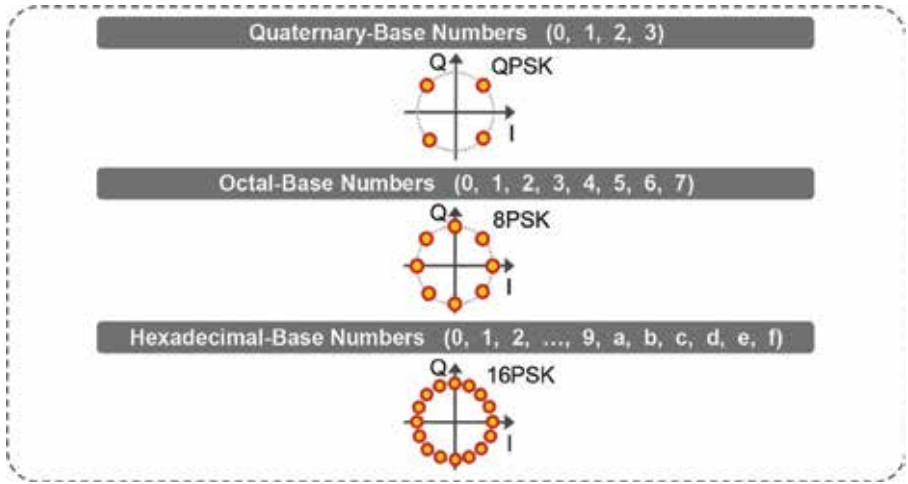

**Figure 2.** Schematic constellations of advanced multilevel modulation formats representing M-ary (quaternary, octal, hexadecimal) numbers (QPSK, 8PSK, 16PSK).

## 2. Binary optical logic

In the last two decades, binary optical computing has been widely studied. Up to now, many schemes have been demonstrated to realize various elementary optical logic operations, including AND, OR, NOT, XOR, XNOR, NAND, and NOR [32, 55, 61, 62, 76–85]. By combining multiple elementary optical logic operations, advanced logic operations such as half-adder, half-subtractor, full-adder, and full-subtractor have also been proposed and demonstrated [36, 86–91]. **Figure 3** shows an example of simultaneous half-adder, half-subtractor, and OR logic gate [36].

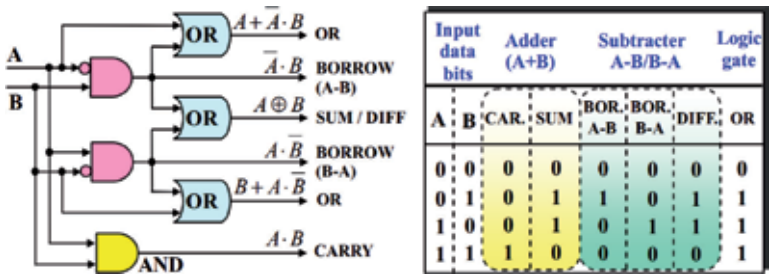

| Input data bits | | Adder (A+B) | | Subtracter A-B/B-A | | | Logic gate |
|---|---|---|---|---|---|---|---|
| A | B | CAR. | SUM | BOR. A-B | BOR. B-A | DIFF. | OR |
| 0 | 0 | 0 | 0 | 0 | 0 | 0 | 0 |
| 0 | 1 | 0 | 1 | 1 | 0 | 1 | 1 |
| 1 | 0 | 0 | 1 | 0 | 1 | 1 | 1 |
| 1 | 1 | 1 | 0 | 0 | 0 | 0 | 1 |

**Figure 3.** Digital gate-level diagram and logical truth table for simultaneous half-adder, half-subtractor, and OR logic gate.

Despite favorable operation performance of the binary operation, it still suffers from the limited bit rate and low spectral efficiency. Owing to the great success of advanced modulation format and coherent detection in optical communication, the implementation of M-ary optical computing becomes possible. Since the multiple constellation points in the complex plane of multilevel modulation format can be used to represent M-ary numbers, it is easy to extend binary optical computing to M-ary.

## 3. M-ary optical computing using HNLF

We propose and demonstrate M-ary optical computing of advanced multilevel modulation signals based on degenerate/nondegenerate FWM in HNLFs.

We first demonstrate high-speed two-input high-base optical computing (addition/subtraction/complement/doubling) of quaternary numbers using optical nonlinearities and (differential) quadrature phase-shift keying ((D)QPSK) signals. **Figure 4** illustrates the concept and operation principle of the proposed quaternary addition/subtraction/complement/doubling. Four-phase levels of (D)QPSK signal represent quaternary numbers. Three nondegenerate FWMs and three degenerate FWMs in an HNLF are exploited to simultaneously implement multiple arithmetic functions. The input of the HNLF contains two (D)QPSK signals (*A*, *B*) and one continuous wave (CW) pump. Six converted idlers (idlers 1–6) are generated by three nondegenerate FWMs (idlers 1–3) and three degenerate FWMs (idlers 4–6). The relationships between the electrical field (E) and optical phase (Φ) under non-depletion approximation are expressed as

$$E_{i1} \propto E_A \cdot E_B \cdot E_{CW}^*, \quad \Phi_{i1} = \Phi_A + \Phi_B - \Phi_{CW} \tag{1}$$

$$E_{i2} \propto E_A \cdot E_B^* \cdot E_{CW}, \quad \Phi_{i2} = \Phi_A - \Phi_B + \Phi_{CW} \tag{2}$$

$$E_{i3} \propto E_A^* \cdot E_B \cdot E_{CW}, \quad \Phi_{i3} = \Phi_B - \Phi_A + \Phi_{CW} \tag{3}$$

$$E_{i4} \propto E_{CW} \cdot E_{CW} \cdot E_A^*, \quad \Phi_{i4} = 2\Phi_{CW} - \Phi_A \tag{4}$$

$$E_{i5} \propto E_{CW} \cdot E_{CW} \cdot E_B^*, \quad \Phi_{i5} = 2\Phi_{CW} - \Phi_B \tag{5}$$

$$E_{i6} \propto E_B \cdot E_B \cdot E_{CW}^*, \quad \Phi_{i6} = 2\,\Phi_B - \Phi_{CW} \tag{6}$$

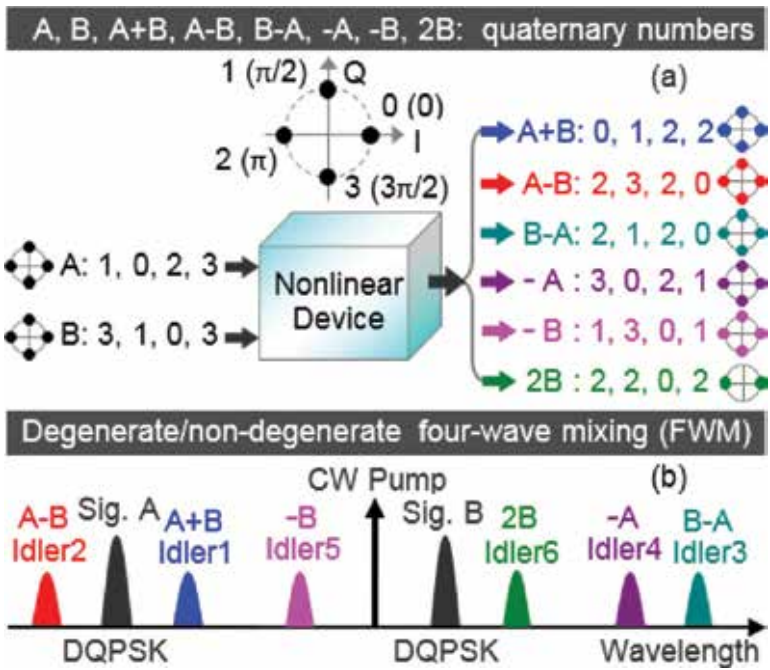

**Figure 4.** (a) Concept and (b) principle of two-input high-base optical computing (quaternary addition/subtraction/complement/doubling) using a single nonlinear device and (D)QPSK signals.

Owing to the phase wrap characteristic with a periodicity of $2\pi$, it is implied from Eqs. (1) to (6) that idlers 1–6 carry out modulo four operations of quaternary addition (A + B), dual-directional subtraction (A − B, B − A), complement (−A, −B), and doubling (2B), respectively.

Shown in **Figure 5** are measured spectra. Two 100-Gbit/s $2^7$-1 RZ-(D)QPSK signals (A, 1546.6 nm; B, 1555.5 nm), and a CW pump (1553.2 nm), are launched into a 460-m HNLF. The low and flat dispersion of HNLF enables multiple FWM processes, and thus six idlers are obtained. The six idlers correspond to addition (A + B), subtraction (A − B, B − A), complement (−A, −B), and doubling (2B) of quaternary numbers (A, B), respectively.

We measured waveforms and balanced eyes of the demodulated in-phase (I) and quadrature (Q) components of two input 100-Gbit/s (D)QPSK signals and six converted idlers. The 100-Gbit/s (D)QPSK signal is demodulated using a delay-line interferometer (DLI) with a 20 ps delay difference between two arms. The obtained results are shown in **Figures 6** and **7**, which confirm the successful implementation of 50-Gbaud quaternary addition (A+B), dual-directional subtraction (A−B, B−A), complement (−A, −B), and doubling (2B) based on FWM in an HNLF.

**Figure 8** shows the bit error rate (BER) curves. The power penalty is about 4 dB for addition, while 3 dB for subtraction, 2 dB for complement, and 3.1 dB for doubling at a BER of $10^{-9}$. The measured constellations using an optical complex spectrum analyzer are shown in **Figure 9**. One can clearly see that addition (A+B) and subtraction (A−B, B−A) have four-phase levels (0, $\pi/2$, $\pi$, $3\pi/2$), while doubling (2B) has only two-phase levels (0, $\pi$).

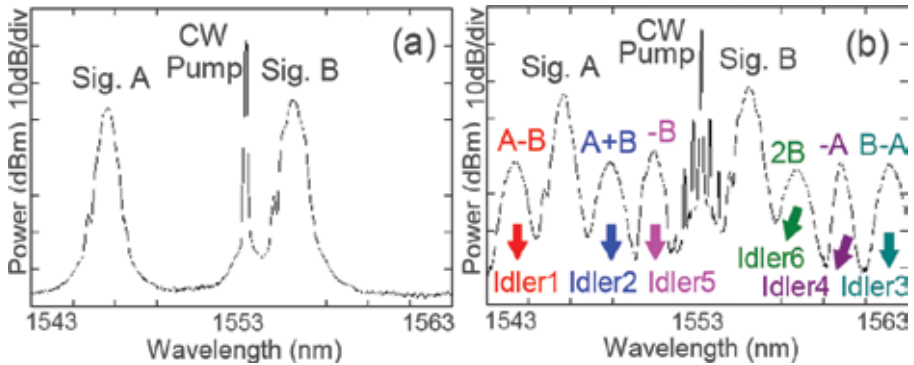

**Figure 5.** Measured spectra for 50-Gbaud two-input quaternary addition/subtraction/complement/doubling (a) before and (b) after HNLF.

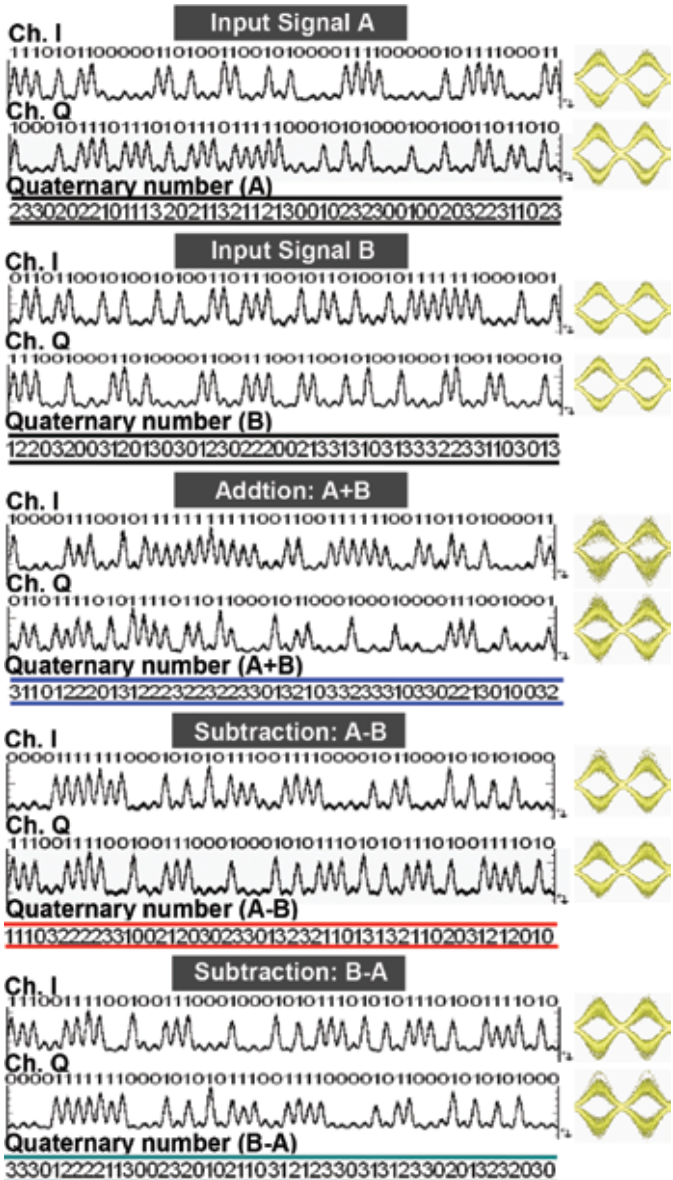

**Figure 6.** Demodulated waveforms and balanced eyes for 50-Gbaud two-input quaternary addition and dual-directional subtraction using 100-Gbit/s (D)QPSK signals.

**Figure 7.** Demodulated waveforms and balanced eyes for 50-Gbaud quaternary complement and doubling using 100-Gbit/s (D)QPSK signals.

**Figure 8.** Measured BER curves for input/output signals (A, B), quaternary addition (A+B), dual-directional subtraction (A−B, B−A), complement (−A, −B), and doubling (2B).

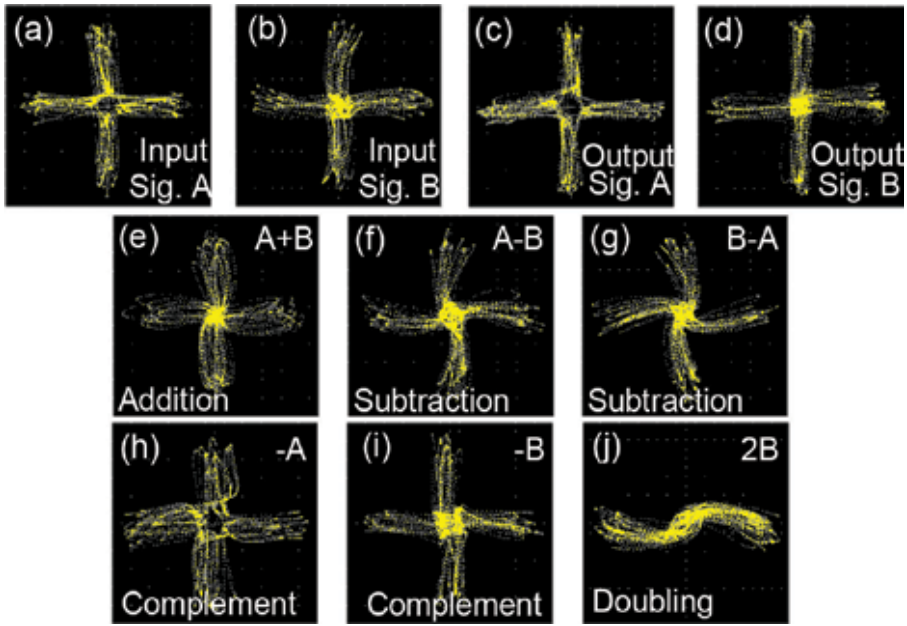

**Figure 9.** Measured constellations for 50-Gbaud two-input quaternary addition, dual-directional subtraction, complement, and doubling using 100-Gbit/s (D)QPSK signals.

## 4. Graphene-enhanced optical nonlinearity for M-ary optical computing

Graphene as a purely two-dimensional material with only one-carbon-atom thickness has received great interest since it features many interesting and useful electrical, optical, chemical, and mechanical properties [92, 93]. Over the last decade, many remarkable optical properties of graphene have been discovered, such as self-luminosity, tunable optical absorption, strong nonlinearity, saturable absorption, etc. [94–96]. Recently, optical nonlinearities have been observed in graphene in various configurations, e.g., slow-light graphene-silicon photonic crystal waveguide [97], graphene optically deposited onto fiber ferrules [98], and graphene based on microfiber [99]. The large absorption and Pauli blocking effect in graphene, together with the ultrafast carrier dynamics and strong optical nonlinearity with a fast response time, make graphene-based photonic devices suitable for performing efficient nonlinear functions. Very recently, an experimental observation of FWM-based wavelength conversion of a 10-Gb/s non-return-to-zero (NRZ) signal was reported [100]. In this section, we introduce our recent progress in optical M-ary computing functions using a graphene-assisted nonlinear optical device.

**Figure 10** illustrates the fabrication process of the graphene-assisted nonlinear optical device. First, a monolayer graphene was grown on a Cu foil by the chemical vapor deposition (CVD) method. Poly(methyl methacrylate) (PMMA) film was next spin coated on the surface of the graphene-deposited Cu foil, and the Cu foil was etched away with 1 M FeCl₃ solution. The resultant PMMA/graphene film (5 mm × 5 mm) was then washed in deionized water several times and transferred to deionized water solution or Si/SiO₂ substrate. Then, the floating PMMA/graphene sheet was mechanically transferred onto the fiber pigtail cross section and dried in a cabinet. After drying at room temperature for about 24 hours, the carbon atoms could be self-assembled onto the fiber end facet. The PMMA layer was finally removed by boiling acetone. By connecting this graphene-on-fiber component with another clean and dry fiber connector, the nonlinear optical device was thereby constructed for nonlinear optical signal processing applications.

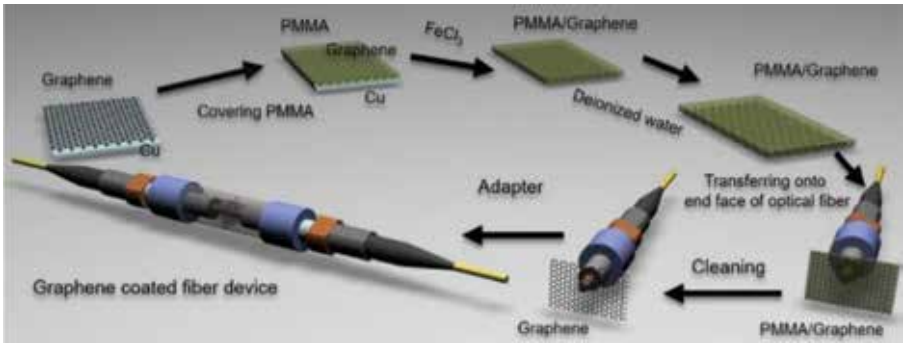

**Figure 10.** Fabrication process of the graphene-assisted nonlinear optical device.

**Figure 11(a)** depicts the optical microscope (OM) image of the grown graphene film transferred on a 300-nm SiO₂/Si substrate. **Figure 11(b)** shows a scanning electron microscopy (SEM) image of the graphene sheet transferred on silicon-on-insulator (SOI). One can clearly see the evidence of the uniformity of the graphene. The Raman spectrum of the graphene, as displayed in **Figure 11(c)**, shows a weak D peak and a strong 2D peak. The D to G peak intensity ratio is ~0.08, which indicates that the graphene formed on a SiO₂/Si substrate was almost defect-free.

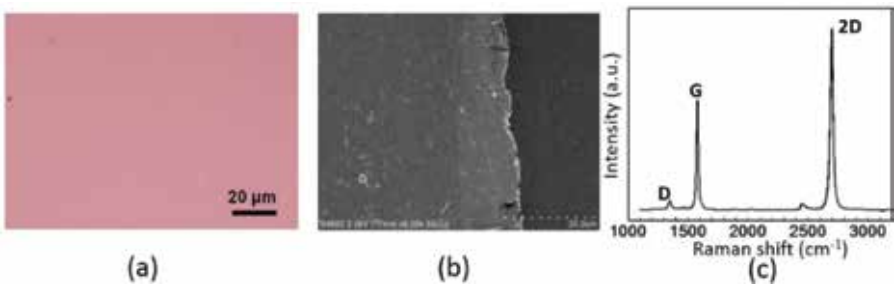

**Figure 11.** (a) Optical microscope (OM) image of graphene transferred on a SiO₂/Si substrate. (b) SEM image of graphene transferred on silicon-on-insulator (SOI). (c) Typical Raman spectrum of single-layer graphene on a SiO₂/Si substrate (excitation wavelength: 532 nm).

We first examine the wavelength conversion of the graphene-assisted nonlinear optical device. **Figure 12(a)** shows a typical output FWM spectrum obtained after the CVD single-layer graphene-coated fiber device. In the experiment, the signal wavelength is fixed at 1550.12 nm. A newly converted idler at 1546.88 nm is generated when the pump is set to be 1548.49 nm. We also measure the output spectrum without graphene for reference under the same experimental conditions. As clearly shown in the inset of **Figure 12(a)**, the power of converted idler without graphene is observed to be ~5.5 dB lower than the one with graphene. That is, under the same experimental conditions, the converted idler without graphene is ~71.9% lower than the one with graphene. Hence, the degenerate FWM in graphene contributes more in the wavelength conversion process. The insets of **Figure 12(a)** also depict measured QPSK constellations of the converted idler and the input signal. We also present a comparison of the FWM conversion efficiency as a function of the pump power with and without graphene. As shown in **Figure 12(b)**, the pump wavelength is fixed at $\lambda_{pump}$ = 1548.49 nm and the signal is $\lambda_{signal}$ = 1550.12 nm. One can clearly see that the conversion efficiency increases with the pump power. When the pump power varies from 23 dBm to 33 dBm, the enhanced FWM conversion efficiency by graphene changes from 4.7 dB to 7.5 dB.

**Figure 13(a)** plots the converted idler wavelength as a function of the pump wavelength when the pump power is fixed at 31 dBm. A linear wavelength relationship between the converted idler and pump is observed. The measured FWM conversion efficiency of tunable wavelength conversion with and without graphene is shown in **Figure 13(b)**. The signal wavelength is fixed at 1550.12 nm and the pump wavelength is tuned from 1547 to 1553 nm. When using graphene-coated fiber device, the conversion efficiency varies about 1.7 dB within a ~6 nm wavelength range. By comparing the measured pump wavelength-dependent conversion efficiency with and without graphene, one can clearly see that the FWM conversion efficiency with graphene is enhanced more than 5 dB within the tuning range of pump wavelength.

To characterize the performance of QPSK wavelength conversion, we further measure the BER curve as a function of the received observed OSNR for B-to-B signal and converted idler.

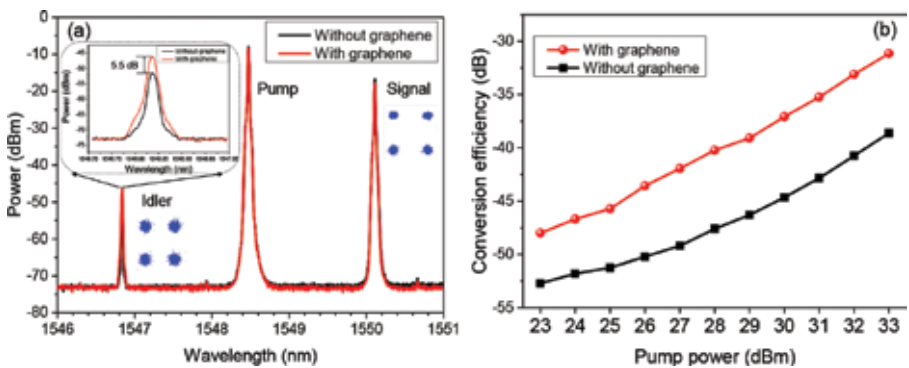

**Figure 12.** (a) Measured FWM spectra with (circle) and without (square) graphene. (b) Measured conversion efficiency of FWM with and without graphene when pump power is tuned from 23 to 33 dBm.

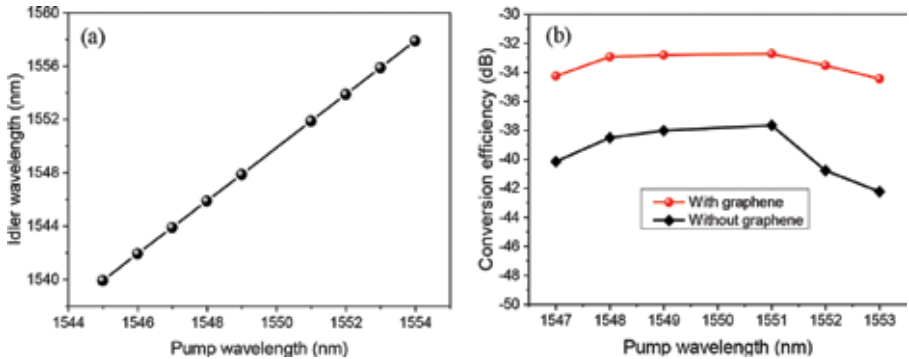

**Figure 13.** (a) Converted idler wavelength versus pump wavelength. (b) Measured FWM conversion efficiency with and without graphene when pump wavelength is tuned from 1547 to 1553 nm. Pump power: 31 dBm.

**Figure 14** plots measured BER performance for tunable QPSK wavelength conversion with the converted idler generated at 1546.88, 1539.92, and 1557.90 nm, respectively. The measured conversion efficiencies for converted idlers at 1546.88, 1539.92, and 1557.90 nm are −36.2, −48.2, and −39.8 dB, respectively. As shown in **Figure 14**, the observed OSNR penalty is around 1 dB at a BER of $1 \times 10^{-3}$ (7% forward error correction (FEC) threshold) for QPSK wavelength conversion with the converted idler at 1546.88 nm. The received OSNR penalties of ~2.2 dB at a BER of $1 \times 10^{-3}$ are observed for converted idlers at 1539.92 and 1557.90 nm. The increased OSNR penalty is mainly due to the reduced conversion efficiency for converted idlers at 1539.92 and 1557.90 nm. The right insets of **Figure 14** depict corresponding constellations of the B-to-B signals and converted idlers. The obtained results shown in **Figures 11–14** imply favorable performance achieved for tunable wavelength conversion of QPSK signal using a fiber pigtail cross section coated with a single-layer graphene.

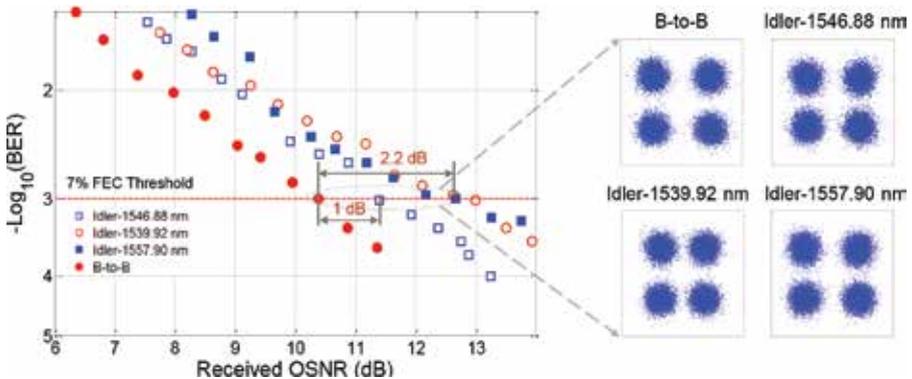

**Figure 14.** Measured BER versus received OSNR for wavelength conversion of QPSK signal. Insets show constellations of QPSK.

We then show the results of optical computing based on the fabricated graphene-assisted nonlinear optical device. **Figure 15** illustrates the concept and principle of two-input hybrid quaternary arithmetic functions. From the constellation in the complex plane (**Figure 15(a)**), it is clear that one can use four-phase levels ($\pi/4$, $3\pi/4$, $5\pi/4$, $7\pi/4$) of (D)QPSK to represent quaternary base numbers (0, 1, 2, 3). To implement two-input hybrid quaternary arithmetic functions, the aforementioned graphene-assisted nonlinear optical device is employed. Two-input quaternary numbers (A, B) are coupled into the nonlinear device, and then two converted idlers (idler 1, idler 2) are simultaneously generated by two degenerate FWM processes. **Figure 15(b)** illustrates the degenerate FWM process. We derive the electrical field (E) and optical phase ($\Phi$) relationships of two degenerate FWM processes under the pump non-depletion approximation expressed as

$$E_{i1} \propto E_A \cdot E_A \cdot E_B^*, \quad \Phi_{i1} = \Phi_A + \Phi_A - \Phi_B \tag{7}$$

$$E_{i2} \propto E_B \cdot E_B \cdot E_A^*, \quad \Phi_{i2} = \Phi_B + \Phi_B - \Phi_A \tag{8}$$

where the subscripts A, B, i1, and i2 denote input signal A, signal B, converted idler 1, and idler 2, respectively. Owing to the phase wrap characteristic with a periodicity of $2\pi$, it is implied from the linear phase relationships in Eqs. (7) and (8) that idler 1 and idler 2 carry out modulo 4 operations of hybrid quaternary arithmetic functions of doubling and subtraction (2A−B, 2B−A).

**Figure 16** depicts measured typical spectrum obtained after the CVD single-layer graphene-coated fiber device. Two 10-Gbaud NRZ-(D)QPSK signals at 1550.10 (A) and 1553.60 nm (B) are employed as two inputs. The power of two input signals (A, B) is about 32 dBm. The conversion efficiency is measured to be around −36 dB. One can clearly see that two converted idlers are obtained by two degenerate FWM processes with idler 1 at 1546.60 nm (2A−B) and idler 2 at 1557.20 nm (2B−A). The resolution of the measured spectrum is set to 0.02 nm. The steps in the measured spectrum are actually the modulation sidebands of two NRZ-(D)QPSK carrying signals. In order to verify the hybrid quaternary arithmetic functions, we measure the phase of symbol sequence for two input signals and two converted idlers, as shown in **Figure 17**. By carefully comparing the quaternary base numbers for two input signals and two converted idlers, one can confirm the successful implementation of two-input hybrid quaternary arithmetic functions of 2A−B and 2B−A.

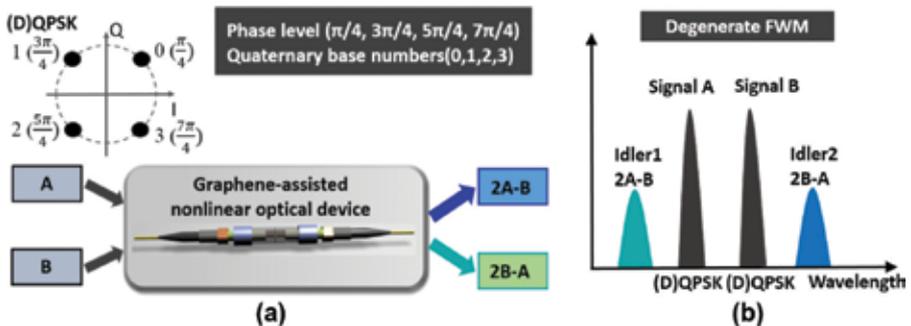

**Figure 15.** (a) Concept and (b) principle of hybrid quaternary arithmetic functions (2A−B, 2B−A) using degenerate FWM and (D)QPSK signals.

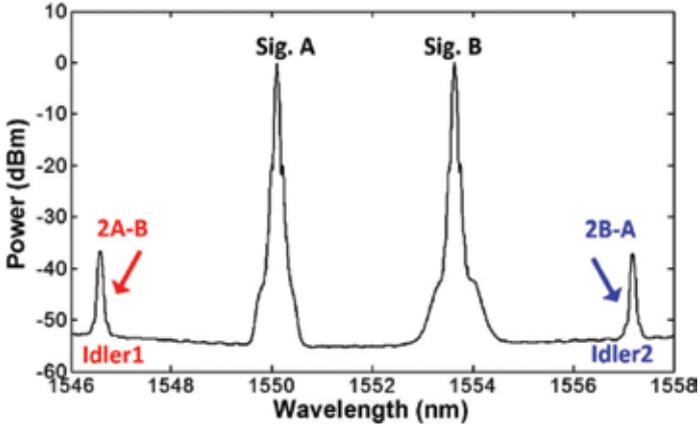

**Figure 16.** Measured spectrum for 10-Gbaud two-input hybrid quaternary arithmetic functions.

We further investigate the BER performance for the proposed optical two-input hybrid quaternary arithmetic functions. The OSNR penalties at a BER of 2×10⁻³ for hybrid quaternary arithmetic functions are measured to be about 7.4 dB for 2A−B and 7.0 dB for 2B−A. The insets in **Figure 18(a)** show constellations of the last point of the BER curves of output Sig. B and 2A−B. The constellation of Sig. B is measured under an OSNR of 12.6 dB, while the constellation of 2A−B is observed under an OSNR of 19.6 dB. To clearly show the differences between these two constellations, we also assess the EVM of these two constellations, i.e., EVM = 27.61% for output Sig. B and EVM = 30.09% for output 2A−B. The significant performance degradations for the two-input hybrid quaternary arithmetic functions (2A−B, 2B−A) might be ascribed to the relatively low conversion efficiency for two converted idlers at 1546.60 nm and 1557.20 nm and accumulated distortions transferred from two input signals (A, B). It is possible to further enhance the conversion efficiency by appropriately increasing the number of graphene layers employed in the experiment. **Figure 18(b)** depicts the BER performance as a function of the relative time offset between two signals (signal offset) under an OSNR of ~20 dB. It is found that the BER is kept below enhanced forward error correction (EFEC) threshold when the signal offset/symbol time is within 15 ps, which indicates a favorable tolerance to the signal offset.

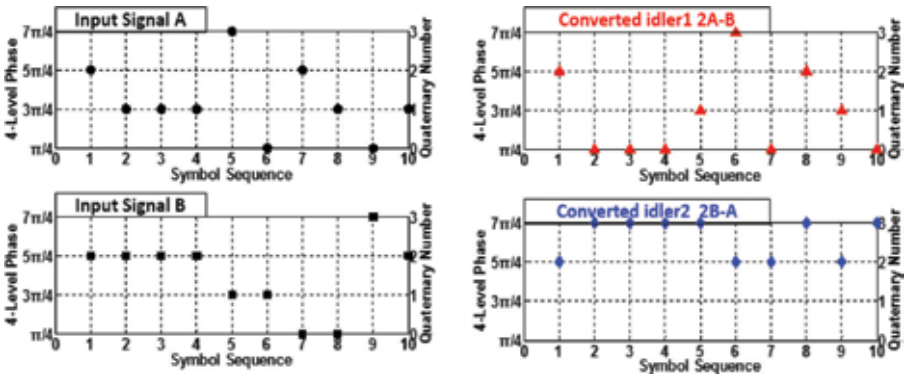

**Figure 17.** Measured phase of symbol sequence with coherent detection for 10-Gbaud two-input hybrid quaternary arithmetic functions.

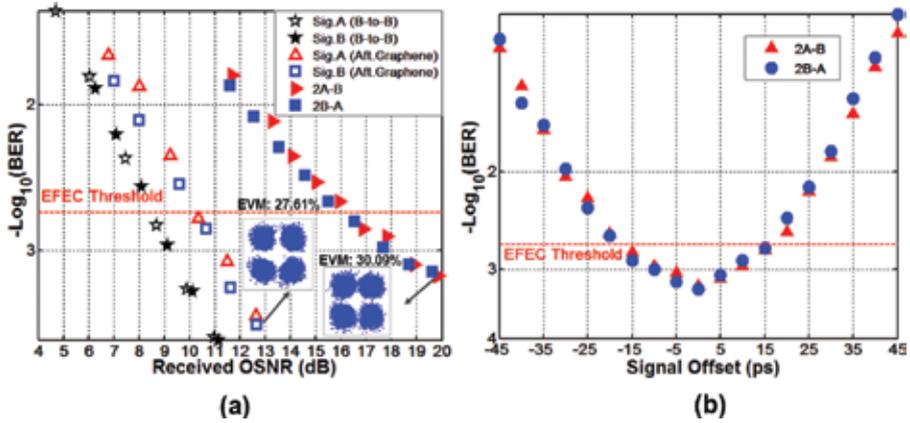

**Figure 18.** Measured BER curves for two-input hybrid quaternary arithmetic functions of 2A−B and 2B−A; (b) BER versus signal offset.

We also propose an approach to performing three-input optical addition and subtraction of quaternary base numbers using multiple nondegenerate FWM processes based on graphene-assisted device.

**Figure 19** illustrates the concept and working principle of the proposed graphene-assisted three-input high-base optical computing. Three input (D)QPSK signals (A, B, C) are launched into the nonlinear device, in which three converted idlers (idler 1, idler 2, idler 3) are simultaneously generated by three nondegenerate FWM processes. Quaternary hybrid addition and subtraction of A+B−C, A+C−B, and B+C−A are obtained simultaneously.

In the experiment, the wavelengths of three input signals A, B, and C are fixed at 1548.52, 1550.12, and 1552.52 nm, respectively. **Figure 20** depicts measured typical optical spectrum obtained after the single-layer graphene-coated fiber device. One can clearly see that three converted idlers are generated by three nondegenerate FWM processes with idler 1 at 1546.13 nm (A+B−C), idler 2 at 1550.92 nm (A+C−B), and idler 3 at 1554.13 nm (B+C−A), respectively. The conversion efficiencies of three nondegenerate FWM processes are measured to be larger than −34 dB. In order to verify the quaternary optical computing functions, we measure the phase of symbol sequence for three input signals and three converted idlers, as shown in

**Figure 21**. By carefully comparing the quaternary base numbers for three input signals and three converted idlers, one can confirm the successful implementation of graphene-assisted three-input quaternary optical computing (i.e., quaternary hybrid addition and subtraction) functions of A+B−C, A+C−B, and A+C−B.

To characterize the performance of the proposed graphene-assisted three-input high-base optical computing functions, we further measure the BER curves as a function of the received OSNR for B-to-B signals and three converted idlers. **Figure 22** depicts measured BER curves for 10-Gbaud three-input quaternary hybrid addition and subtraction of A+B−C, A+C−B, and B+C−A. As shown in **Figure 22**, the observed OSNR penalties of three-input quaternary hybrid addition and subtraction are accessed to be less than 7 dB at a BER of 2×10⁻³ (7% EFEC threshold). The increased OSNR penalties might be mainly due to the relatively low conversion efficiency for converted idlers and accumulated distortions transferred from three input signals (A, B, C). The insets in **Figure 22** depict corresponding constellations of the B-to-B signals and converted idlers. The BER curves and constellations of three output signals (A, B, C) after graphene are also shown in **Figure 22** for reference.

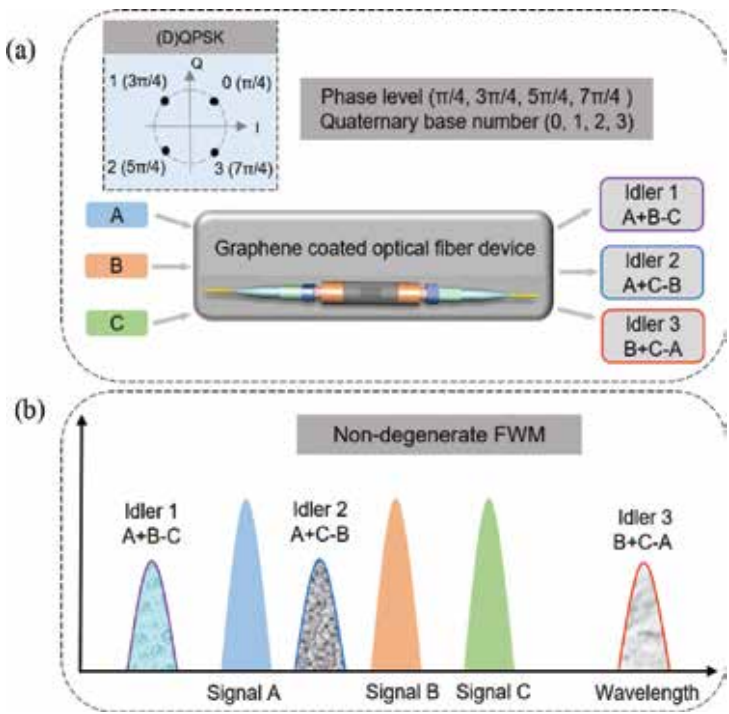

**Figure 19.** (a) Concept and (b) principle of graphene-assisted three-input (A, B, C) quaternary hybrid addition and subtraction (A+B−C, A+C−B, B+C−A) using nondegenerate FWM and (D)QPSK signals.

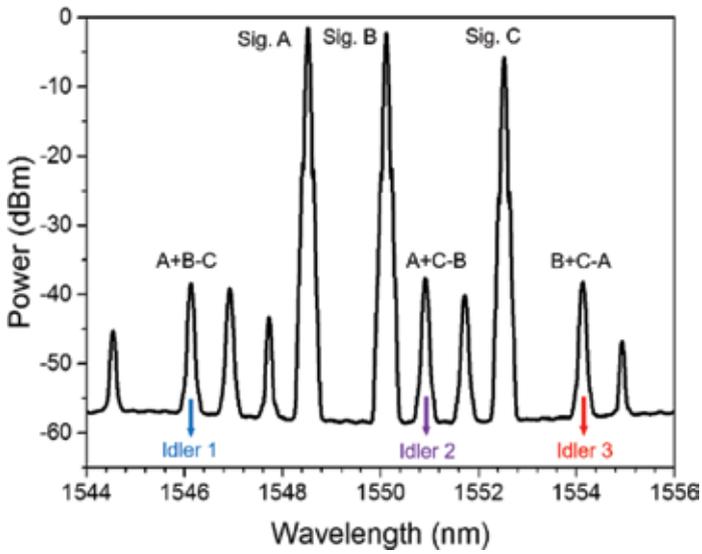

**Figure 20.** Measured spectrum for 10-Gbaud three-input quaternary hybrid addition and subtraction.

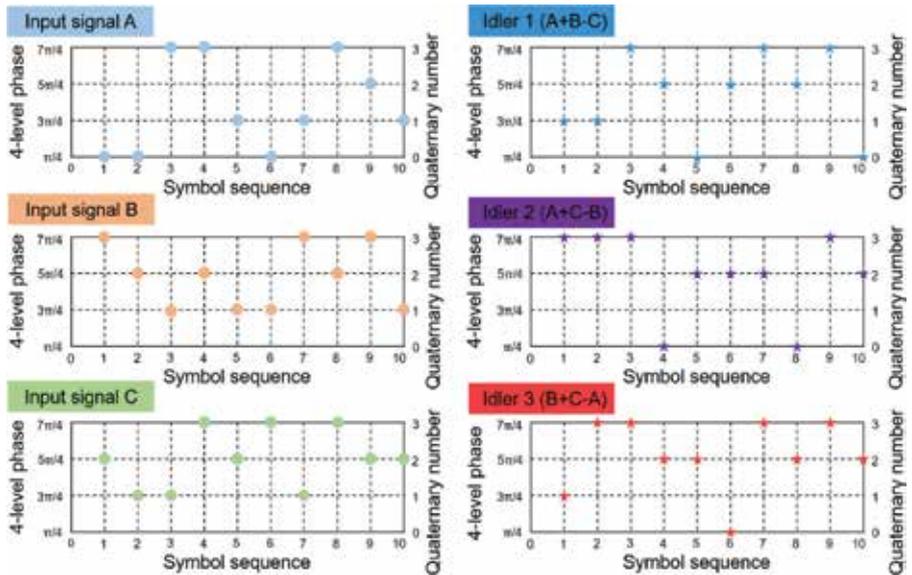

**Figure 21.** Measured phase of symbol sequence by coherent detection for 10-Gbaud three-input quaternary hybrid addition and subtraction.

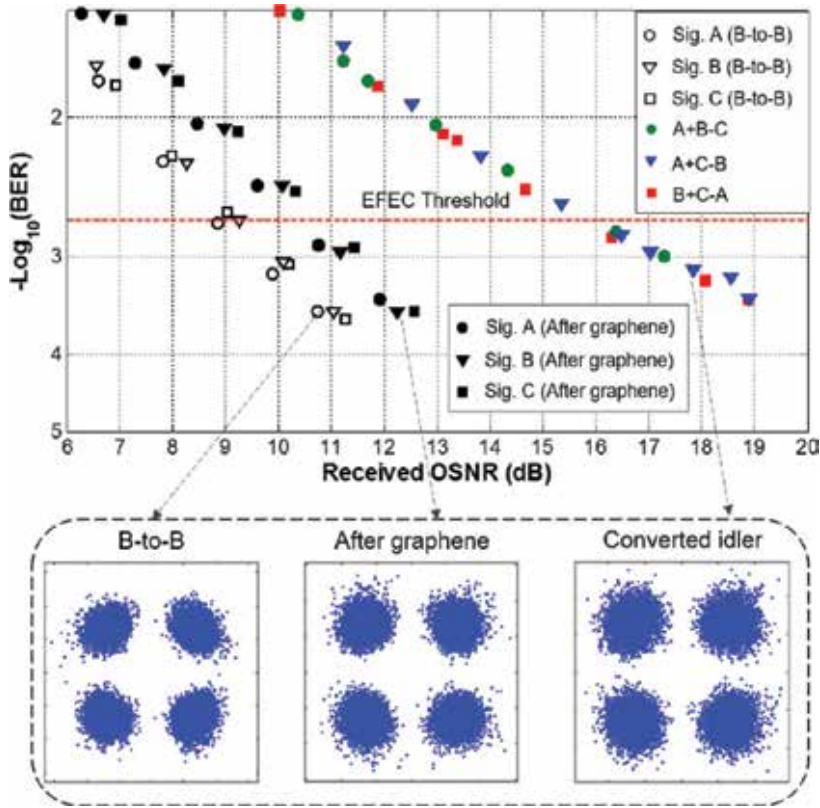

**Figure 22.** Measured BER curves for 10-Gbaud three-input quaternary hybrid addition and subtraction of A+B−C, A−C−B, and B+C−A. Insets show constellations of (D)QPSK signals.

## 5. On-chip M-ary optical computing

Silicon photonics has become one of the most promising photonic integration platforms for its ultrahigh level of integration, low power consumption, and CMOS compatibility. In addition, nonlinear interaction will also be enhanced in silicon waveguides due to its tight light confinement. Thus, SOI is also considered to be a favorable nonlinear optics platform. To minimize the footprint of the computing building block and lower the power consumption, we demonstrate on-chip M-ary optical computing by adopting silicon photonics technology.

We first experimentally demonstrate all-optical two-input (A, B) optical quaternary doubling/subtraction (2A−B, 2B−A) using a silicon waveguide. The silicon waveguide used in the experiment is shown in **Figure 23**.

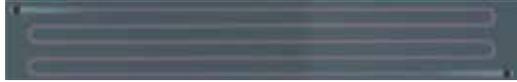

**Figure 23.** Photomicrograph of the silicon waveguide.

**Figure 24** shows the measured symbol sequence for two-input optical quaternary hybrid doubling/subtraction. It can be confirmed from **Figure 24** that simultaneous quaternary hybrid doubling/subtraction (2A−B, 2B−A) are successfully implemented using QPSK, degenerate FWM, and coherent detection.

We also experimentally demonstrate three-input (A, B, C) optical quaternary addition/subtraction (A+C−B, A+B−C, B+C−A) using such a silicon waveguide. **Figure 25** shows the measured symbol sequence for three-input optical quaternary addition/subtraction.

It is relatively difficult to experimentally demonstrate higher-order computing using a pure silicon waveguide due to the large OSNR penalty. Thus, we simulate hexadecimal optical computing using nonlinear interactions in a silicon-organic hybrid slot waveguide [101]. **Figure 26(a)** shows the structure of a silicon-organic hybrid slot waveguide. It features a sandwich structure with a low-refractive-index PTS [polymer poly (bis para-toluene sulfonate)] layer surrounded by two high refractive index silicon layers. The TM mode profile and its power density along x/y directions are depicted in **Figure 26(b)–(d)**. Tight light confinement is observed in the nanoscale nonlinear organic slot region, which offers high nonlinearity and instantaneous Kerr response. We assess the effective mode area and nonlinearity to be $7.7 \times 10^{-14}$ m$^2$ and 5500 w$^{-1}$m$^{-1}$, which can potentially facilitate efficient optical signal processing (e.g., hexadecimal addition/subtraction).

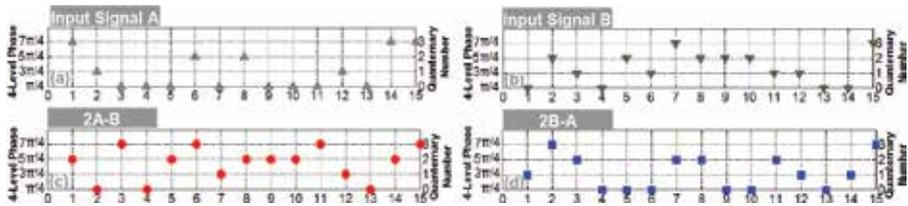

**Figure 24.** Measured symbol sequence for two-input optical quaternary addition/subtraction (2A−B, 2B−A).

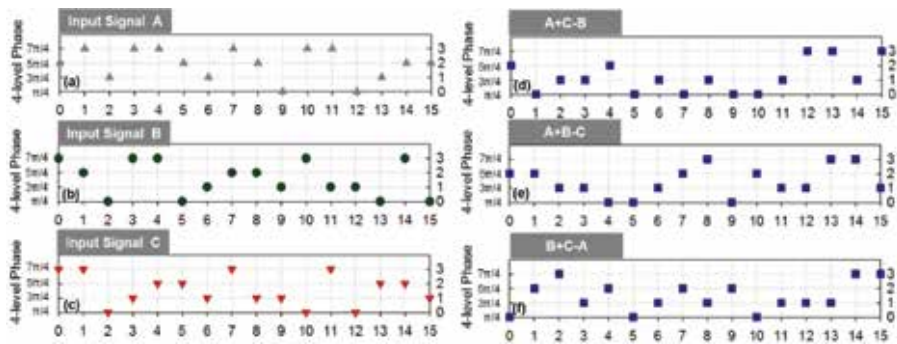

**Figure 25.** Measured symbol sequence for (a)–(c) three-input optical quaternary signal and their (c)–(e) addition/subtraction.

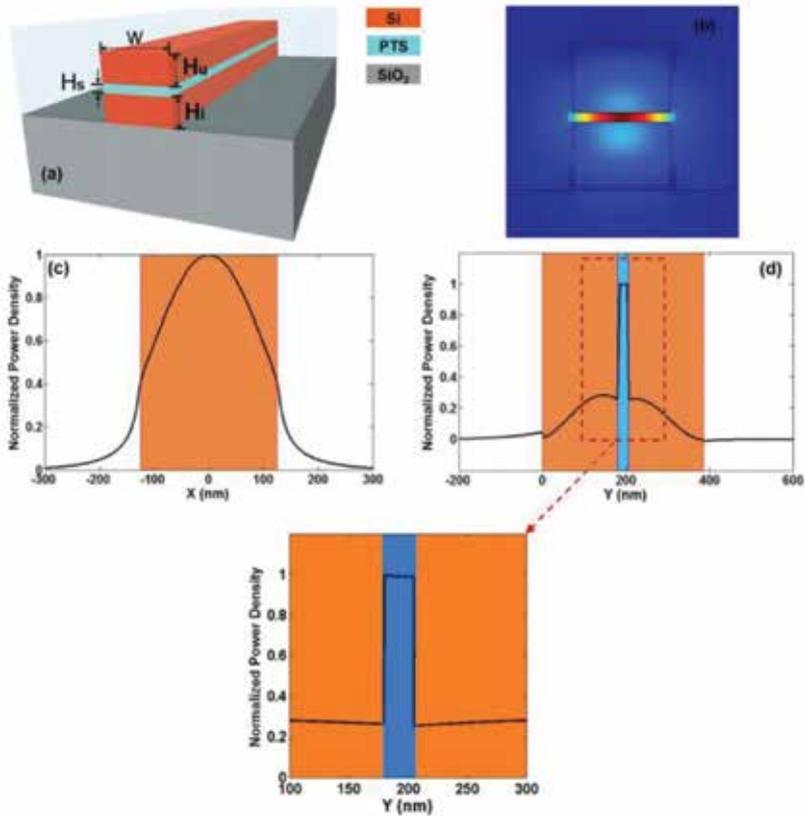

**Figure 26.** (a) 3D structure, (b) mode distribution, (c) and (d) normalized power density along x and y directions of a silicon-organic hybrid slot waveguide.

**Figure 27** depicts simulation results for three-input multicasted 40-Gbaud (160-Gbit/s) hexadecimal addition/subtraction. Twenty symbol sequences are plotted in **Figure 27**, which confirms the successful implementation of three-input hexadecimal addition/subtraction (A + B − C, A + C − B, B + C − A, A + B + C, A − B − C, B − A − C). The constellations are also shown in **Figure 28**.

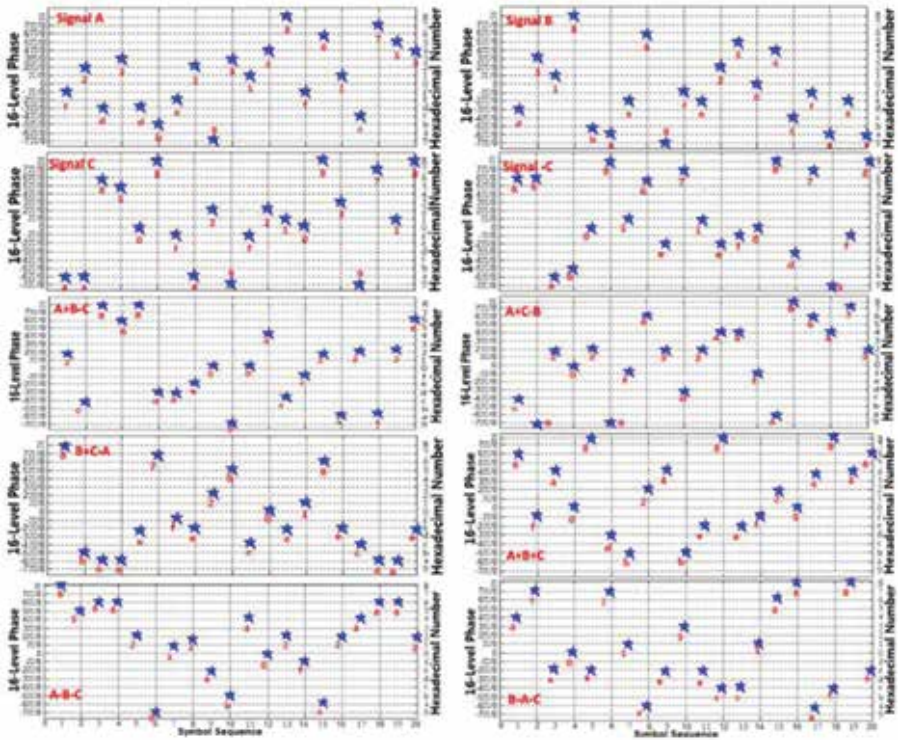

**Figure 27.** Simulated symbol sequence for three-input multicasted 40-Gbaud (160-Gbit/s) hexadecimal addition/subtraction using a silicon-organic hybrid slot waveguide.

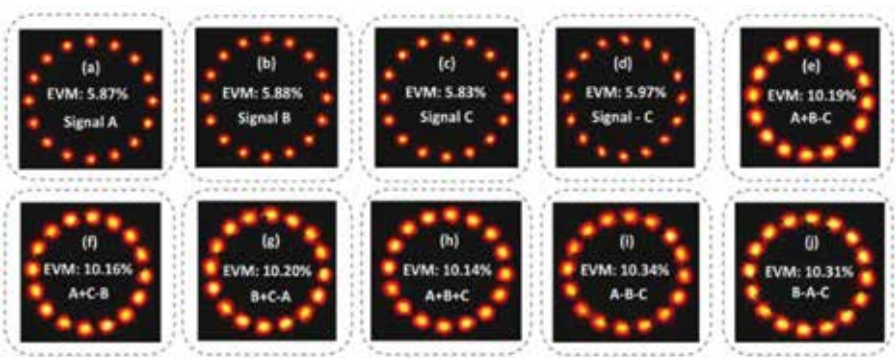

**Figure 28.** Simulated constellations for three-input multicasted 40-Gbaud (160-Gbit/s) hexadecimal addition/subtraction using a silicon-organic hybrid slot waveguide.

We further investigate the EVM of input signals and output idlers as functions of the OSNR of input signals. The results are shown in **Figure 29(a)** and **(b)**. The EVM penalties are less than 4.5 for hexadecimal addition/subtraction under a 28-dB OSNR. EVM of hexadecimal addition/subtraction as a function of input signal power are shown in **Figure 30**. EVM increases slightly (<0.8 dB) with input signal power <50 mW, which implies a large available dynamic range (~27 dB).

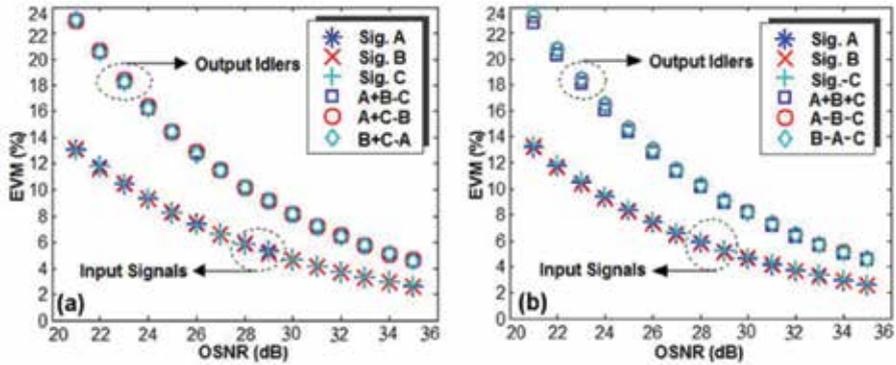

**Figure 29.** Simulated EVM versus OSNR for 40-Gbaud (160-Gbit/s) hexadecimal addition/subtraction using a silicon-organic hybrid slot waveguide.

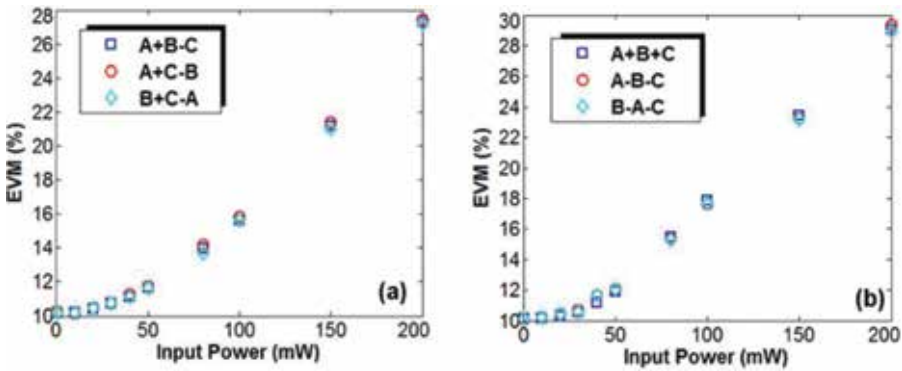

**Figure 30.** Simulated dynamic range of signal power for 40-Gbaud (160-Gbit/s) hexadecimal addition/subtraction using a silicon-organic hybrid slot waveguide.

## 6. Conclusion

In this chapter, we have reviewed recent research efforts toward M-ary optical computing by adopting multilevel modulation signals and exploiting optical nonlinearities.

1. M-ary optical computing using HNLF: By adopting 100-Gbit/s two-input (D)QPSK signals (A, B) and exploiting three degenerate FWM processes and three nondegenerate FWM processes in an HNLF, simultaneous 50-Gbaud two-input quaternary addition (A+B), dual-directional subtraction (A−B, B−A), complement (−A, −B), and doubling (2B) have been demonstrated in the experiment.

2. Graphene-enhanced optical nonlinearity for M-ary optical computing: We experimentally demonstrated hybrid two-/three-input quaternary addition/subtraction optical computing in a graphene-assisted nonlinear devices.

3. On-chip M-ary optical computing: To minimize the footprint of the computing building block and lower the power consumption, we demonstrate on-chip M-ary optical computing by adopting silicon photonics technology. We experimentally demonstrated on-chip quaternary addition/subtraction optical computing in a silicon waveguide. On-chip hexadecimal addition/subtraction is also numerically investigated using a silicon-organic hybrid slot waveguide.

Addition and subtraction are considered to be fundamental building blocks of digital signal processing. Optical signal processing technology opens a new world for ultrahigh-speed arithmetic operations. With future improvements, other different optical nonlinearities on various nonlinear optical device platforms would also be employed to flexibly manipulate the amplitude and phase information of advanced multilevel modulation signals. In addition, more complicated computing functionalities can be introduced, which might open diverse interesting applications in robust optical computing operation.

## Acknowledgements

This work was supported by the National Program for Support of Top-notch Young Professionals, the National Natural Science Foundation of China (NSFC) under grants 61222502, 11574001, and 11274131, the Program for New Century Excellent Talents in University (NCET-11-0182), the National Basic Research Program of China (973 Program) under grant 2014CB340004, the Wuhan Science and Technology Plan Project under grant 2014070404010201, the Fundamental Research Funds for the Central Universities (HUST) under grants 2012YQ008 and 2013ZZGH003, and the seed project of Wuhan National Laboratory for Optoelectronics (WNLO). The authors thank the Center of Micro-Fabrication and Characterization (CMFC) of WNLO for the support in the manufacturing process of silicon waveguides. The authors also thank the facility support of the Center for Nanoscale Characterization and Devices of WNLO.

## Author details

Jian Wang* and Yun Long

*Address all correspondence to: jwang@hust.edu.cn

Wuhan National Laboratory for Optoelectronics, School of Optical and Electronic Information, Huazhong University of Science and Technology, Wuhan, Hubei, China

# Networking Solutions for Integrated Heterogeneous Wireless Ecosystem

Roman Florea, Aleksandr Ometov, Adam Surak, Sergey Andreev and Yevgeni Koucheryavy

Additional information is available at the end of the chapter

**Abstract**

As wireless communications technology is steadily evolving to improve the offered connectivity levels, additional research on emerging network architectures is becoming timely to understand the applicability of both traditional and novel networking solutions. This chapter concentrates on the utilization of cloud computing techniques to construct feasible system prototypes and demonstrators within the rapidly maturing heterogeneous wireless ecosystem. Our first solution facilitates cooperative radio resource management in heterogeneous networks. The second solution enables assisted direct connectivity between proximate users. The contents of the chapter outline our corresponding research and development efforts as well as summarize the major experiences and lessons learned.

**Keywords:** traffic offloading, heterogeneous network, SDN, LTE, WiFi, 5G

## 1. Introduction

### 1.1. Motivation

This chapter describes application of cloud computing techniques to improve technology research process with examples in the area of modern wireless networks and in various environments built around these networks. Cloud computing toolbox proved to be very useful in augmenting research in domains of heterogeneous networks (HetNets) [1] with prototypes, testbeds and demonstrators. These research domains include centralized radio resource management in emerging cellular network architectures, network assistance role in device-to-device (D2D) communications [2] and study of prospective services in these networks.

**INTECH**

open science | open minds

Applying virtualization and programmability principles to the research infrastructure resulted in a flexible environment providing the researchers with tools to easily orchestrate various scenarios and network layouts. Virtualization platform allowed to quickly set up new architecture entities while programmable and scriptable networking allowed to build overlay networks interconnecting the entities. Using these tools facilitated the processes of designing and assessing future network architectures demonstrating certain connectivity and functionality interesting for the research.

First use case demonstrated in this chapter elaborates on designing a platform that would mimic a device connected through a heterogeneous network, allowing researchers to experiment with traffic flow optimization in an environment close to the envisioned next-generation network architecture. Prototype solution and testbed were designed building on software defined network principles of automation, abstraction and software-based flow switching and were implemented using overlay networks and virtual network functions.

Second demonstrated use case is within D2D communications research, where the task was to design architecture demonstrating feasibility of traffic offloading from infrastructure network onto direct links. Prototype was implemented with automated routing control in overlay network.

Resulting research environment supports the envisioned use of cloud computing in production networks with such technology trends as Heterogeneous Cloud Radio Access Network and Virtual Evolved Packet Core, as well as encourages researchers to adopt cloud tools in their work.

## 1.2. Potential solutions

Fortunately, the evolution of information technology (IT) and communications is not all about challenges. Over the last decade, the development of technology in various directions such as cloud computing, virtualization, high speed fiber optics communications, efficient signal processing, advances in data center IT architectures, etc., has provided researchers and engineers with powerful tools for elaborating future-proof solutions to modern challenging demands.

HetNets are modern solution for service providers, brought to address emerging connectivity demands by hierarchically adding smaller and smaller cells [3]. The resulting setup allows user device to interact with the infrastructure via multiple radio access technologies (RATs). While more and more user devices are being equipped with multiple radio transceivers, this architecture allows mobile network operators to significantly improve network capacity by efficiently utilizing spectra of these radio technologies, thus offering higher quality of experience to their customers.

Thorough study of integration of new RATs and understanding of required intelligence sharing between user equipment (UE) and infrastructure, for efficient use of these technologies, are envisioned to be fundamental enablers for future 5G networks. A confirmed example of benefits brought by HetNet architectures could be performance improvements in an unlicensed-band network, like WiFi, by leveraging centralized control from designated entity in mobile network core, like 3GPP LTE [4].

The agility and efficiency of modern software development processes have made software elaboration, operation and evaluation principles very attractive to other technology fields. Software-defined networking (SDN) paradigm comprises a set of automation and abstraction concepts aimed to bring networking closer to its ultimate goal of interconnecting users and applications in the most efficient way, by providing entities that use the network with tools to shape network services to their needs [5]. Within the scope of this work, SDN vision aligns well with HetNet architectures in desire to optimize resource allocation through centralized decisions, based on end-to-end view of traffic flows, and in attempt to involve both user and network sides of communication into optimization process, while allowing UE to efficiently use all the available communication technologies. While the term SDN is relatively new and there is a lot of debate around SDN concepts going among researchers, engineers and equipment vendors, the principles of network control automation, abstraction from underlying network functionality and software-based flow switching laid foundation for prototyping within this research work.

Another networking solution proved useful in research and development of new architectures is the concept of *overlay* networks [6]. Building another logical network layer on top of existing transport network is useful in highly dynamic environments like data centers, or in cases, where major changes to *underlay* transport network are very expensive, like in large scale operator networks. Within this work, configuring tunnels over different provider networks and different technologies allowed to achieve desired connectivity without modifying underlying network protocols, which otherwise would require open access to a live cellular installation. The drawback is increased overhead in the network, but it does not interfere with research objectives and is acceptable for the scope of this work.

### 1.3. Overview of tools

– **Open VPN** is a network tunneling software. Open VPN is widespread due to its openness and availability for a variety of platforms. Particularly, interesting for this project is its capability to dynamically deploy and execute custom applications or scripts on connecting client, based on triggered events, and capability to run as remote gateway interconnecting multiple private networks.

– **Open vSwitch** is a multilayer software switch. Important feature for the project is its ability to expose forwarding functions to remote entities for programmatic extension and control via protocols like OpenFlow.

– **VMware vSphere** Easy to use and well-documented virtualization solution. Used to abstract and share physical server resources to multiple virtual machines. Important feature is its robustness.

– **Docker** is software packaging solution, allowing to isolate an application at OS kernel level together with all required dependencies, files and network interfaces in a standardized unit container. Used for persistent and automated deployment and runtime across various environments.

– **General Routing Encapsulation** (GRE) is simple tunneling protocol used to build overlay networks on top of IP networks. Important feature is its capability to encapsulate various protocols including Ethernet.

– **Network Functions Virtualization** (NFV) is a concept complementary to SDN. NFV suggests new way to implement and operate network functions, by abstracting them from hardware appliances and using virtualization techniques to package these functions as virtual machines running on generic hardware.

### 1.4. Goals and structure of this work

This work addresses several prototyping solutions produced with the utilization of cloud computing techniques. These resolve several important challenges in the streamlined delivery of connectivity with higher service experience over the multi-radio heterogeneous deployments to a variety of modern handheld and wearable mobile devices. The two major goals of the work are as follows:

1. Developing a framework for the centralized radio resource management in multi-radio heterogeneous networks with highly configurable and integrative methods.

2. Implementing network-assisted direct communications mechanisms over converged LTE and WiFi radio access infrastructure with flexible and dynamic performance.

Sections 2 and 3 cover the summarized work within the research on the enabling technologies, namely, Heterogeneous Networks and Device-to-Device Communications. The subsequent section describes the challenges faced during system design, the limitations of the current architecture and the proposed solutions to resolve these issues.

## 2. Centralized radio resource management in HetNets

### 2.1. Introduction and motivation

5G communications ecosystem targets to reach higher rates never challenged before, and to keep up with these rates, modern heterogeneous network architectures have already developed improved ways to integrate various RATs. Seeking new vectors for evolution, emerging paradigm of heterogeneous cloud radio access network (H-CRAN) merges RAT integration with advanced cloud infrastructures [7, 8]. This approach enables improved management on the network-wide scale, allowing to implement cross-cell radio resource allocation in a coordinated way. Recent research addressed the gaps in theoretical performance analysis and provided assessment and mathematical methodology for *real-time* optimization of cooperative radio resource management in H-CRAN [4]. Resulting algorithms allow to balance between throughput and fairness metrics in a flexible way, as well as might align with network operator's development plans. Also, this approach demonstrated some advantages over state-of-the-art multi-radio resource allocation schemes.

## 2.2. Realistic testbed implementations

The next phase of our research is to design the considered algorithms in a practical testbed, to then address their expected operation. In this subsection, we account for realistic demands while developing the proposed methodology. To this matter, it is required to integrate radio resource management solutions into the stack of protocols for contemporary LTE and WiFi systems.

### 2.2.1. Implemented R&S demonstrator

Recently, our research group at Tampere University of Technology (TUT) has completed underlying HetNet testbed configuration, based on LTE eNodeB emulator, CMW500, by Rohde and Schwarz[1]. In this experiment, CMW500 was readily set up for a scenario of offloading traffic from LTE to WiFi, to assess capabilities of CMW500 equipment for 5G research. Integration example demonstrated a simple handover scheme for WiFi-LTE data offloading case, where WiFi signal strength was main deciding factor [9]. CMW500 was configured with LTE FDD cell signaling on frequency band 7, downlink 2645 MHz, uplink 2525 MHz, total bandwidth = 20 MHz. WiFi AP and CMW500 were bridged at the same server, which was also configured to lease IP addresses to UE via DHCP. An RF Shield Box model CMW-Z10 was used to control WiFi link quality. The shield box was connected to CMW500 unit via RF1 COM port and served as antenna for cellular link. This way, UE placed into open shield box connected to gateway server via both LTE and WiFi and was able to use the link with a better quality. Closing shield box lid introduced over 80 dB of digital attenuation for WiFi, thus triggering UE to switch its data transmission to LTE. When the lid was open again, WiFi link quality became better than that of LTE, and traffic was offloaded back to WiFi. CMW500 capabilities to emulate LTE provider's network allowed to quickly setup a testbed for improved offloading logic design, for the purposes of 5G research. Further research targeted intelligent data offloading, by employing advanced LTE link parameters. **Figure 1** demonstrates a short video presenting the setup. The video is available at http://winter-group.net/rohde-schwarz-tutorial/.

### 2.2.2. H-CRAN architecture prototype

Current prototype implements a multi-radio scenario, where a cellular network is coupled with a WiFi access point, thus providing UE with a possibility to seamlessly alternate both RATs, or to effectively employ two of them simultaneously. Clearly, development of the considered technique demands access to the side of the cellular BS. It is also required that the UE is capable of communicating the relevant control information on the appropriate radio channels. However, development kits for contemporary mobile platforms that are available on the market offer only limited support for controlling the respective data flows and interfaces.

---

[1]Video presentation of the experiment with R&S CMW500: http://winter-group.net/rohde-schwarz-tutorial/.

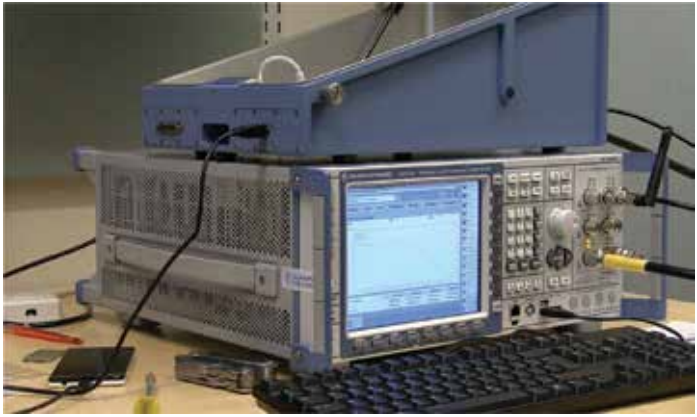

**Figure 1.** Setup with Rohde and Schwartz CMW500.

In this framework, the UE was connected to a traditional cellular system and a stand-alone AP of WiFi, both offering Internet connectivity with their corresponding ISPs. Simulating an integrated network for the two considered radio links, the UE starts two VPN tunnels to cellular and WiFi radio interfaces, respectively, while terminating at the aggregating node (the latter may, e.g., be placed onto the Internet). The said aggregator, establishing two links, mimics a packet gateway module residing in the LTE system.

Aggregating both radio links at the side of the UE, the Open vSwitch (OVS)[2] was utilized. Accordingly, vSwitch daemon, running on the UE, maintains links that correspond to WiFi and cellular connections in addition to a virtual local interface, which becomes a binding point as far as the outgoing traffic is concerned. The latter is generated by the applications with the help of *sockets* API. Here, the OVS specification requires that all of the connections in the virtual switch are capable of working with Ethernet headers. Note that it may not be the case in a cellular system exposed to the network as a point-to-point RmNet interface.

However, the discussed aspect might be resolved by upgrading vSwitch to enable the type of interfaces at hand, or using a dedicated module for skipping the processing of the Ethernet headers with an offset. Our testbed follows the lines of adding an extra tunneling layer on top of the current VPN with a Generic Routing Encapsulation (GRE) tap tunnel (known as Ethernet GRE or EGRE as well) because it offers implementation simplicity. For the sake of consistency, the GRE tap layer has also been implemented for WiFi VPN. The respective stack of the protocols for the UE is summarized in **Figure 2**. A similar approach is followed at the aggregating node on the side of the operator (simulated). Correspondingly, on its highest level of abstraction, our considered architecture comprises two switches connected by two redundant links. The core topology of our system is displayed in **Figure 3**.

---

[2]Open vSwitch website: http://openvswitch.org/.

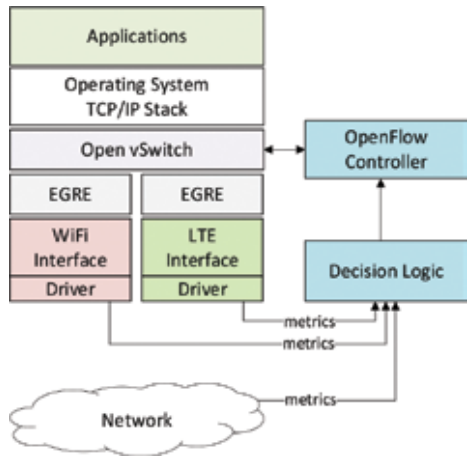

**Figure 2.** Proposed testbed topology for multi-RAT HetNets.

Our considered design allows for developing a controller that is capable of maintaining the forwarding table of vSwitch by relying on the OpenFlow protocol[3], and hence deliver the required levels of per-flow resource allocation (named here traffic steering) based on dynamic demands. A stand-alone module aims to collect information regarding the state of the underlying physical links from the involved RATs, such as received signal strength indicator (RSSI) and radio signal strength. The controller uses these data to efficiently assign new flows to either of outgoing interfaces. Coming back to the analytical algorithm implementation, this means that it is possible to transfer implemented resource allocation logic (e.g., relative fairness scheme) into the OpenFlow controller and specify control channel interface (which is LTE in this case).

### 2.2.3. Infrastructure for the prototype

The network side entities were installed as virtual machines (VMs) running on virtualization platform, built with vSphere[4] from VMWare. This approach allowed to deploy and interconnect architecture elements in a highly flexible and dynamic fashion.

For the research scenario, WiFi access point was connected to the Internet through university network and configured as a router between wireless and wired sides. UE was connecting to both WiFi and LTE networks and had Internet connectivity through either network. Using specific static routes, the UE was contacting particular VPN gateways over certain technology, that is, IP address of the gateway terminating LTE side VPN was routed via LTE network, with similar behavior for WiFi side. On the network side, traffic to the VPN aggregator node was chained by the platform through an entry node switch that was splitting traffic destined to LTE VPN and WiFi VPN, and then passed through WAN link emulator node. Details of such service chaining are illustrated in subsection 2.3.

---

[3] OpenFlow website: https://www.opennetworking.org/.
[4] VMWare vSphere solution: https://www.vmware.com/products/vsphere/.

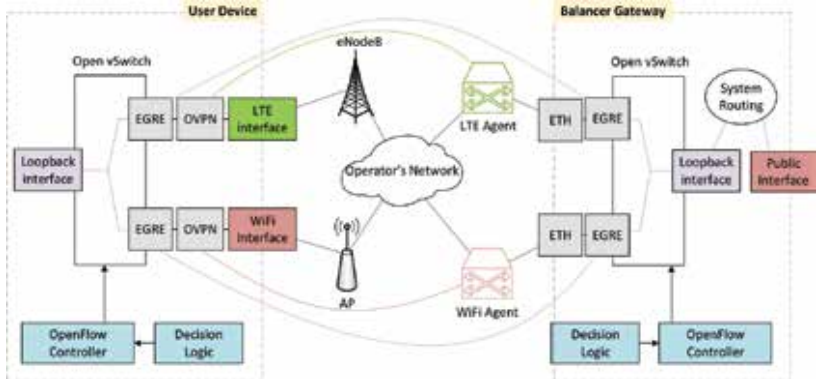

**Figure 3.** Overall architecture of prototype implementation.

## 2.3. Technical details

Most important features of current prototype implementation are as follows: software-based switching, network control automation and *overlay* network over regular provider networks. The following section summarizes how flow-based switching was implemented with Open vSwitch software, and network control was performed with OpenFlow controllers. Next, design of VPN and processes of measuring and influencing link conditions are outlined. Significant work was carried out to prepare UE to handle these tools — the modifications to the UE and procedures to enable it for research scenarios are covered as well. Finally, architecture is summarized with the workflow of a research scenario.

### 2.3.1. Toolchains and custom software on user equipment

An appropriate UE platform for the purposes of implementing our testbed has been the SailfishOS[5] running on Jolla mobile phones[6]. It combines *Linux* and *MER* software[7] as well as enables a powerful software development kit (SDK). The platform in question delivers all of the necessary tools to construct customized kernel modules, together with the generic GNU\ Linux software. Therefore, by installing OVS, GRE and virtual Ethernet (VETH) modules, in addition to OVS database and userspace tools, was rather straightforward. The software in SailfishOS is packaged with RPM Package Manager (RPM) (originally Red Hat Package

---

Manager, now a recursive acronym), which provides powerful tools to integrate third-party software and to install additional components.

Some of the tools used in prototype require integration at OS kernel level, which means that UE kernel has to be built with corresponding functionality enabled. Customization of MER-based Sailfish OS is performed through Platform SDKs provided by the project. These platform SDK tools include compilers, Scratchbox2 cross-compilation toolkit, MIC image creator, Zypper package manager and other instruments to make development easier. In this setup, SDKs are run as an image in a *chroot* environment; however, SDK can also be used as a dedicated virtual machine. SDK was installed on development machine from a *rootfs tarball* containing essential tools for MER platform development. The goal of using platform tools was to compile GRE, Open vSwtich, and VETH kernel modules for SailfishOS running on UE. As a basis for the new kernel configuration, authors extracted configuration file from running system of a Jolla device. Configuration file was edited to include appropriate modules and was used for the build process, resulting in corresponding kernel object (.ko) files compiled. These module files were transferred to the UE and loaded into running kernel as part of prototype implementation. Another feature of the Platform SDK that was useful for us is that provided *cross-compiler tool-chain* can be used to build custom userspace software, which is otherwise not available in official repositories. Using this *toolchain*, authors were able to compile userspace part of OVS software.

### 2.3.2. OpenFlow software switch

OVS is a multilayer virtual switch, designed to target at multi-server virtualization deployments. It provides a lot of important features for highly dynamic environments at large scales; however, for this prototype, the most important feature of OVS is that it adheres to the emerging SDN paradigm and thus supports forwarding based on flow tables and can use OpenFlow as a method of exporting remote access to control traffic forwarding. These features enable flexible placement of traffic control entity (at UE itself, on network side, or combination of both), and per flow control over traffic forwarding, meaning that certain sessions can be selectively placed on WiFi or LTE interfaces in a dynamic way. OVS consists of kernel side module, responsible for actual packet forwarding in the *data-path*, and userspace daemon, interacting with network state database and exposing control functions to external entities.

OpenFlow-based SDN vision implies that forwarding tables of a device should be maintained by a separate entity—the controller. This controller is responsible for making forwarding decisions and programming them into the flow tables along the data path, using Open Flow protocol. The solution in this installation was based on POX controller—a networking software platform written in Python[8]. Controller assumes the management of UE's forwarding plane consisting of 3 interfaces—WiFi interface, LTE interface and internal system interface. Main task for controller is to run one of the optimization algorithms supplied by researchers. Algorithm implementation receives metrics fed from the software, performs measurements of network conditions on both radio interfaces, and reallocates incoming and outgoing flows accordingly.

---

[8]POX controller: https://openflow.stanford.edu/display/ONL/POX+Wiki.

### 2.3.3. Network side

In the absence of access to an operational cellular network installation, researchers had to simulate joint control of WiFi and LTE networks for UE using VPN tunnels. This testbed used Open VPN owing to its simplicity and wide platform adoption. Clients would connect to dedicated VPN anchors through WiFi and LTE links, this way allowing the laboratory network to control setup of tunnel interfaces in a similar way as real operator network would control setup of physical links.

The authentication in VPN network is based on Rivest-Shamir-Adleman (RSA) certificates, as opposed to pre-shared key-based authentication. Using EasyRSA tools, authors have created certificate hierarchy where every user device would receive its own certificate to authenticate in the VPN network, and, based on information in that certificate, network would uniquely identify the user and configure it with predefined features.

Network-wise, VPN anchor was configured for *subnet* layout. Unlike in *point-to-point* layout, where a dedicated subnet is allocated for every client-server link, in *subnet* layout, all clients share common address space which slightly facilitates routing on the server side for prototype purposes. During start-up process, VPN agent places policy-based routing (PBR) rules into routing subsystem, to selectively route traffic from VPN network to the gateway facing the Internet to provide external connectivity.

To separate operation domains of WiFi and LTE side VPN servers, without the need to run them on different virtual or physical nodes, both processes are executed on the same VM, isolated in dedicated Docker[9] containers. Docker is a set of tools that provide convenient way to manage Linux container (LXC) environments. Every container running *openvpn* process has in its namespace a dedicated host network interface, facing the client, and a VETH link bridged with host network. This way, *openvpn* container acts as a router between host side network, leading to internal operator network, and client network behind VPN network. Detailed network layout of the VPN server is presented in **Figure 4**. Container mounts into its internal filesystem a directory with corresponding configuration files and certificates (e.g., LTE directory for container running LTE side) and executes *openvpn* process according to these configuration files.

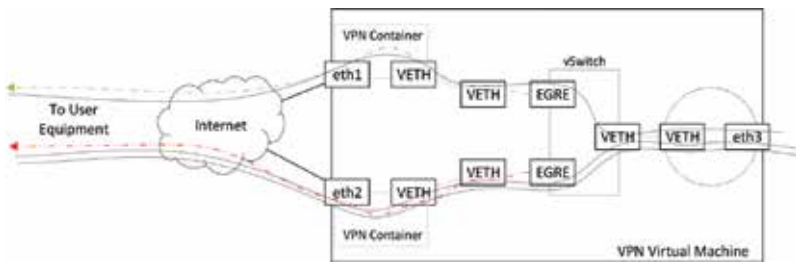

**Figure 4.** Detailed network architecture of VPN server.

---

[9]Docker platform: https://docs.docker.com.

Most of commercial-off-the-shelf (COTS) WiFi access points are shipped with proprietary, closed source operating systems, and are built with network chips run by closed drivers from hardware vendors. This way, access point becomes a "black box", exposing enough control to set up the network, but prohibiting any new functionality on top of the hardware. For an access point to be useful in such research, both operating system and wireless interface firmware must be open, to allow modifications and give access to internal variables. A good example of such device is a dual band WiFi router from Linksys model WRT1900AC. For this research, the device was installed with OpenWRT[10] system, custom compiled to include WiFi drivers as modules, rather than built into the kernel, making it easier to dynamically add changes into driver's code. One of the most important changes to the operating system was to enable Linux *debugfs* filesystem that exposes multiple system values, like status of numerous transmit queues and current transmission rates.

This way, AP was prepared to send measurements of interest to the process implementing forwarding optimization logic, where the latter would combine WiFi state data with LTE state data, or any other information found useful by researchers, like prioritization of users and RANs, and would yield new forwarding rules for UE side and operator side.

To simulate various network conditions, all traffic to the emulated aggregator node on network side was routed through a dedicated VM running WANem—a wide area network emulator[11]. This allowed us to impose various WAN characteristics common to LTE networks, like network delay, packet loss, packet re-ordering and jitter to the links that are in fact running in research laboratory LAN. Such service chaining of traffic dedicated to one VM through another VM is not natively supported by basic feature set of available vSphere installation, so several safety policies had to be disabled, and in such cases, certain broadcast and unknown traffic should be suppressed by virtual switch, and greater caution should be paid when routing traffic, to avoid forwarding loops. Network scheme of traffic flows in the setup with WAN emulator node is demonstrated in **Figure 5**.

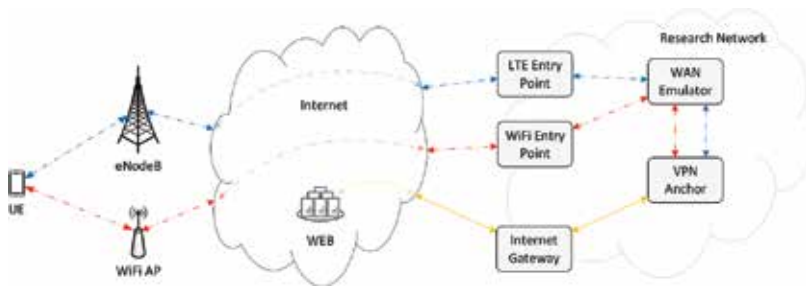

**Figure 5.** Setup with WAN emulation node.

*2.3.4. Architecture summary and workflow*

Network side of the deployment consists of a number of VMs serving as VPN anchor, WAN Emulation and Internet gateway. When VPN anchor VM boots up, it performs the following steps

1. starts VPN processes isolated in containers;

2. moves dedicated network interfaces to corresponding container namespaces;

3. creates a VETH pair for each container;

4. moves one end of every VETH pair to corresponding container namespace;

5. runs GRE tap tunnel to client over second end of VETH pair when client connects;

6. adds GRE tap interface to OVS;

7. creates another VETH pair for host networking;

8. adds one end of host VETH pair to vSwitch;

9. sets second end for routing to internal operator network.

When UE side and network side virtual forwarding planes are ready, both devices start measuring link qualities. Depending on coordination scheme scenario, controller software is started on either side and both virtual switches connect to it. Measurements from UE, AP and operator side are also sent to the controller. This brings the setup to final operational state, where controller would use all available metrics, to run resource optimization algorithm provided by researchers, and would re-allocate traffic sessions across available links accordingly.

## 2.4. Some numerical results

In this subsection, we provide a comparison of prospective H-CRAN deployments with full UE-centric network control methods as well as network-assisted multi-radio integration architectures that have a lower degree of manageability. Focusing on the illustrative example of the UE-centric solution, we analyze the max-usage control scheme. Here, each UE attempts to occupy all of the available radio resources on all possible RANs that it may connect to. Such a strategy is greedy, but it does not impose tight signaling requirements on neither the network nor the UE, as well as performs reasonably well from the individual user throughput perspective. Most importantly, this option is very easy to implement.

The second scenario that we have taken into account is a network-assisted resource allocation scheme. Basically, it exploits the concept of cell range expansion. The main idea behind it is to manage the efficient thresholds of association to WiFi and LTE cells for those UEs that reside under macro coverage. The approach in question allowed for incorporating the load variation effects across the individual RANs by effectively increasing/deceasing the UE numbers that are connected to the respective cells. At the same time, a separate decision-making entity implemented at eNB together with the corresponding signaling to supply UEs with the relevant assistance information is required when utilizing this method.

In this implementation, we force the UE to prefer a single RAN at a time [4]. Such a decision was made mainly due to its practical feasibility, since otherwise said technique would lose to the UE-centric approach by assigning the maximum radio resources to the users with better connectivity conditions, while potentially not allowing the users with worse channel to even attempt the WiFi/small cells. In a nutshell, the description of the considered algorithm is the following: first, the user connects on the WiFi layer, but should the resultant signal turn out to be poor, the UE will then probe the pico connectivity layer and, if there remains no other appropriate opportunities, the user will then be handled at the macro layer.

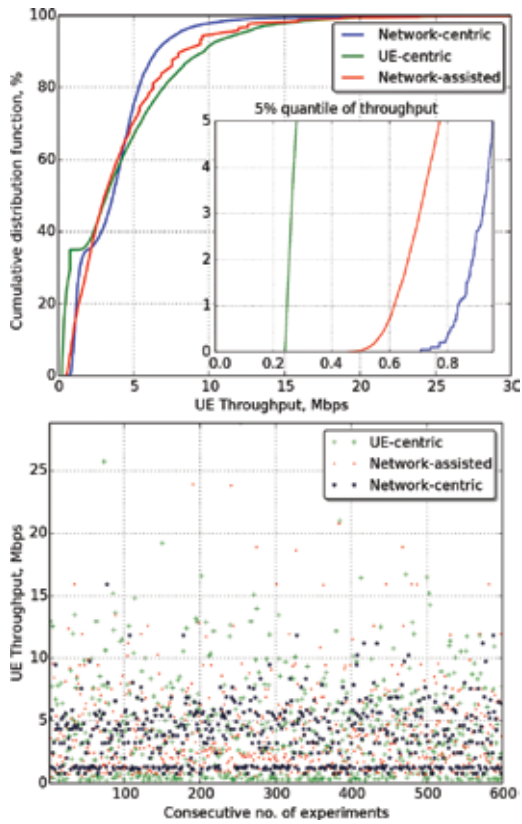

**Figure 6.** Per-UE throughput CDF.

In addition, we take into account a network-centric scheme made available by the H-CRAN that centrally manages the cross-RAT radio resources subject to the actual connection quality of the user. In this case, both network-assisted and network-centric techniques offer more attractive performance in the H-CRAN scenarios. In addition, the centralized solution offers better control flexibility.

We analyze all the three scenarios from the per-UE throughput point of view and the results are given in **Figure 6**, top subplot. The threshold for the network-assisted scheme for both LTE and WiFi small cells has been set to the maximum sensitivity level. Concerning the network-centric solution, resource management method takes as an input the vector of spectral efficiencies for each user first without any modifications. As a result of this default configuration, all three solutions demonstrate comparable results in terms of the average throughput, but the network-centric mechanism performs much better at the cell edge.

Next, we have a closer look at the results for the per-UE throughput given in the bottom sub-plot of **Figure 6**. It is clearly visible that the UE-centric scheme is characterized by lower fairness as compared to the network-centric/assisted methods based on 5%-quantile sub-figure embedded into **Figure 6**, top right subplot.

The last analyzed scenario is the network-assisted case. Here, the previously discussed effect is not visible as the UEs have only one active connection at all times. At the same time, we may observe another interesting effect: step-wise behavior may be seen in the upper part of the curve. It could be explained based on the difference between the UEs with very good WiFi link qualities—various numbers of UEs associate with different APs causing variations in throughput if the number of connected UEs is low and the SNR of each user is high.

Summarizing, the centralized resource optimization and control technique is suitable for real-time resource allocation across multiple RANs in H-CRAN environment. It brings high level of flexibility and adjustable balance between the overall H-CRAN system throughput and fairness of the resulting allocations.

## 3. Network assisted device-to-device communications

### 3.1. Introduction and motivation

The increase in mobile data traffic pushes network operators to seek ways to relieve congestion on their infrastructures. A natural way to mitigate the shortage of available radio resources is to deploy an increasing number of various sized BSs; however, such approach is costly and faces some practical challenges. An alternative strategy would be to leverage the capability of modern UEs to establish simultaneous connections via different radio links, and to enable traffic offloading from cellular network onto D2D connections in unlicensed bands like WiFi [10]. The problem with WiFi is that there is no fast and efficient method of service or device discovery built into the protocol, as well it lacks functionality to efficiently manage multiple D2D links [11]. Recent research [12] demonstrates that these limitations can be solved by a certain amount of network assistance made available to D2D communications.

The offloading principles, discussed in Section 2, are not limited to be used only in network operator's infrastructure. Similar techniques could be utilized to improve network performance, by offloading communications between proximate users from infrastructure to direct link between users, or D2D links. This section covers research and prototyping activities performed in exploring improvements to user traffic offloading from infrastructure network onto direct D2D links brought by assistance from cellular network.

## 3.2. Implementing traffic offloading prototype

Within proof-of-concept implementation, authors of this work were responsible for making required changes to UE operating system, elaborating offloading algorithm based on IP routing, setting up and maintaining infrastructure to run prototype components and implementing UE side of the prototype [12]. Next subsections elaborate on implementation of these prototype components.

### 3.2.1. Android networking subsystem

Android, as a Linux-based system[12], allows to have simultaneous connections over more than one radio interface. Once the LTE and the WiFi links are activated, the UE operates on two interconnected networks, while having a single default gateway to transfer its traffic outside these systems. In the considered case, it may be feasible to connect to another peer on the WFD channel only when the IP header's address of the destination node is the WFD address of the said peer, whereas the address of the source is that of the originating UE. Also, WFD link uses private address range that is not reachable through anything else than WFD link; once the link is disconnected, the peer becomes unreachable. For the suggested D2D architecture, a device would be required to be capable of reaching the public IP address of its peer on the LTE interface via the WFD connection.

As a major target of our constructed solution, we aimed at its design transparency for the currently available applications. Accordingly, we contribute to the user adoption levels of the proposed technical implementations. Along these lines, any changes at the physical layer may not be appropriate, as these are tightly coupled with radio interface driver. The same applies to link layer—implementing new functionality, there would make the solution vendor-specific. Since existing applications heavily rely on current transport layer protocols, modifications in transport layer are also not considered. On the other hand, system offers various tools to make changes in IP layer (or network layer). This would allow to leave underlying radio interfaces as-is and would enable application developers to use same routines to obtain network access as before. IP addresses are in a way bound to the physical interfaces, but selection process is made without direct interaction with interfaces. This would allow us to create an interface independent solution without modifications of upper layers.

The default configuration of an Android system allows to have routes with multiple gateways, but one of them is inserted into routing table with lower cost than the others. This way no load-balancing is performed and only one route is used. In case of WiFi link and LTE link, or cellular link in general, the route through LTE gateway is preferred for the Internet connectivity because WiFi link and, especially, WFD do not guarantee Internet connectivity at all. Therefore, changing cost of default route might cause unreachability of the Internet for all applications in the mobile device.

---

[12]Android OS: http://www.openhandsetalliance.com/android_overview.html.

### 3.2.2. Traffic offloading based on routing

The proposed solution is based on allowing mobile device to route IP traffic as usual, and then inject more specific routes for particular peers into routing table. As Android system is based on Linux kernel, it is possible to enable routing functions by modifying the value of system variable *net.ipv4.conf.all.forwarding* from 0 to 1. After this change, Android mobile device gains capabilities of a generic router, known from computer networks. This way, mobile device can forward packets from one interface to another. An interesting point to note here is that it allows to send IP packets with source address being LTE interface public IP address and destination address being peer's WFD private IP address, and IP layer of Android system will send them through WiFi interface.

There are, however, several issues with such approach to link selection that need to be solved. The first issue arises with application use of network sockets. When an application needs network connectivity, it requests the service from operating system using sockets API, and operating system chooses to use one of the available IP addresses as source address. Source IP address, source transport layer port, destination IP and destination port must remain the same throughout communication, which cannot be guaranteed in case session is bound to WiFi interface IP addresses, because interface can be disabled at any moment. Another issue is with routing private networks through operator's infrastructure—if a session is bound to WFD's private IPs, it cannot be switched to LTE interface because the operator's infrastructure prohibits routing packets to this private IP range.

As a workaround for prototype demonstration, a set of OpenVPN tunnels has been elaborated from devices to an anchor point in the network infrastructure, thus allowing routing of private networks through operator's core. In a real world deployment, however, this would not be necessary, as operator could set up internal routing according to D2D link networks, issued by D2D server. Also, the system creates an *overlay* GRE tunnel bound to loopback interfaces on the devices. Loopback interface is meant to be a constant anchor point for applications, and only *overlay* tunnel endpoints will be rerouted, ensuring that session IP addresses remain the same all the time, regardless of *underlay* interfaces used, and thus providing service continuity.

Since only communication with one particular peer needs to be offloaded, a route can be inserted into routing table stating that only the peer's loopback IP address is reachable through peer's WFD private IP address. Insertion is performed by the command iprouteaddPEER_LOOPBACK_IP/32viaPEER_WFD_IP. Once this command is accepted by IP routing layer of Android, all traffic with destination IP address PEER_LOOPBACK_IP will be forwarded to WFD interface and this way *overlay* tunnel traffic will be sent over WFD.

The insertion and removal is performed by management application that is running in the background. Both actions are invoked upon a corresponding command from D2D server, but route removal can additionally rely on local channel quality measurements. Within this implementation, the only channel quality metric considered is the RSSI. When reaching a particular threshold in RSSI, route can be inserted or removed, thus selecting D2D or infrastructure channels, and reaching even higher threshold can trigger complete link disband.

This injection of specific routes into routing table does not have to be performed symmetrically on both devices participating in D2D offloading. Also, route injection scheme is not limited to be used only with a single peer.

### 3.2.3. Infrastructure for the prototype

The service relies on three parts: (i) Client side phone application; (ii) Content register service; and (iii) D2D server, as it is shown in **Figure 7**.

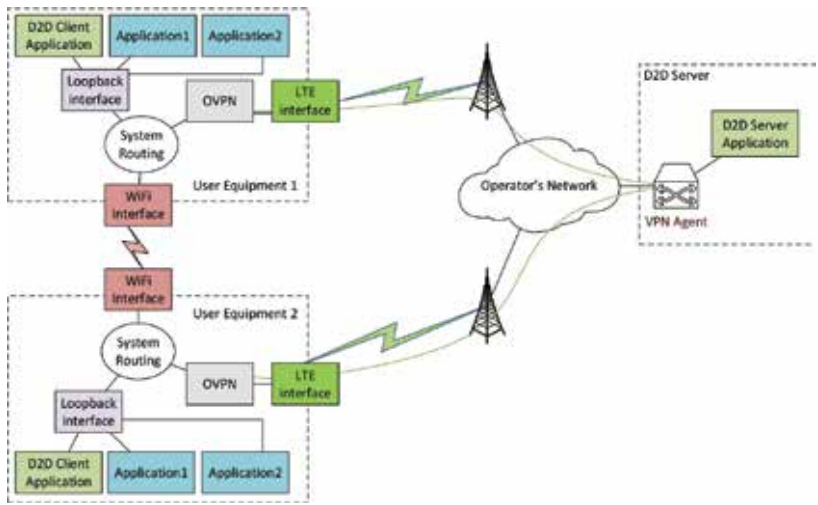

**Figure 7.** Prototype setup for D2D architecture.

The content register is assumed to be any platform providing data sharing services to its users e.g., social networks like Facebook, Google+, etc., with a difference that in this case only metadata about content is shared in form of a hyper-reference, instead of actual data [13]. D2D server, on its turn, is designed to be run by network service provider. It is worth noting that besides suggested architecture layout, the service is flexible enough for both entities to be run by a single authority or even on the same host.

To emulate integration of the service with providers, the two applications at the side of the server have been designed within the cloud infrastructure of the research group. Accordingly, a pair of virtual machines have been deployed to assume the roles of the Content Register and the D2D server, respectively. Here, the Content register component has been developed in PHP and accommodated by the web server running Apache. Hence, our constructed application is a generic webpage, which offers any pre-registered UE an opportunity of posting its willingness to share certain information, or of identifying the desired locations of thus shared information.

User posts only information required to access data, rather than its location, that is, sharing protocol and port number, while IP address that locates the data is handled by D2D server. The part that glues service sharing and service discovery is the introduced in the URI protocol identifier, which is returned by the Content register module—"d2d://". The UEs are made capable of interpreting the protocol scheme in question in terms of a request to initiate the Client application. The latter is able to interact with the D2D server to convert the username of the peer node that serves information into an appropriate IP address. In turn, the D2D server has been realized in Python in a form of a standalone application by utilizing HTTPS as a means of transport for the control messages. The system considers that the mobile data connection of a user is active all the time throughout the use of the service. Our proposed solution indicates that the content register as well as the D2D services are easy to integrate seamlessly into the current web service infrastructure.

End user devices used in service demonstration were Sony Xperia ZL phones provided by Sony Mobile. Android, being an open source mobile platform, provides needed flexibility in configuration and available tools to fulfill requirements demanded from user devices. One of the main features user devices should have is the ability for both mobile data link and WiFi link to be up simultaneously. Due to energy consumption constraints, most of consumer devices on the market restrict their network connectivity to use only one of the available connections at a time. To enable both interfaces in the system, authors had to bypass native Android service controlling WiFi and interact with WiFi driver directly. As Android system to some extent is derived from GNU\Linux, it easily gives us needed tools—*wpa_supplicant* interface controlled via *wpa_cli* utility. Another issue is that in order to integrate with these utilities Android has to be compiled with *userdebug* feature enabled, and stock firmware provided by Sony for their devices does not have this property. Therefore, another option to consider was to build the system based on source code from Android Open Source Project, using proprietary binary drivers released to public by the vendor. However, resulting system at that time lacked radio drivers, needed for intended operation of mobile data connection. Final choice for the user device platform was popular aftermarket firmware Cyanogenmod[13], based on Android and maintained by FreeXperia group.

Another requirement for the mobile system of the end user is to be able to receive the incoming links on a mobile data connection. Accounting for the fact that the network operators utilize private IPv4 address pools to distribute across user devices as well as provide Internet access by means of a NAT, using the services that run on the user devices externally from the outside of the local connection may not be feasible with such a setup. One of the possible solutions to overcome this issue would be using IPv6 addresses, but local service providers at the time being did not provide IPv6 connectivity options to the extent suitable for practical demonstration of D2D service. Another tested option for demonstration was encapsulating mobile data link of both communicating devices inside a VPN tunnel to a common VPN server, thus moving both devices into the same IP subnet. Also, discussing the issue with local service provider—TeliaSonera Finland Oyj—researches were offered a certain access point name that would provide user devices with a publicly routable IPv4 address.

---

[13]Cyanogenmod distribution: https://wiki.cyanogenmod.org/w/Main_Page.

### 3.3. Prototype implementation: technical details

This subsection covers deeper technical details of implementation for service architecture prototype described above. As mentioned before, the whole system comprises three entities—UE, D2D server and Content register. Besides that, prototype infrastructure contains a node serving as anchor point for VPN layer, and a node serving as gateway for Internet connectivity.

#### 3.3.1. User equipment

User side of the prototype was implemented in two layers—main network assisted offloading logic, implemented as native Android application, and functional layer performing all interactions with Android OS implemented in Bash shell scripts.

Bash scripts implement functions like inserting/withdrawing routes, enabling WFD links and establishing tunnels to the infrastructure. Also, they provide an abstraction layer to D2D Client application, while client application in turn establishes control channel with D2D server and interacts with the server using HTTPS to execute offloading algorithm steps.

The workflow of suggested solution starts with running D2D client Android application. The first phase for application is to invoke initialization *shell* script to bring UE to a state, where it would be ready to interact with infrastructure and D2D server. As mentioned in subsection 3.2, system WiFi service will not allow to have both radio interfaces up simultaneously, so the initialization script will stop this service and will run a new copy of *wpa_supplicant*, with configuration files and parameters needed to prepare WiFi *p2p* interface. Also, initialization script loads necessary additional kernel modules. The last phase of initialization script is to start OpenVPN tunnel, and this phase creates a blocking dependency, where initialization script in order to complete needs OpenVPN tunnel to successfully establish connection with infrastructure. This is resolved by implementing a *trap* in the initialization script, so that it can safely *sleep* after starting OpenVPN tunnel, and be resumed by *openvpn* process after tunnel is up. Finally, initialization script will mark its completion in a process *pid* file, so that other system parts can later check if initialization has already been performed.

After *init* phase is performed, mobile device has established a control channel to D2D server via the tunnel and is ready to setup WiFi direct connections, when instructed to do so. D2D Client application regains control, registers itself to D2D Server and UE becomes available for offloading. Depending on the role of UE as publisher or consumer, D2D Client application is invoked either manually or by a link from Content register (see subsection 3.3), respectively. Since components responsible for the network-assisted offloading features are completely decoupled from application layer, in case user has published some content, they must ensure that the application that is actually serving it (e.g., a file server, game server or multimedia streaming application) is running and is awaiting incoming connections. In case user followed the link from Content Register, D2D Client application subscribes to server with data from the link. At this stage, network knows that one user is ready to serve content and the other one is willing to receive it with offloading features enabled.

The next step to prepare devices for traffic offloading is to start *loopback* interfaces and *overlay* GRE tunnels. When both peers public LTE (or OpenVPN in case of this implementation) IP addresses are known, D2D server instructs both devices to start a *loopback* interface, to add routes to peer's *loopback* interface over cellular network and to bring up a GRE tunnel bound to those *loopback* interfaces. These functions to operate *loopbacks* and GRE are implemented in corresponding *shell* scripts, as well. *Overlay* tunnel endpoints will be rerouted to switch traffic from one radio network to another providing service continuity as described in subsection 3.2. Addressing for the *loopback* networks and *overlay* network is carried out by coupling pre-defined prefixes with host portion of device's LTE IP address.

When D2D server detects that two users, who have previously subscribed to be assisted in D2D communication, are now in appropriate conditions to use offloading, server will generate a group name and a pre-shared key for WFD and will instruct one device to start WFD by invoking a Bash *shell* script implementing start, join, leave, or remove functions. The script, in its turn, will use *wpa_cli* tool to interact with *wpa_supplicant* process started at initialization phase. As in case with *loopback* and *overlay* network addressing, for this prototype implementation, when WFD link is established, it is configured in a stateless manner with IP addresses from 10.1.1.0/30 range with 0.1 assigned to group owner and 0.2 to WFD client.

At this point, both devices have two radio interfaces up and are ready to offload traffic, by rerouting *overlay* tunnel endpoints through either of these interfaces. When instructed by the network, D2D Client application will insert or withdraw routes by invoking corresponding *shell* scripts. D2D server keeps track of UE location, and when the distance and RSSI metrics reach predefined levels, server will instruct devices to use WFD, to fallback to using cellular network while keeping WiFi radio on or to completely shutdown WiFi interface.

Upon users' request D2D server can de-register them and stop following their proximity.

### 3.3.2. D2D server

Server is running Python application listening for HTTPS requests on port 8099. All the relevant data from clients are passed as HTTP parameters and server is running an internal database to store registered users and their associated data like public IP addresses and current location. In present prototype implementation, UE will periodically poll the server about status changes and server will reply with corresponding instructions according to current network state.

### 3.3.3. Content register

Content register is a PHP application, running embedded database to authenticate users and store metadata about their shared content. When users wish to post some service available for access and ready to be offloaded to D2D link when appropriate, they log in to Content register and publish access protocol and port number for the service, e.g., http/80. When another user sees this intent to share, it can request shared content, and Content Register will present the metadata in form of a URI with d2d://scheme and embedded information about peer username, target D2D server, shared service protocol and port *d2d://amie@d2d.winter.rd.tut.fi@simhost.winter.rd.tut.fi/webcam:8080*.

Web browser in UE is setup to recognize d2d://URI scheme as bound to D2D Client local Android application. Hence, following the link given by Content register will start D2D Client application that will establish control channel to its D2D server and will register its intent to engage into D2D connection with the named peer, when network conditions are favorable to do so. Optionally, user may choose to start using shared content straight away and be offloaded when possible, or to wait before content is available via direct link.

### 3.3.4. Utility nodes

All server side components are running in virtual environment, implemented with vSphere from VMWare. Both D2D server and Content register are running on dedicated virtual machines. To terminate OpenVPN connections, another dedicated VM was installed. The VPN VM is selectively routing traffic from VPN connections, either to other peers using *openvpn* "subnet" type of configuration, or to the Internet gateway using Policy Based Routing. The Internet gateway is another VM, running Brocade Vyatta[14] software router and performing NAT to provide Internet connectivity to internal networks.

## 3.4. Some numerical results

In this subsection, some results on the performance evaluation of the discussed network-assisted D2D system are given. These comprise the measurements executed on the real deployment (discussed above), as to facilitate the future D2D implementations. The primary goals of said evaluation are as follows.

– To indicate the bottlenecks that could potentially hinder the adoption of the D2D connectivity by the future wireless technology.

– To establish the appropriate performance bounds and limitations for the D2D technology and outline which services could be most suited for direct communications in contemporary and near-future markets.

In order to provide the best available accuracy, the considered D2D protocol was split into individual messages and we further performed our measurements on a significantly large sample set [13]. First, we measured latency between the client device equipped with a USB LTE dongle and the D2D served deployed behind the UGW. The distance between the client and the eNodeB has been fixed to 10 m for the entire time of measurements. We utilized Iperf tool to assess the maximum throughput of a particular radio access technology. Later, it has also been used to create a preset load on both uplink and downlink channels, thus emulating the required measurement conditions.

The measured network latencies are displayed in **Table 1**. We may notice that if the overall loading on the LTE network does not exceed 90%—the round-trip times between the client and the D2D server range from 13 to 25 ms. Hence, cell load of up to 90% does not have any evident effect on the D2D signaling procedure.

---

[14]Vyatta Router: http://vyos.net/wiki/Main_Page.

| Cell load | Idle | 50% | 90% | 99% |
|---|---|---|---|---|
| Measured RTT (ms) | 13.00 | 25.00 | 25.00 | 1248.00 |
| UL latency (ms) | 10.40 | 16.67 | 16.67 | 832.00 |
| DL latency (ms) | 7.00 | 8.33 | 8.33 | 416.00 |
| Stage 2 (ms) | 27.00 | 41.67 | 41.67 | 2080.00 |
| Stage 3 (ms) | 33.00 | 58.33 | 58.33 | 2912.00 |
| Stage 4—path switch (ms) | 7.00 | 8.33 | 8.33 | 416.00 |
| Full procedure (ms) | 67.00 | 108.33 | 108.33 | 5408.00 |

**Table 1.** Network latency measurements for different cell loads.

If the cell load is increased to the extreme conditions of 99%, that is, the input UE queues are overfilled, the resulting values range on the order of seconds. Hence, without the appropriate QoS support in LTE, the obtained latency would dramatically influence the proximity detection, as well as the overall user adoption of the considered D2D technology.

Further, we analyzed the delay values for different types of cell load, as it is shown in **Figure 8**. While these values remain negligibly small for up to 90% loading, the saturated UE scenario yields a major increase in the observed delay. In this case, it becomes hardly acceptable and ranges in seconds. In this extreme case, the implementation of QoS on the UE side is crucial to prioritize the D2D signaling messages. While there is some additional delay introduced, in almost all of the cases, the connection will be established on time for the users to reliably benefit from it before they come into physical contact, even when walking towards each other.

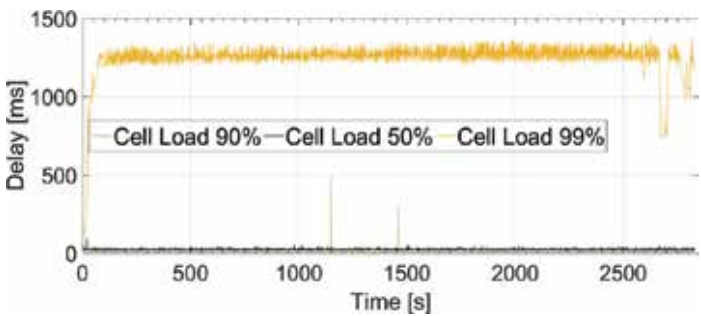

**Figure 8.** Network latency for alternative cell loads.

# 4. Summary of this work

## 4.1. Key lessons learned

This subsection highlights the challenges faced in the system design, the limitations of the current architecture and the network deployment choices.

Regarding the centralized radio resource management in **HetNets**, most of today's UE devices have both cellular and WiFi interfaces on board. At the same time, the need to reduce energy consumption has led equipment manufacturers to impose a limitation on the mobile UE's operating system to have at most one radio interface active at a time (except for the Bluetooth). Consequently, the emerging demand for improved wireless connectivity has convinced major equipment vendors to assist the developers in access and effective utilization of the heterogeneous wireless connectivity. A promising example of a research-friendly UE platform is Jolla phones running Sailfish OS. It was verified that the devices utilized in our trial have a flexible and open architecture, augmenting modern hardware with capable developer tools to allow for intended modifications on system level and thus achieve the desired degrees of connectivity.

As the next point, we were looking for the feasible options to integrate cellular and WLAN RATs to advance the vision of H-CRAN. In reality, an integration of a WiFi access point into the operator's network would require access to an open cellular BS, which is hard to acquire in the research environment today. Hence, we decided to encapsulate the corresponding radio links into separate OpenVPN tunnels utilizing two independent wireless technologies simultaneously.

Our prototype architecture operates as an advanced SDN solution and specifically employs the OpenFlow protocol to manage effectively the UE connectivity. The operational radio channels on a particular device are considered to be connected to a common forwarding plane on the smartphone that is mimicked by utilizing the Open vSwitch software. At the same time, actually occurring forwarding decisions are made by the controller software based on a set of active measurements of the radio link conditions. In our trial, the controller is deployed on the device side but it could be easily moved to the operator's environment.

From the network assisted **D2D** communications perspective, the proposed solution follows the general D2D implementation guidelines considered in the standards. However, it has encountered several deployment issues, which have forced deviations from the reference descriptions. The existing architectural principles in the LTE system prevent from implementing the cellular assistance mechanisms in the manner consistent with the engineering expectations. Therefore, it was necessary to implement a number of careful steps to provide a working system based on the contemporary technology.

The resulting trial implementation is developed to offer the required packet-switched data services together with important options, including VoIP over the integrated LTE and WiFi deployment. The EPC has been dimensioned to allow for high data rate applications with the relevant QoS provisioning. For the calls (both voice and video), the switching function-ality is realized by utilizing the high capacity IP Multimedia Subsystem together with its key components. This is carried out for mobile as well as fixed users and enables connectivity with outside telephone/teleconferencing networks.

The ultimate target of the LTE test network implementation is to leverage a holistic and customizable mobile system allowing for prompt implementation and demonstration of the emerging D2D communications technology. We therefore efficiently utilized this important tool to confirm that LTE-assisted WiFi-Direct paradigm has already matured sufficiently to be implemented in the commercial-grade integrated LTE/WiFi networks.

## 4.2. Conclusions

This subsection concludes the chapter with a review of presented networking solutions for integrated heterogeneous wireless ecosystem. Wireless networks are constantly evolving in offered connectivity levels, thus strongly consolidating in our lives as a necessity. More and more devices are joining the network requesting continuous high quality service, which brings unprecedented challenges to network design in upcoming 5G era. Transformation of mobile user experience requires complex changes in both network infrastructure and device operation, where user experience is optimized taking into account surrounding network context.

Advances in modern cloud technologies offer a set of enablers for next generation networks, consolidating management and resource allocation under a single abstraction level exposed to automation and orchestration frameworks for live optimizations. Primary focus of this work was to support current research on novel integrated multi-radio network architectures, by elaborating a set of prototypes and testbeds.

Solutions in Section 2 support research performed on optimization problems in cooperative radio resource management in H-CRAN.

– Implementation is using COTS equipment making said resource management available already today and demonstrating that no hardware changes are required.

– Prototype allows user device to be connected to both LTE and WiFi at the same time. Setting battery life issues aside, this feature significantly extends user connectivity.

– Resource allocation is controlled using multidimensional view on the network, employing nonconventional parameters like cross RAT loading.

– System efficiently utilizes available radio connections by properly assigning traffic flows, e.g., streaming high resolution video over WiFi while downloading important software update over LTE.

– Deployment is flexible and loosely coupled to enable portability, automation and scalability of the solution.

Solution in Section 3 enables direct communication between proximate users to offload traffic from network infrastructure and create new proximate services.

– Implemented using COTS equipment making the solution available without additional modifications to hardware.

– Allows to leverage peer proximity to offer richer set of user applications and services.

– Increases user awareness about proximate peers, services and content. With this feature, the user is not limited to only consume the services offered by the network provider.

– Network assistance ensures efficient use of short range radio interfaces by facilitating discovery and authentication functions. This significantly relaxes energy constraints currently seen in technologies like WiFi Direct.

– Deployment using conventional web mechanisms enables prompt integration of new entities and connectivity solutions.

This chapter demonstrated application of novel and traditional networking concepts in implementing prototypes and demonstrators for emerging wireless network architectures and ecosystems. Resulting implementations supported research in centralized radio resource management and network assisted traffic offloading. Proposed approach demonstrated efficient prototyping for research scenarios using automation and abstraction concepts, decoupling network functions from hardware and modifying open source components. Use of modern software packaging and delivery techniques, like hardware level virtualization, OS kernel level virtualization and automated configuration, significantly accelerated implementation process and made resulting architecture components highly reusable for future projects.

## Author details

Roman Florea[1], Aleksandr Ometov[1], Adam Surak[2], Sergey Andreev[1]* and
Yevgeni Koucheryavy[1]

*Address all correspondence to: sergey.andreev@tut.fi

1 Tampere University of Technology, Tampere, Finland

2 Algolia, France

# CLOUD COMPUTING
## ARCHITECTURE AND APPLICATIONS

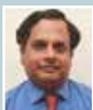

Edited by Jaydip Sen

Prof. Jaydip Sen has around 25 years of experience in the field of communication network protocol design, network analysis, cryptography, and network security and analytics. He worked in reputed organizations like Oil and Natural Gas Corporation Ltd., India; Oracle India Pvt. Ltd., India; Akamai Technology Pvt. Ltd., India; Tata Consultancy Services Ltd., India; National Institute of Science and Technology, India; and Calcutta Business School, India. Currently, he is associated with Praxis Business School, Kolkata, India, as a professor in the Department of Analytics and Information Technology. Prior to this, he was a lead scientist in the Innovation Lab of Tata Consultancy Services, India, where he led the research and development activities in wireless communication, embedded systems and ubiquitous applications, design, and development. He has over 16 years of research and development experience. His research areas include security in wired and wireless networks, intrusion detection systems, secure routing protocols in wireless ad hoc and sensor networks, secure multicast and broadcast communication in next-generation broadband wireless networks, trust- and reputation-based systems, quality of service in multimedia communication in wireless networks and cross-layer optimization-based resource allocation algorithms in next-generation wireless networks, sensor networks, and privacy issues in ubiquitous and pervasive communication, big data analytics, R, Python, Hadoop and MapReduce programming. He has more than 120 publications in reputed international journals and referred conference proceedings and 6 book chapters in books published by internationally renowned publishing houses, e.g., Springer, CRC Press, IGI Global, etc. He has delivered expert talks and keynote lectures in various international conferences and symposia. He is a senior member of ACM and IEEE. He is also an active member of the security group of IEEE 802.16 standard body and has submitted a number of proposals for the evolving 802.16m standard and ETSI. His biography has been listed in Marquis Who's Who in the World every year since 2008. Prof. Sen obtained Bachelor of Engineering (BE) in Mechanical Engineering with honors from Jadavpur University, Kolkata, India, in the year 1988 and Master of Technology (MTech) in Computer Science with honors from the Indian Statistical Institute, Kolkata, in 2001.

In the era of Internet of Things and with the explosive worldwide growth of electronic data volume, and associated need of processing, analysis, and storage of such humongous volume of data, it has now become mandatory to exploit the power of massively parallel architecture for fast computation. Cloud computing provides a cheap source of such computing framework for large volume of data for real-time applications. It is, therefore, not surprising to see that cloud computing has become a buzzword in the computing fraternity over the last decade. This book presents some critical applications in cloud frameworks along with some innovation design of algorithms and architecture for deployment in cloud environment. It is a valuable source of knowledge for researchers, engineers, practitioners, and graduate and doctoral students working in the field of cloud computing. It will also be useful for faculty members of graduate schools and universities.

# INTECH
## open science | open minds

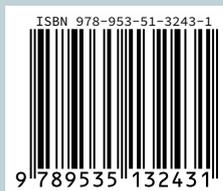

**INTECHOPEN.COM**